\documentclass[review]{elsarticle}
\journal{Journal of \LaTeX\ Templates}

\usepackage{lineno}
\usepackage{graphicx}
\usepackage{caption}
\usepackage{subcaption}
\usepackage{color}
\usepackage{amsmath}
\usepackage{amssymb}
\usepackage{cleveref}
\usepackage{stmaryrd}
\makeatletter
\renewcommand{\fnum@figure}{Fig. \thefigure}
\makeatother

\graphicspath{{./figures/}}

\journal{Journal of Composites Part B: Engineering}

\begin{document}

\begin{frontmatter}

\title{Creep rate based time to failure prediction of adhesive anchor systems under sustained load}

%% Group authors per affiliation:
\author{Ioannis Boumakis$^{1}$, Kre\v{s}imir Nin\v{c}evi\'{c}$^{1}$, Jan Vorel$^{1,2}$, Roman Wan-Wendner$^{1,3}$}

\address{%
$^{1}$ \quad Christian Doppler Laboratory, University of Natural Resources and Life Sciences, Peter-Jordanstr. 82, 1190 Vienna, Austria\\
$^{2}$ \quad Dept. of Mechanics, Faculty of Civil Engineering, Czech Technical University in Prague, Czech Republic \\

$^{3}$ \quad Magnel Laboratory for Concrete Research, Ghent University, Tech Lane Ghent Science Park - Campus A, Technologiepark-Zwijnaarde 904, 9052 Ghent, Belgium \\

}

%\author[mysecondaryaddress]{Global Customer Service\corref{mycorrespondingauthor}}
\cortext[mycorrespondingauthor]{Corresponding author}
\ead{roman.wanwendner@ugent.be}

%\address[mymainaddress]{1600 John F Kennedy Boulevard, Philadelphia}
%\address[mysecondaryaddress]{360 Park Avenue South, New York}

\begin{abstract}
This contribution studies a well-known failure criterion and its application to the life-time prediction of adhesive anchor systems under sustained load. 
The Monkman-Grant relation, which has previously been applied to a wide range of materials, is now applied to adhesive anchors installed in concrete. 
It postulates a linear relationship between the logarithm of stable creep rate and time to failure. 
In this paper the criterion is evaluated first on a large experimental campaign on one concrete involving two chemically different adhesives and then by several experimental data sets reported in literature. 
In all cases the data is well represented and highly accurate predictions are obtained. 
The second part of the paper focuses on the relationship between stable creep rate and relative load level of the remote constant stress based on the Norton-Bailey and the Prandtl-Garofalo creep laws.  
The latter was found to perform better on fitting the experimental data. 
Finally, the combination of the Monkman-Grant criterion and the aforementioned creep laws allows the prediction of stress versus time to failure curves including uncertainty bounds, that are in very good agreement with all experimental data sets, making it an interesting alternative to existing test methods for adhesive anchor systems under sustained loads. 
  
\end{abstract}


\end{frontmatter}

%\linenumbers

\section{Introduction}
\label{sec:State:INTRO}
Fastening systems represent an important element of modern structural engineering. Their role as connectors between different structural elements makes them a crucial component for (i) fast modular construction based on prefabricated / precast elements \cite{EligenCook, EligenMalle} but also (ii) for strengthening, repair and rehabilitation of existing structures \cite{fib_code_2006}. 
Next to cast-in connectors, especially post-installed mechanical and adhesive anchors can be found in various forms on the market \cite{EligehausenSilva}.
The potential consequences of anchorage failure are quite severe and outweigh by far the costs of the connection elements themselves. 
For this reason, all fastening products have to undergo rigorous tests \cite{aci_aci_2011, eota_etag_2008, ASTME1512, AC58} covering the short-term response in different load situations as well as durability aspects and the long-term response, e.g. under sustained load.
In the recent past two accidents occurred in which the ceiling panels of two tunnels collapsed with tremendous economical damage and even the loss of human life \cite{NTSB_2007_02} due to failed adhesive anchors. 
This raised the awareness towards the sustained load response of adhesive anchors, although the reason for the collapse was found in bad installation or wrong product selection in both cases.

Current approval standards in the United States and Europe \cite{aci_aci_2011, eota_etag_2008, ASTME1512, AC58} utilize a displacement criterion derived from short-term pull-out tests that must not be exceeded by extrapolated displacement histories of sustained load tests.
More specifically, first, the tensile pull-out capacity, $N_{pull-out}$ is measured in confined tests, i.e. pull-out tests with close support. 
From this data the displacement threshold, $\delta_{s}$, also called displacement at loss of adhesion, is derived.
The long-term performance under sustained load is checked by creep tests at constant load, $N_{sust}<N_{pull-out}$, and constant temperature that last for typically $1000-2000$hours during which the evolution of displacements are recorded. 
This data is extrapolated by a suitable regression model in order to obtain long-term displacement estimates $\delta_{t}$. 
As long as the predicted displacement does not exceed the displacement limit, i.e. $\delta_{t}\leq \delta_{s}$, the check is passed.
There are different types of creep function used in codes, e.g. power-laws \cite{aci_aci_2011, eota_etag_2008}, or logarithmic functions \cite{ASTME1512, AC58}.
This check is used to determine the save sustained load level, $N_{sust}/N_{pull-out}$, at standard but also elevated temperature for a given service life of 50 years.

The aforementioned method has the advantage of simplicity as a sole pass/fail method. 
The method is also conservative since the expected failure displacement would be larger than $\delta_{s}$, since it follows the post-peak softening branch of the short term test as showed in \cite{Muciaccia, BouTert18, diluzio-nonloinearcreep}. 
On the other side, the extrapolated displacement $\delta_{t}$ is sensitive to the selected creep function as pointed out in \cite{EligehausenSilva, cook_nchrp_2013}. 
Additionally, in the case of the power function, Wan-Wendner and Podrou\v zek \cite{WendPoda,WendPodb} found a significant sensitivity of $\delta_{t}$ on the initial displacement. 
Furthermore, the extrapolation method fails to predict failure times as function of different parameters, e.g. load level, temperature, anchor geometry and type of adhesive as discussed by Kr\"{a}nkel et al. \cite{KRANKEL2015458}.

An alternative method introduced by Cook et al. \cite{cook_nchrp_2009} attempts to overcome the aforementioned limitations. 
In this method the adhesive anchor system is tested under relatively high sustained load levels, compared to the short term pull-out capacity, and the failure times, $t_f$ are measured. 
Then a stress versus time-to-failure plot is constructed and a power-law relation is fitted, yielding a straight line if plotted against the logarithm of time.
The extrapolation of this fit allows the selection of the range of the load levels that will not lead to failure during a given life-time. 
The stress versus time to failure method has the advantage of actually investigating creep failure instead of merely limiting long-term displacements that are obtained by extrapolation with models that only apply to a stable creep process, i.e. without failure \cite{WendPoda, WendPodb}. 
On the other side, the time to failure method suffers from large scatter of typically two orders of magnitude in failure times for a given load level and the required long testing times for low load levels. 
This situation could be improved by better knowledge concerning the functional form of stress versus time to failure curves -- still an ongoing research problem due to the aforementioned problems.
An attempt of predicting failure times using a numerical framework that accounts for damage was carried out by Kr\"ankel and Gehlen \cite{KRANKEL2015458}. 
Nevertheless, no conclusion on the shape of the stress versus time to failure curve was made.
Nin\v{c}evi\'{c} et al. \cite{Nincevic} followed the findings of Boumakis et al. \cite{BouTert18} and proposed a sigmoid stress versus time-to failure curve based asymptotic matching, i.e. imposing the physically reasonable behavior for high loads approaching the short-term capacity or low loads.
The last is an improvement in comparison to the simple power-law, since long-term extrapolations do not predict failure for unloaded systems. 
Most importantly, the proposed function can be used to directly obtain an asymptotic load level under which no failure will be obtained. 
This value can also be fixed at a desired (conservative) load level, e.g $0\%$. 

The short-comings of both approaches can be overcome by the application of the Monkman-Grant criterion (MG) \cite{MonkmanGrant}, for the estimation of failure times of bonded anchor systems as will be demonstrated in this paper. 
The criterion was first introduced to predict creep failure times in metals. 
Despite its simplicity it succeeds to predict failure times, with a certain accuracy \cite{Povolo1985}, for a wide range of materials, including also polymers \cite{Reifsneider,GUEDES2006703}. 
Later Dobe\v{s} and Milicka, \cite{Dobe1976382}, proposed the modified Monkman-Grant criterion (MMG) by scaling the failure times with the actual displacements at failure, thus, improving significantly predictions. 

In this contribution variations of the Monkman-grant criterion are applied to time to failure data-sets of bonded anchors for five adhesive mortars, two of them tested by the authors, and the other three obtained from the literature, \cite{cook_nchrp_2013, cook_nchrp_2013A}. After discussing basic features of time to failure tests in section~2 and introducing the formulation of the criterion in section~3, the application of the criterion is investigated using the authors own data only in section~4. 
%This method provides predictions that are validated againts actual failure times. Additionally, it could be used as an approval guideline for bonded anchors under sustained loads. 
Finally, following the Miyano approach \cite{Miyano}, the stress versus time to failure curve is reconstructed in section~6, verifying the two domain curve that was introduced in \cite{Nincevic, BouTert18} in a phenomenological manner. 
Therefore, an appropriate function relating the creep rate to the remote stress is necessary.
This manuscript takes into account typical creep models found in the literature, as investigated in section~5. 
First the power function known as the Norton-Bailey creep law \cite{Norton, BaileyC} is investigated.
This law was used frequently in the past within the scientific community for formulating creep models. 
However, it has the disadvantage of a non-constant exponent for different load levels.
Additionally, a function that accounts for the rate of bond breakage and restoration is studied. 
This function was originally proposed by Prandtl~\cite{Prandtl192885} and afterwards modified  by Garofalo~\cite{Garofalo}. It is commonly known as Prandtl-Garofalo creep law.   
In the last section, all previously discussed findings are validated utilizing three independent literature data sets on different adhesives, concretes, and geometries.

\section{Time to failure tests}
This study uses experimental tests of bonded anchors under sustained loads, with failure, for two different adhesive polymers, referred to as product~1 (P1), and product~2 (P2). 
The anchor rods with diameter of $16$mm, were installed at an embedment depth of $75$mm, for both types of adhesive mortars. 
The experimental configuration and the details are extensively discussed in \cite{Nincevic} and, thus, only a short overview is provided in this manuscript. 

First, confined short-term pull-out tests were performed to define the reference pull-out capacity $N_{pull-out}$ of each system which served as reference for the definition of relative load levels of the sustained load tests. 
The short-term capacity was tested within the sustained load test rig applying the same loading rate of $9.0$~kN/s. 
The relatively high loading rate was chosen such as to minimize the effect of creep during loading. 
The pull-out capacity for the P1 system was determined to be $157.32 ~(\pm 3.39\%)$~kN based on $4$ tests. 
The P2 testes exhibited higher scatter and, thus, $2$ additional tests were added for the characterization of the pull-out capacity, yielding overall a strength of $111.76~(\pm 7.48\%)$~kN. 

\begin{figure}[h!]
  \includegraphics[width=0.5\textwidth]{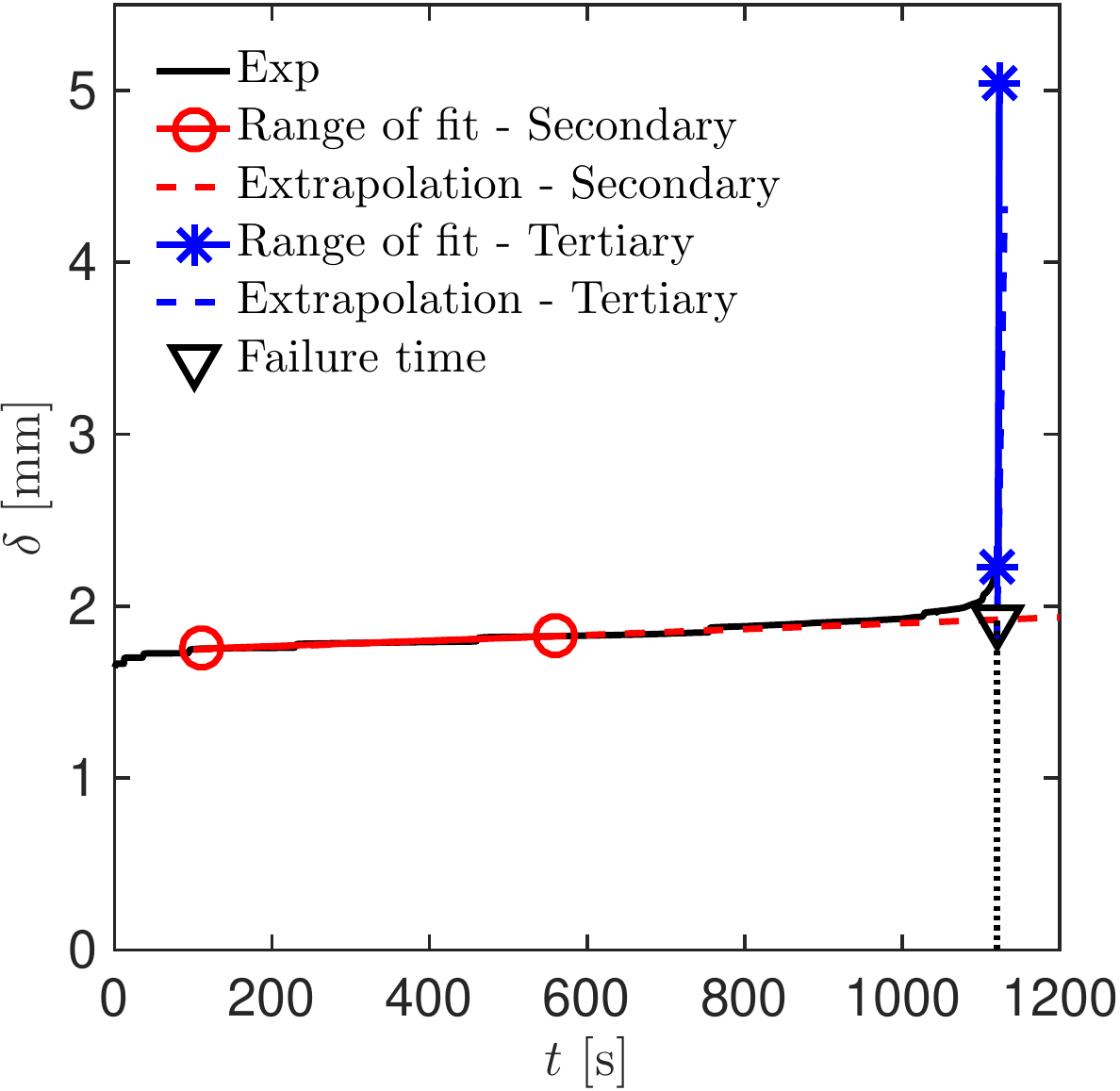}

 \caption{ Definition of failure time  }
 \label{fig:RatesEx1}
 \end{figure}

Then the anchors were loaded with the same loading rate that was used for the short-term tests up to various sustained load levels at which the load was maintained by active pressure feedback control until failure was observed. 
The failure times used for this analysis are estimated as depicted in Fig.~\ref{fig:RatesEx1}.
First, the data are divided into three creep domains and a linear relationship is fitted to both, the secondary and tertiary creep data points. 
The intersection of the two regression lines is termed the failure time, $t_f$. 
It has to be noted that Nin\v{c}evi\'{c} \cite{Nincevic} et al. estimated the failure times from the drop of the hydraulic pressure. 
A comparison of the failure times estimated with the two different approaches revealed no significant differences. 
Other options include criteria based on the local curvature or the maximum deviation from the power-law shape associated with secondary creep.
Table \ref{tab:LLv} shows the tested load levels and the mean failure time of each anchor system. 

\begin{table}[h!]
\centering
	
    \begin{tabular}{p{2.5cm}p{4.6cm}p{4.6cm}}
    \textbf{Load Level} $\mathbf{[\%]}$ & $\mathbf{t_{f}}$ \textbf{[s]} & $\mathbf{t_{f}}$ \textbf{[s]}  \\
		\hline
$~~~~~~~~~~~$ & P1 & P2 \\
\hline
\centering
$95$  & $4.22 ~~~[0.30-9.33]$ & -\\
\centering
$85$ & $18.37 ~~[12.39-28.18]$ & $68.04~ [8.53-179.70]$   \\
\centering
$80$ &  -  & $3,434.38~[584.30-7,912.00]$ \\
\centering
$75$  & $575.39~[56.01-1121]$ & $2,303.66 ~[1,152.00-3,445.00]$  \\
\centering
$70$  & - & Running \\
\centering
$65$ & $3.13 \cdot 10^6~[5.623\cdot 10^4-9.158\cdot 10^ 6]$ & Running  \\
\centering
$60$  & Running & Running  \\
\hline
\end{tabular}
\caption{~Time to failure load levels.}
\label{tab:LLv}
\end{table}  

\begin{figure}[h!]

\begin{subfigure}[b]{0.5\textwidth}
       \includegraphics[width=1\textwidth]{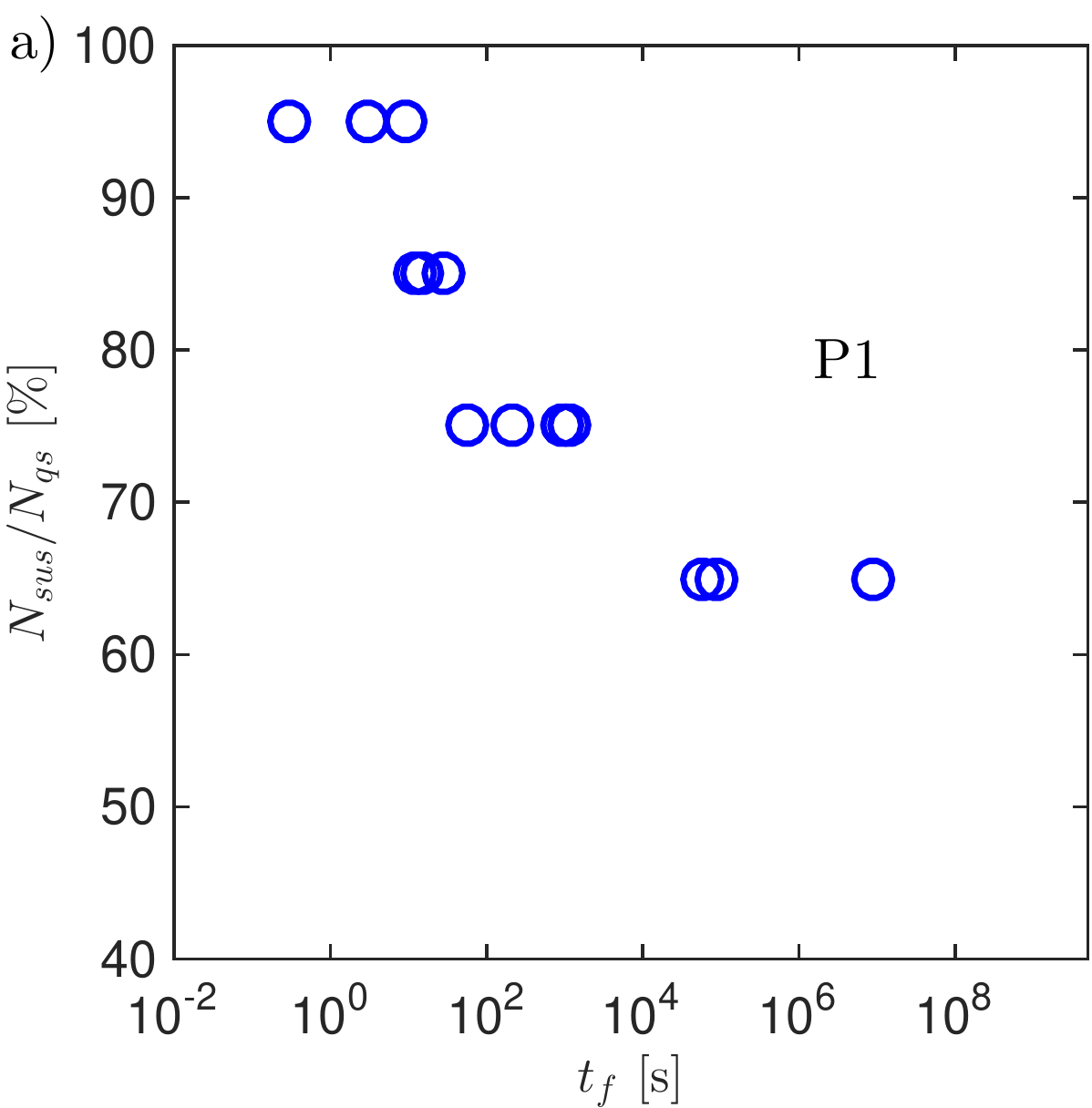}
  \end{subfigure}
   \begin{subfigure}[b]{0.5\textwidth}
      \includegraphics[width=1\textwidth]{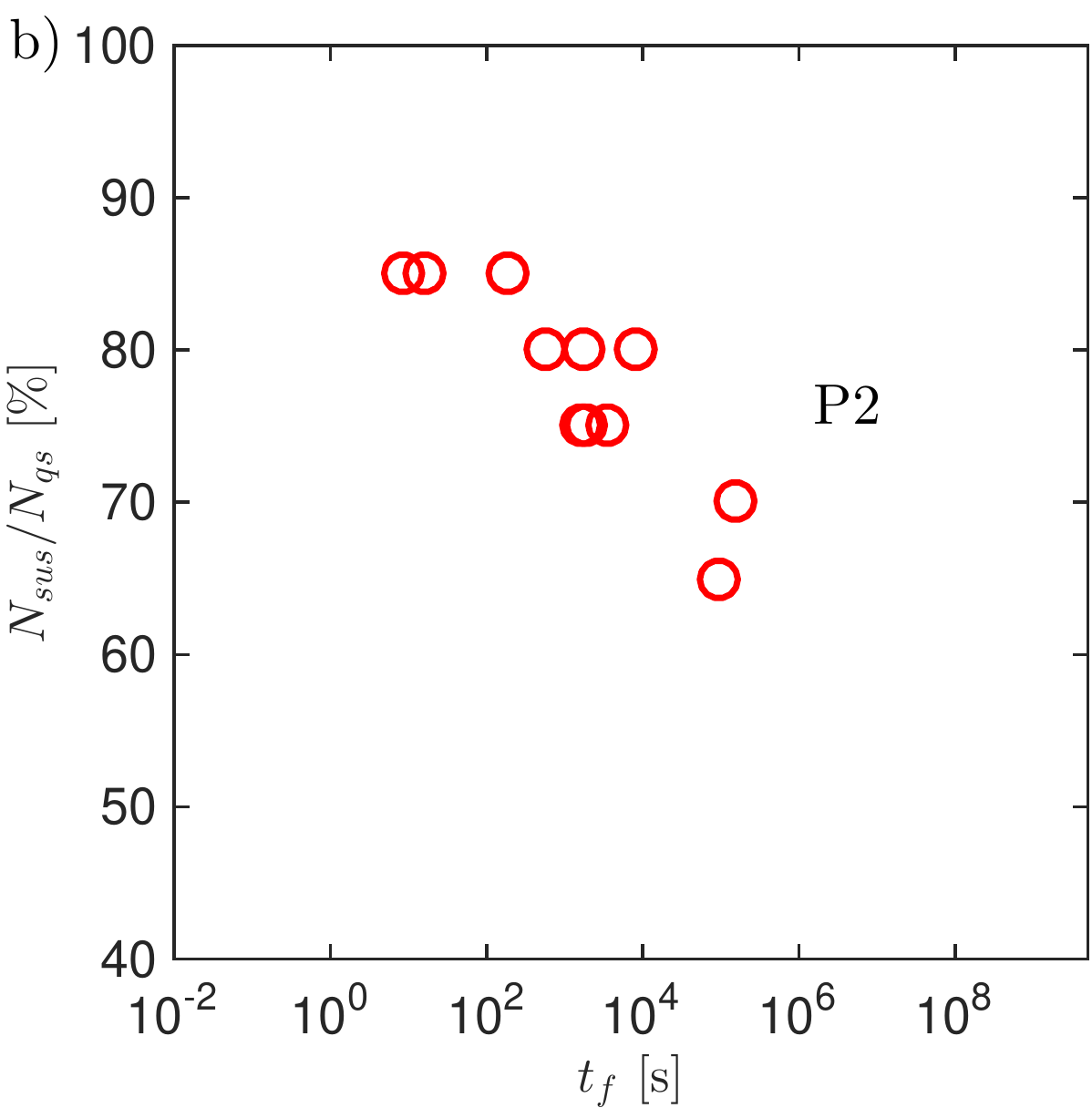}
  \end{subfigure}
 
\caption{Stress versus time to failure curve for tests on adhesive anchors of adhesive P1 (a), and P2 (b).}
\label{fig:TTFData}
\end{figure}

The stress versus time to failure curves for both adhesive anchors are shown in Fig.~\ref{fig:TTFData}.

\section{Monkman-Grant criterion}
\label{sec:MG}

Typically, the creep response of any material is divided in $3$ stages. 
The first is the primary stage, which is governed by a high creep strain rate $\dot{\epsilon}(t)$, that is reducing relatively fast, as a power-law of time. 
The latter was first introduced by Andrade \cite{Andrade} for metals, but it can be observed in various materials, e.g polymers \cite{Ding, Gruber}, rocks \cite{Lockner,Nechad}, and concrete \cite{Wang}. 
After a significant reduction of the creep rate $\dot{\epsilon}(t)$ it approaches a minimum value at which it stabilizes and the secondary creep stage, or steady-state creep, begins.
This stage is characterized by an approximately constant creep rate and it can last from some seconds to many years, depending on the remote stress, and the environmental conditions of the material, e.g. temperature. 
Finally, the tertiary stage of creep occurs when the creep strain rate $\dot{\epsilon}(t)$ starts to increase again after reaching a minimum creep strain rate, this time progressively following also a power-law, as suggested in \cite{Hao}, until failure of the material occurs. 

The onset time of the tertiary creep is called rupture time. 
The prediction of rupture time is a topic of high interest for many researchers. 
As a matter of fact, many models were used for the prediction of rupture times of different visco-elastic materials. 
These can be summarized as the Reiner-Weissenberg criterion, (RW) \cite{RW}, the Zhurkov equation \cite{Zhurkov1984}, the  maximum stress work criterion (MW), \cite{GUEDES2006703}, and the maximum strain work criterion, \cite{GUEDES2006703}. 
The previous criteria are all energy-based, where the total energy is decomposed into stored, $w_{s}$, and dissipated energy, $w_{d}$. 
Both contributions can be estimated e.g. by fitting a rheological model of suitable form to creep displacement data.
Then, failure time prediction models are formulated making use of the components of the rheological model.
A detailed description of all the previous criteria can be found in \cite{GUEDES2006703, MirandaGuedes2004}.

\begin{figure}[h!]
  \includegraphics[width=0.5\textwidth]{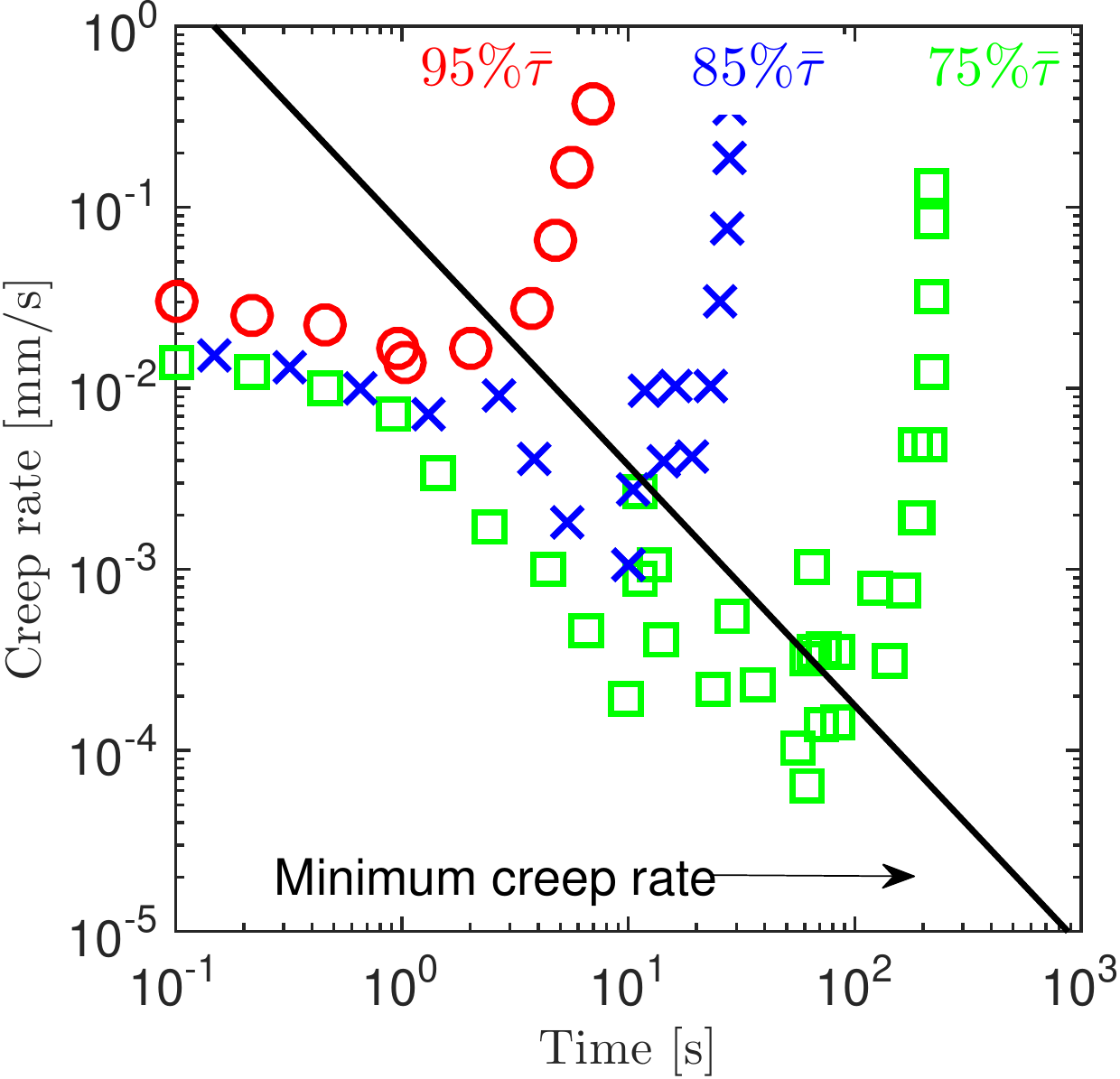}

 \caption{ Displacement rate versus time for different sustained load levels and linear regression on the minimum creep rates.}
 \label{fig:RatesEx}
 \end{figure}

Another failure criterion is the well-known Monkman-Grant (MG) relation, introduced for metals by Monkman and Grant \cite{MonkmanGrant}. 
This is an empirical equation, which relates the minimum creep rate to the failure time.
Although many attempts have been made for a theoretical justification, the mechanism that is described is not yet fully understood. 
However, the Monkman-Grant relation is known for it's highly accurate predictions for different metallic alloys \cite{Dobe1976382, CHOUDHARY20131, ASHBY19843} as well as for other materials like polymers \cite{GUEDES2006703,Castillo}, and ceramics  \cite{Menoncer, Ferbercer}. 
The Monkman-Grant criterion is illustrated in Fig.~\ref{fig:RatesEx}, where the displacement rates of tested bonded anchors under sustained load are shown. 
Later Dobe\v{s} and Milicka \cite{Dobe1976382} proposed the modified Monkman-Grant criterion (MMG) by scaling the failure times with the actual displacements at failure, thus, improving significantly predictions. 

The minimum creep rate can only be determined experimentally if a more or less complete displacement time series including transition to failure is available, see Fig.~\ref{fig:RatesEx1}.
Thus, the MG criterion is limited to actual time to failure data with the corresponding testing effort and cost.
In order to overcome this limitation Evans \cite{Evans2006} proposed an empirical method to predict the minimum creep rate from the creep rate at a given (relatively low) strain level.
With this prediction the MG criterion can be applied to provide access to failure times, reducing in that way the necessary testing time. 
The link between the creep rate at given low strain and failure times is often referred to as generalized Monkman-Grant relation, 

It is obvious that the higher the minimum creep rate is, the shorter the failure time. The equation takes the form  

\begin{equation}\label{eq:MG}
\dot{\epsilon}_{min}\cdot{t}_{f}^{n} = C
\end{equation}
where $\dot{\epsilon}_{min}=$ the minimum creep rate, ${t}_{f}=$ the failure or rupture time, and $n, C=$ material dependent constants. Eq.~\ref{eq:MG} can also be expressed as

\begin{equation}\label{eq:MGL}
\ln{\dot{\epsilon}_{min}} = n\ln{{t}_{f}} + c
\end{equation} 

where $c$ stands for a constant. For easier readability the minimum creep rate is abbreviated in subsequent sections as $\dot{\epsilon}$, or in terms of absolute deformations $\dot{\delta}$.

\begin{figure}[h!]

\begin{subfigure}[b]{0.3\textwidth}
       \includegraphics[width=1\textwidth]{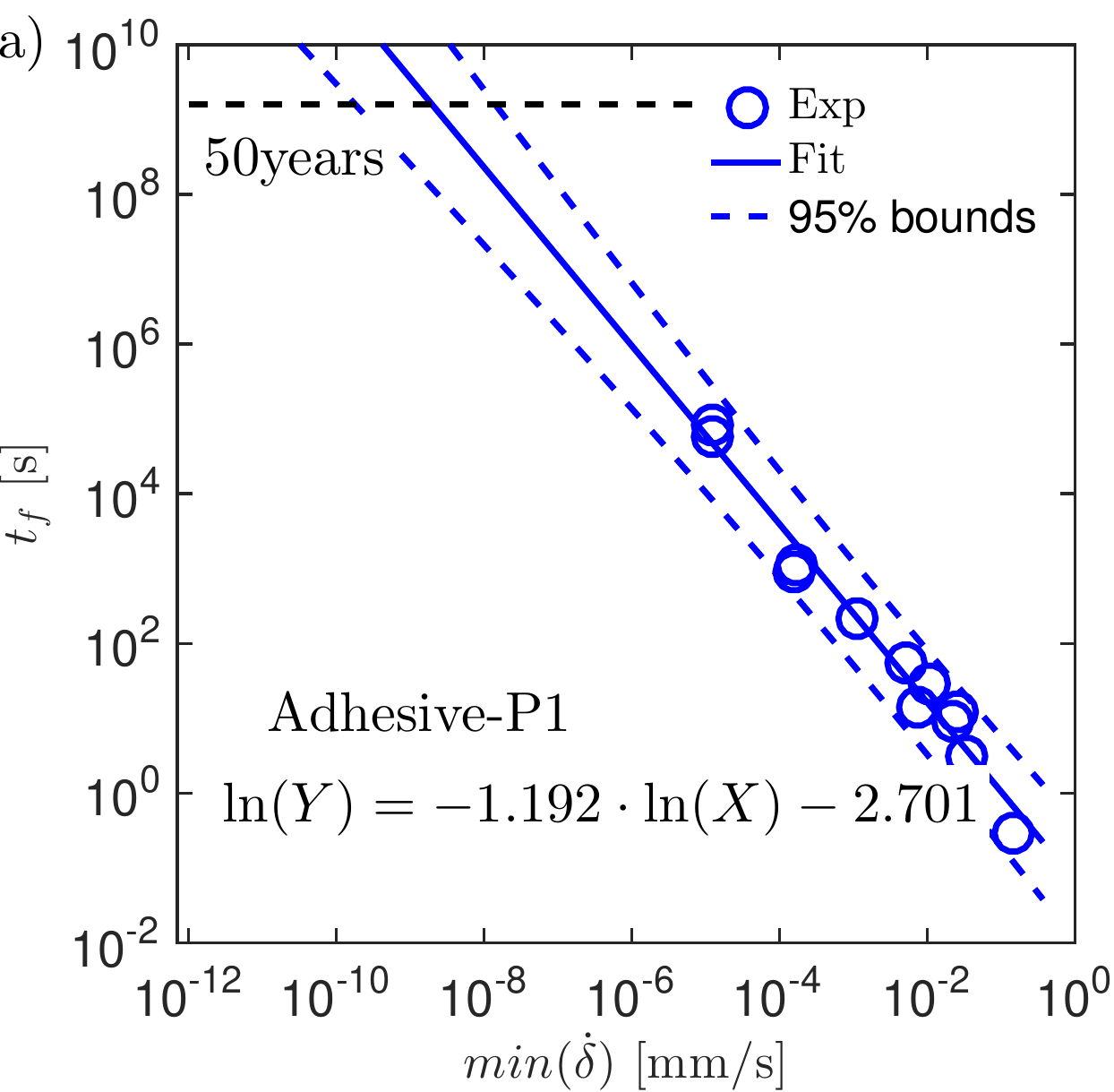}
  \end{subfigure}
   \begin{subfigure}[b]{0.3\textwidth}
      \includegraphics[width=1\textwidth]{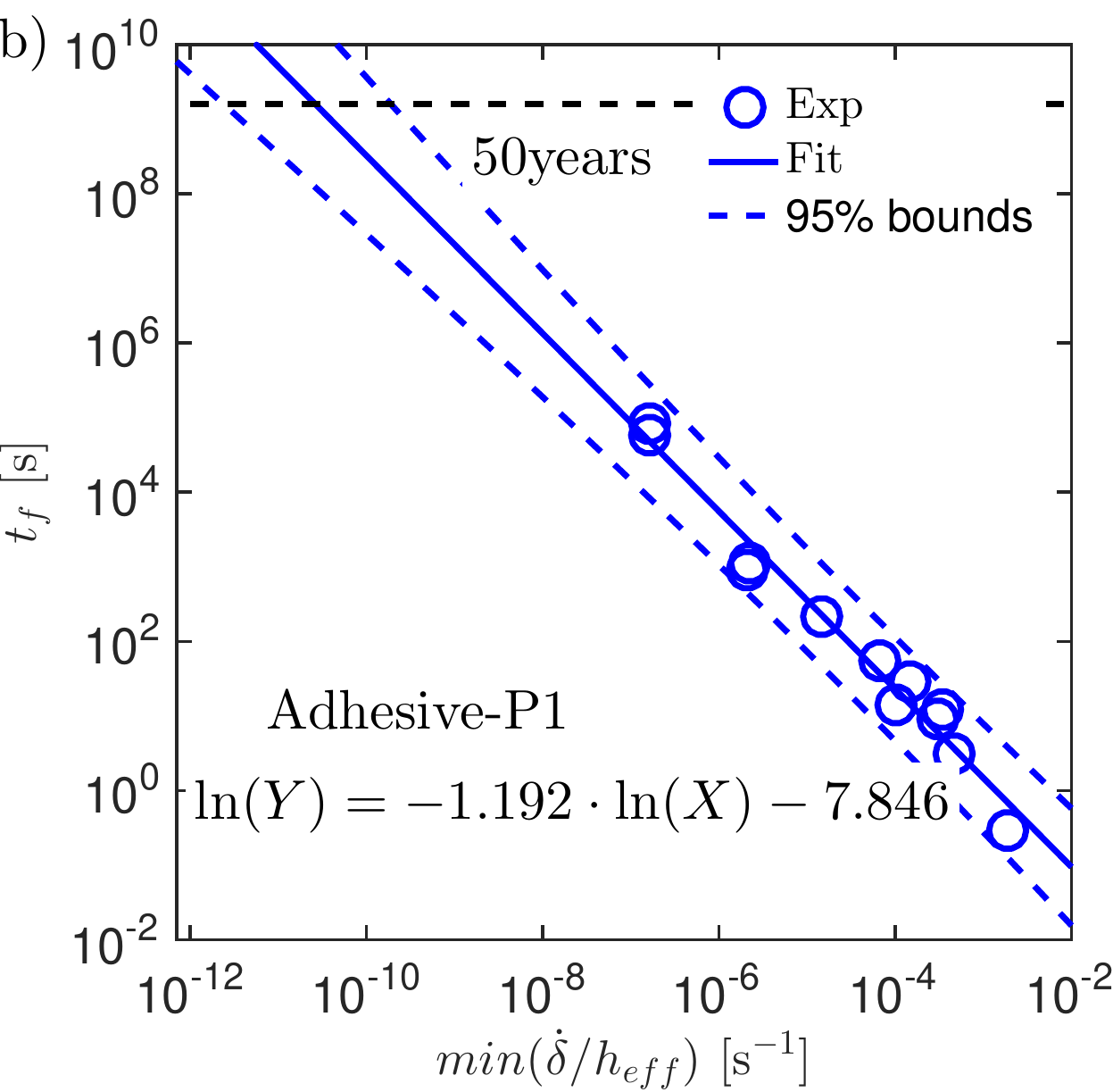}
  \end{subfigure}
  \begin{subfigure}[b]{0.3\textwidth}
      \includegraphics[width=1\textwidth]{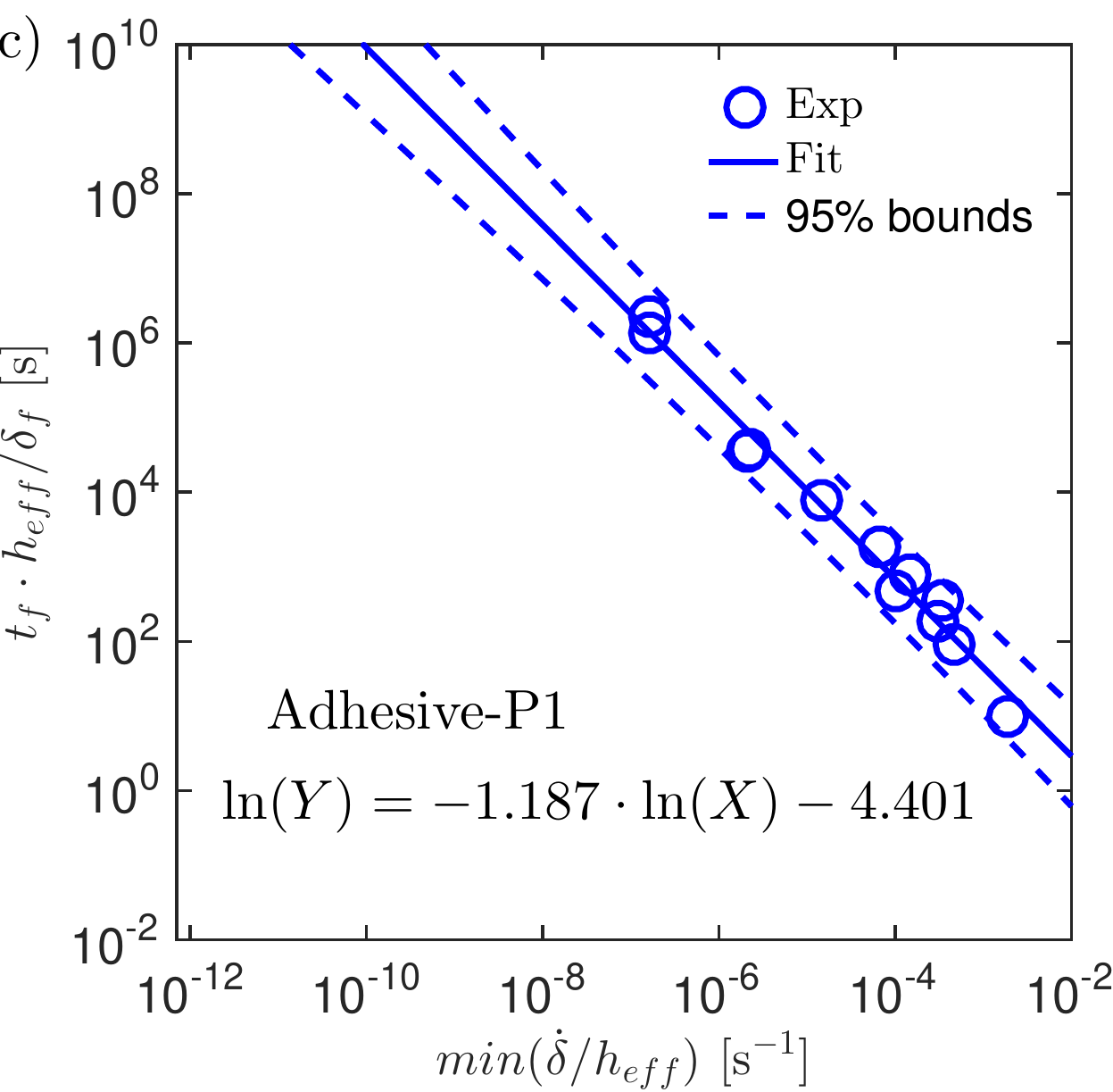}
  \end{subfigure}
  
  \begin{subfigure}[b]{0.3\textwidth}
       \includegraphics[width=1\textwidth]{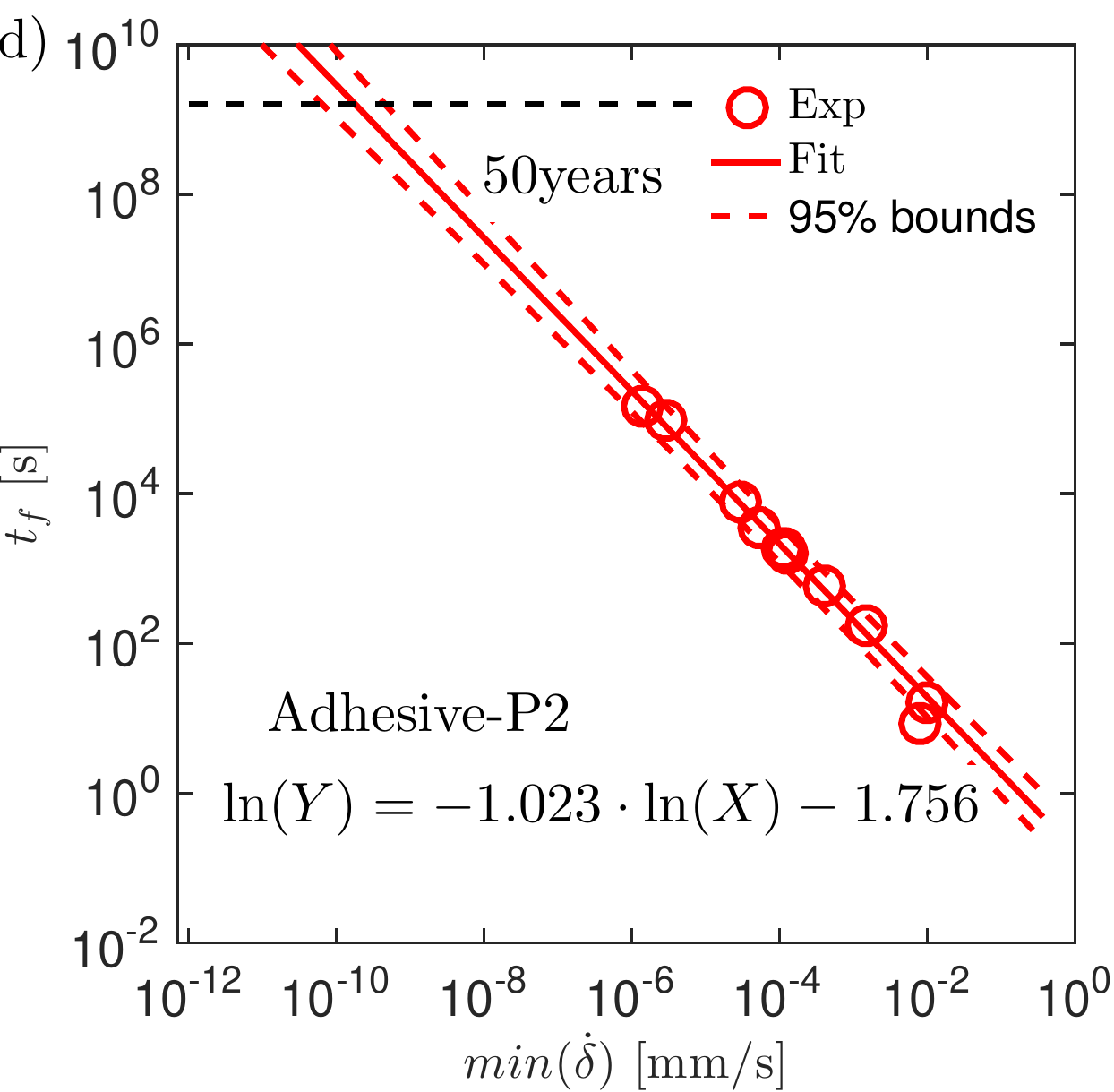}
  \end{subfigure}
   \begin{subfigure}[b]{0.3\textwidth}
      \includegraphics[width=1\textwidth]{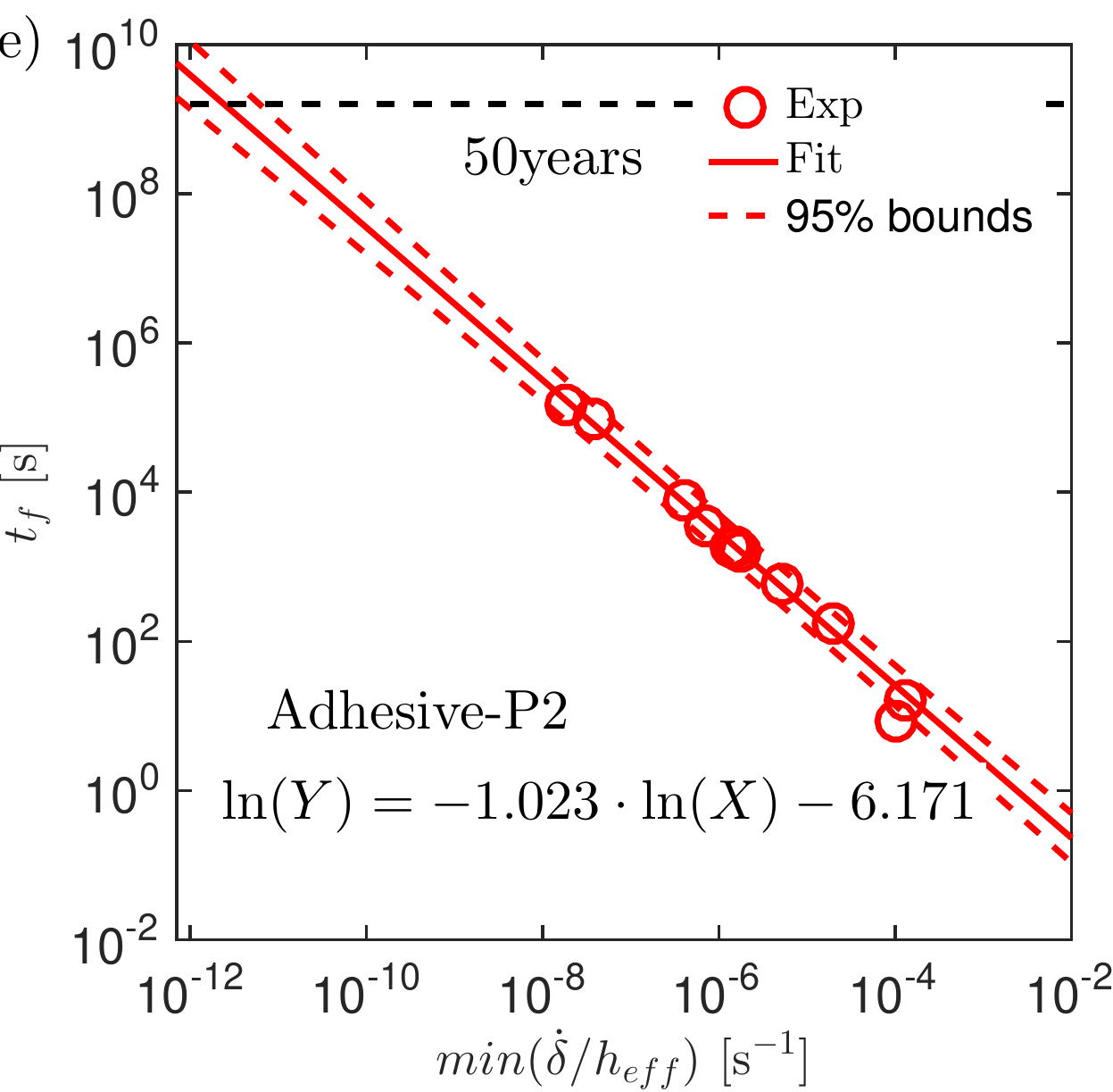}
  \end{subfigure}
  \begin{subfigure}[b]{0.3\textwidth}
      \includegraphics[width=1\textwidth]{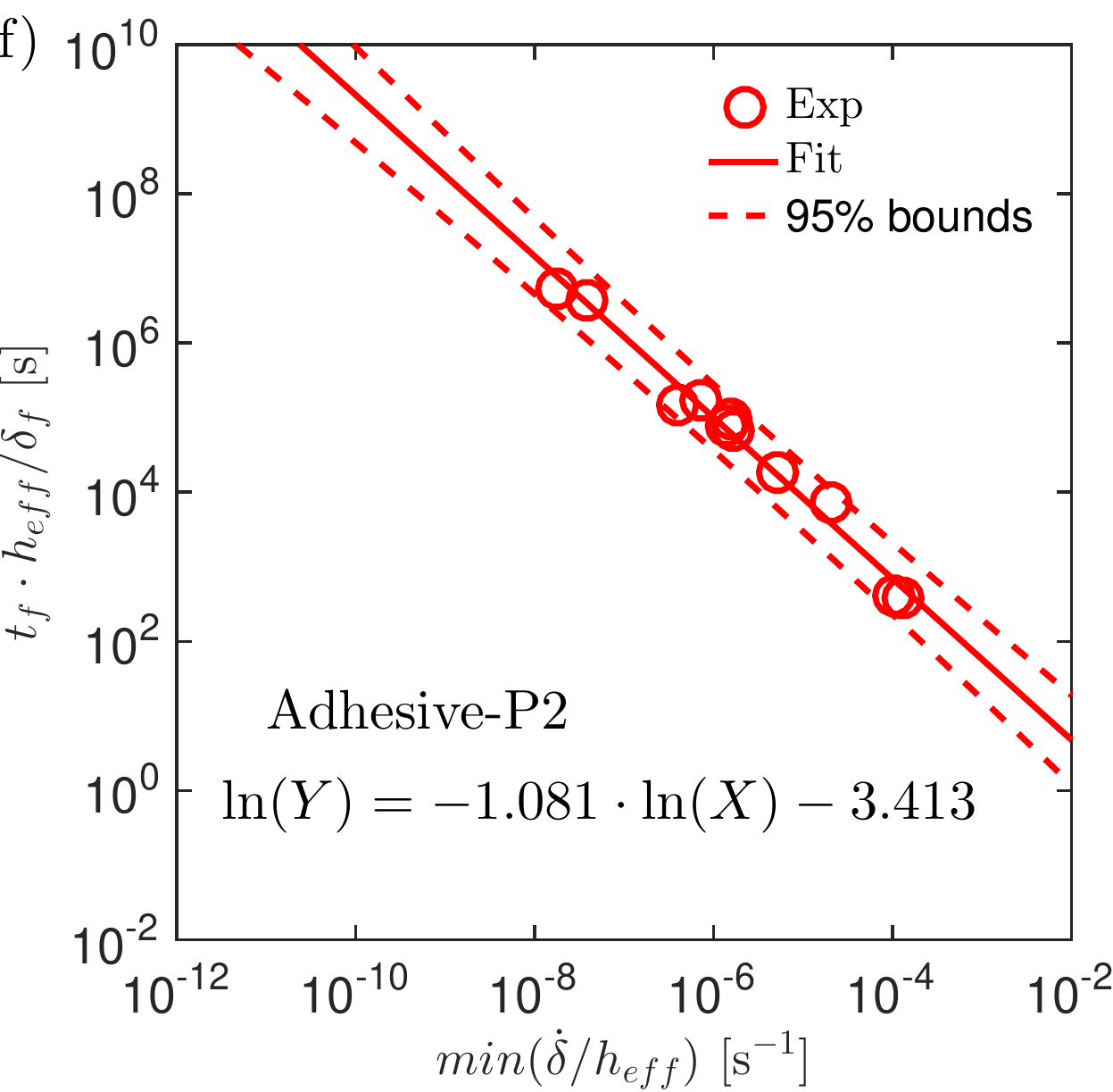}
  \end{subfigure}
  
\caption{Monkman-Grant fit on displacement rates, normalized displacement rates, and modified Monkman-Grant for adhesive P1, (a-c), and adhesive P2 (d-f).}
\label{fig:D3LDPM}
\end{figure}

This paper studies the applicability of the MG criterion to bonded anchor systems. 
Following this approach the secondary creep rates are derived directly from the sustained load experiments, and Eq.~\ref{eq:MGL} is fitted to the data.  
Fig.~\ref{fig:D3LDPM} shows the fit of the Monkman-Grant relation to the experimental data of the P1 and P2 anchors. 

The MG criterion can be directly applied to the displacement measurements at the top of the anchors. 
The fitted equation is shown as solid line in Fig.~\ref{fig:D3LDPM} (a) and (d) for both products, respectively. 
The dashed line denotes the 95\% confidence bounds.
In both cases an almost perfectly linear relationship can be observed with quite narrow confidence bounds, especially in case of P2.
Nevertheless, the formulation of the MG criterion based on the displacement creep rate doesn't allow the comparison of different geometries, i.e. embedment depth.
This obstacle can be overcome by normalizing the displacement rate with the embedment depth. 
The result is an unit-free strain rate based formulation.
This normalization is quite simple, but also consistent with the typically used uniform bond model, as introduced by Cook \cite{cook_behavior_1998}: 
 
\begin{equation}\label{eq:BL}
\tau = \frac{N_{sust}}{\pi\cdot r^2 \cdot h_{ef}}
\end{equation}

where $r$ the anchor diameter. 
It is easily understood that the model implies a constant bond stress distribution along the anchor. 
Thus, also the strain has to be constant along the anchor, yielding a linear increase in deformation with the embedment depth.
The application of this generalized MG criterion to both products is shown in Figs.~\ref{fig:D3LDPM}(b) and (e). 

Finally the MMG relation is studied. 
Accordingly, the failure times are normalized by the failure strains, i.e. $\delta_{f}/h_{eff}$. 
This normalization provides generally better fits and reduces prediction uncertainty. This will be apparent in later analyses. 
Figs.~\ref{fig:D3LDPM} (c) and (f) present the MMG analyses using the actually measured displacements at failure.
Unfortunately, the application of this criterion requires the a-priori knowledge of the exact failure strain which is not available for predictions based on approximating the minimum creep rate with the stable creep rate of ongoing tests.
A conservative estimation of the latter could be the displacement at peak load of the a standard pull-out test, as it is used in the Findley approach, \cite{findley_creep_1976, cook_nchrp_2013}. 
A more realistic estimate would be the displacement of post-peak branch of short-term pull-out tests corresponding to the sustained load level. 
% IB in this example we don't improve. I am not sure if we should keep the next statement.
% RWW fine we can remove it
%In that way the uncertainty of the Monkman-Grant prediction could be low. 

In all previously discussed cases, the Monkman-Grant relation is used to extrapolate to significantly lower creep rates without obtaining excessively large confidence bounds on the extrapolations, e.g. to 50 years. 
It was demonstrated that this method also applies to entire bonded anchor systems and, thus, could serve as an alternative test method. 
Furthermore, by introducing a relation between the minimum creep rate and the constant applied stress, the stress versus time to failure curve can be constructed.
This will be demonstrated in section~5 following the method that was introduced by Miyano \cite{Miyano} for the derivation of the creep strength and the fatigue strength master curves.

%\subsection{Definition of rupture time}

%\subsection{Application of Monkmann-Grant criterion}

\subsection{Sensitivity to data availability and scatter}
\label{sec:State:Robust}
\begin{figure}[h!]
%\vskip -1cm
\begin{subfigure}[b]{0.3\textwidth}
       \includegraphics[width=1\textwidth]{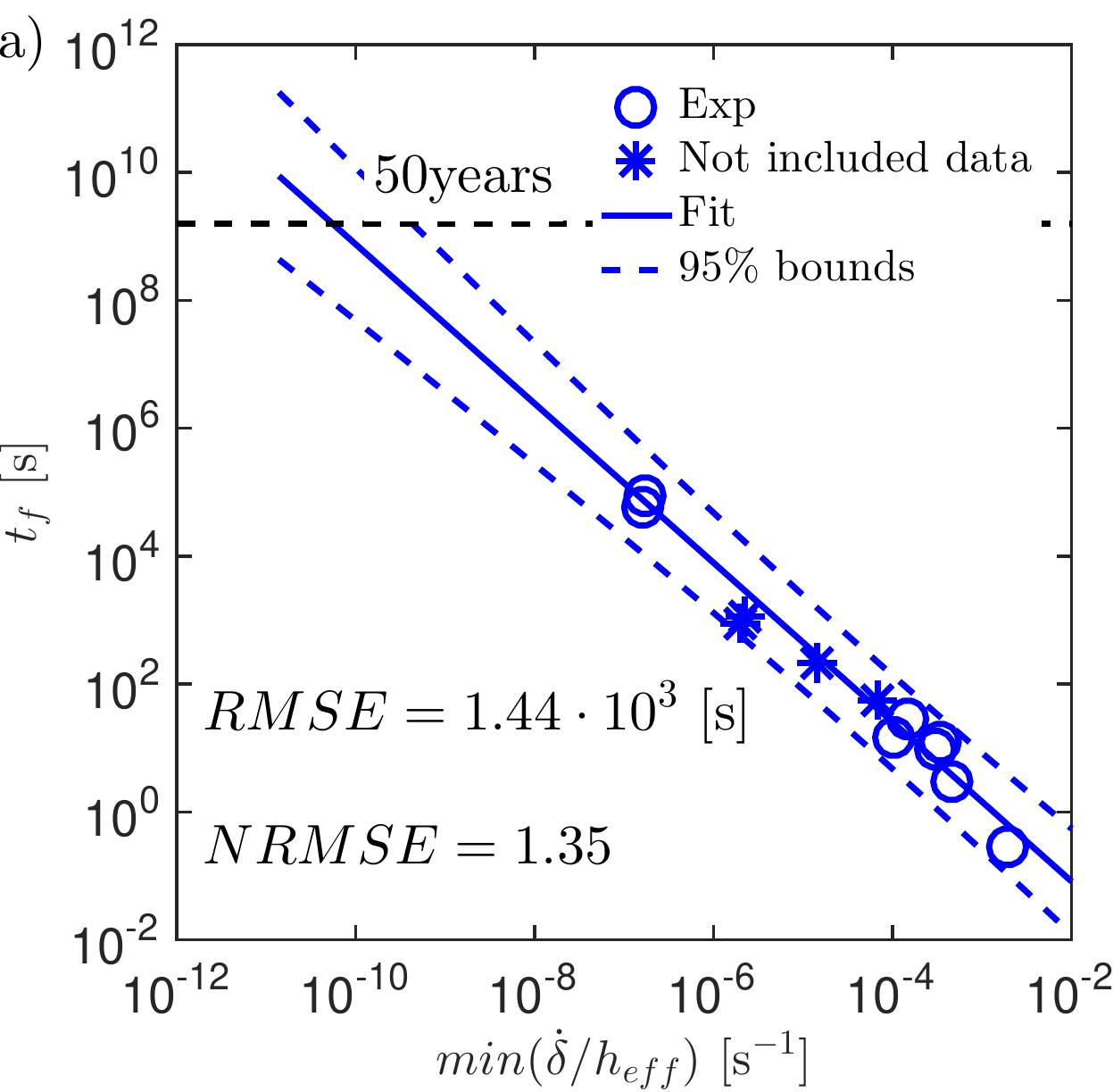}
  \end{subfigure}
%   \begin{subfigure}[b]{0.5\textwidth}
%%   
%      \includegraphics[width=1\textwidth]{./figures/Figxb_190301}
%%       
%  \end{subfigure}
  \begin{subfigure}[b]{0.3\textwidth}
       \includegraphics[width=1\textwidth]{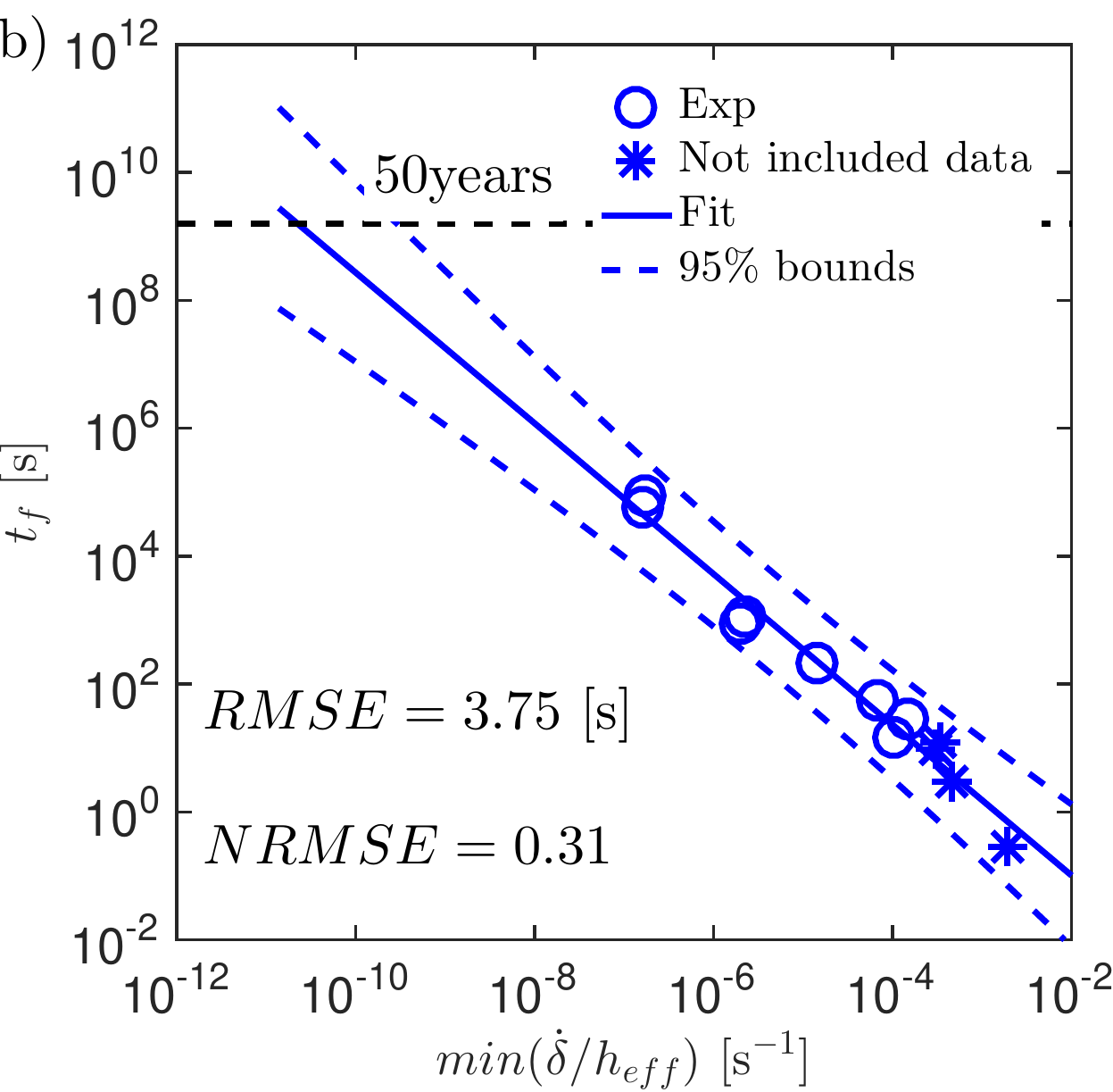}
  \end{subfigure}
    \begin{subfigure}[b]{0.3\textwidth}
       \includegraphics[width=1\textwidth]{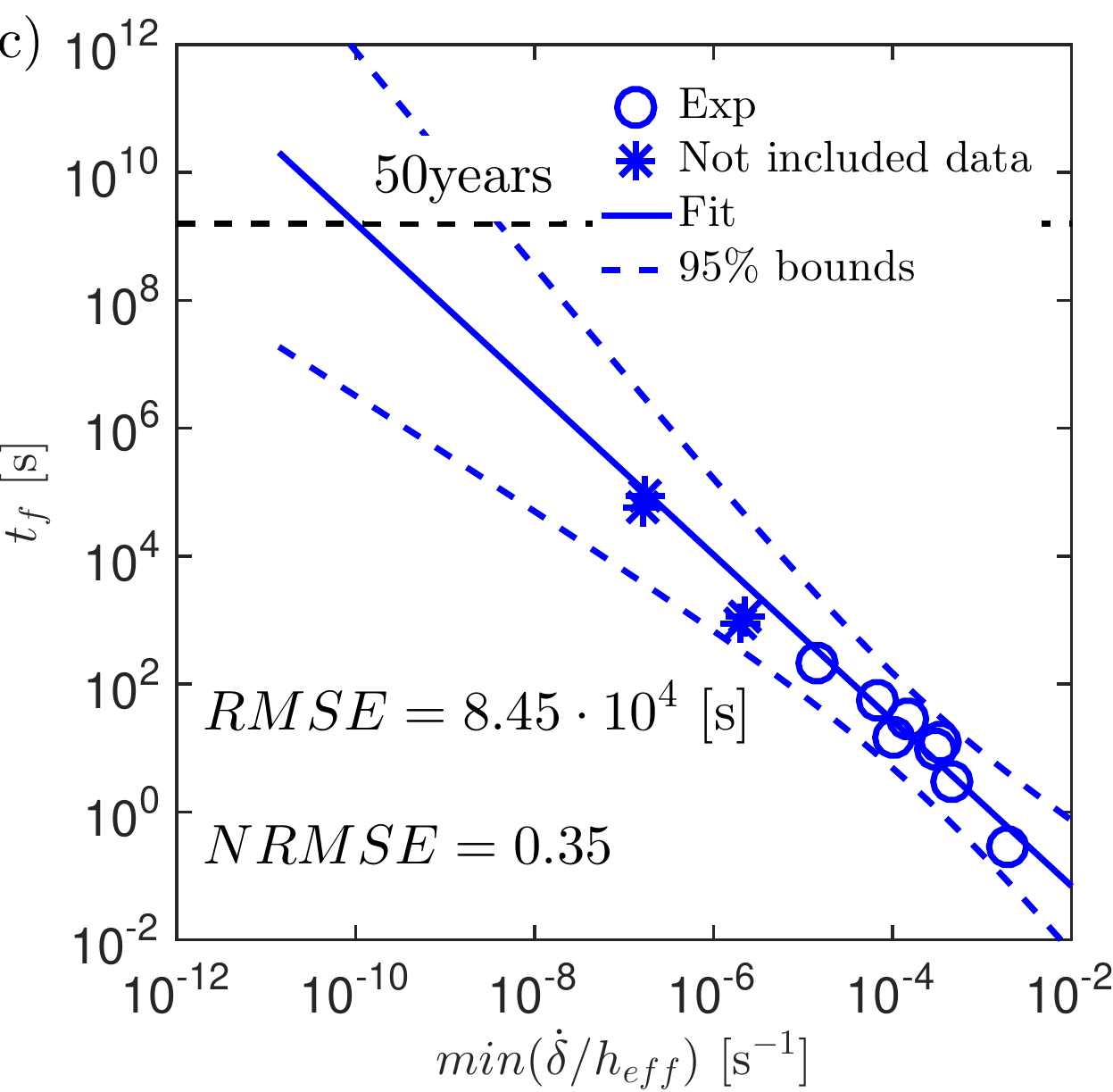}
  \end{subfigure}
%   \begin{subfigure}[b]{0.5\textwidth}
%%   
%      \includegraphics[width=1\textwidth]{./figures/Figxb_190301}
%%       
%  \end{subfigure}

  \begin{subfigure}[b]{0.3\textwidth}
       \includegraphics[width=1\textwidth]{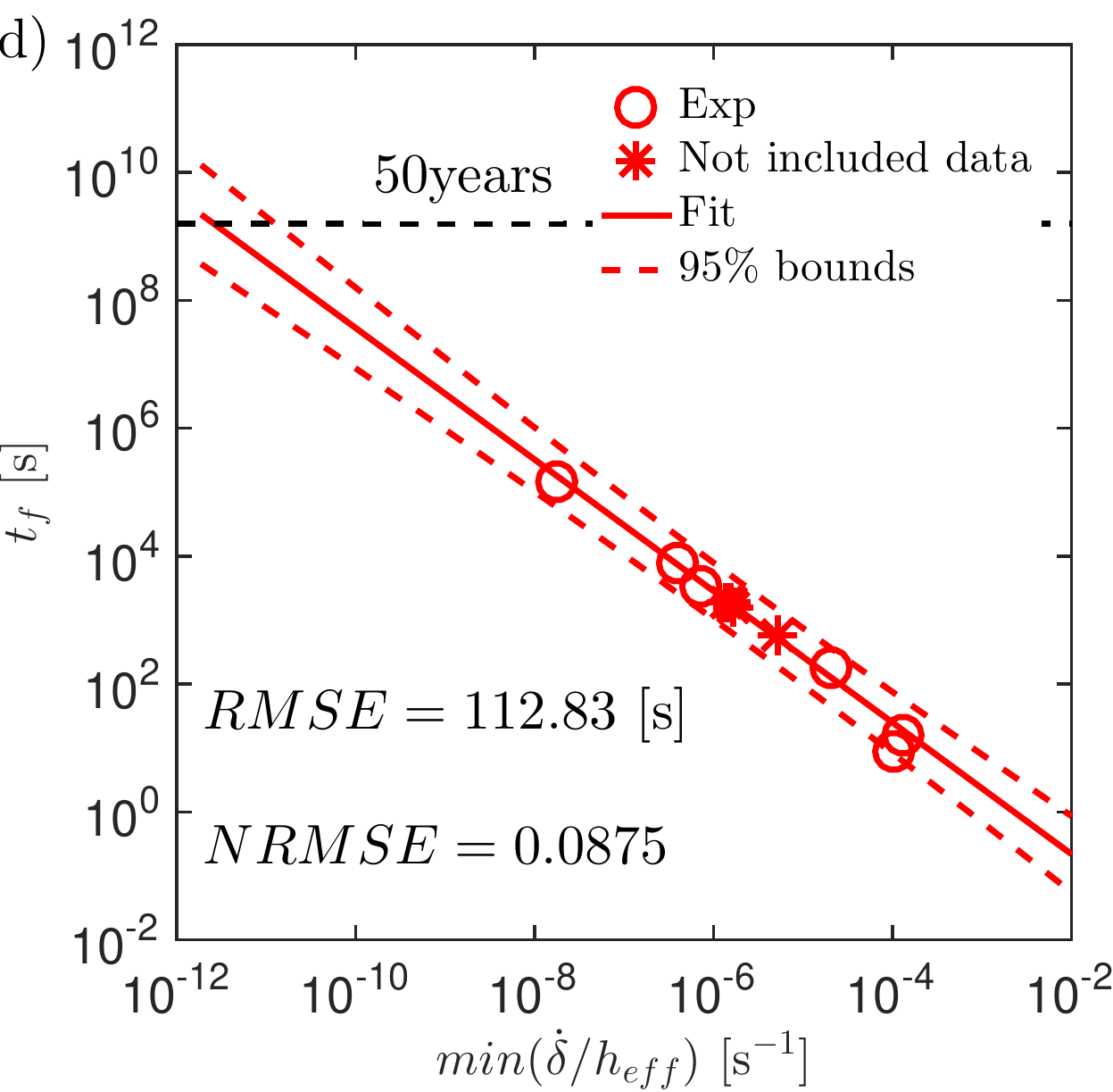}
  \end{subfigure}
    \begin{subfigure}[b]{0.3\textwidth}
%   VELowRates_190315
       \includegraphics[width=1\textwidth]{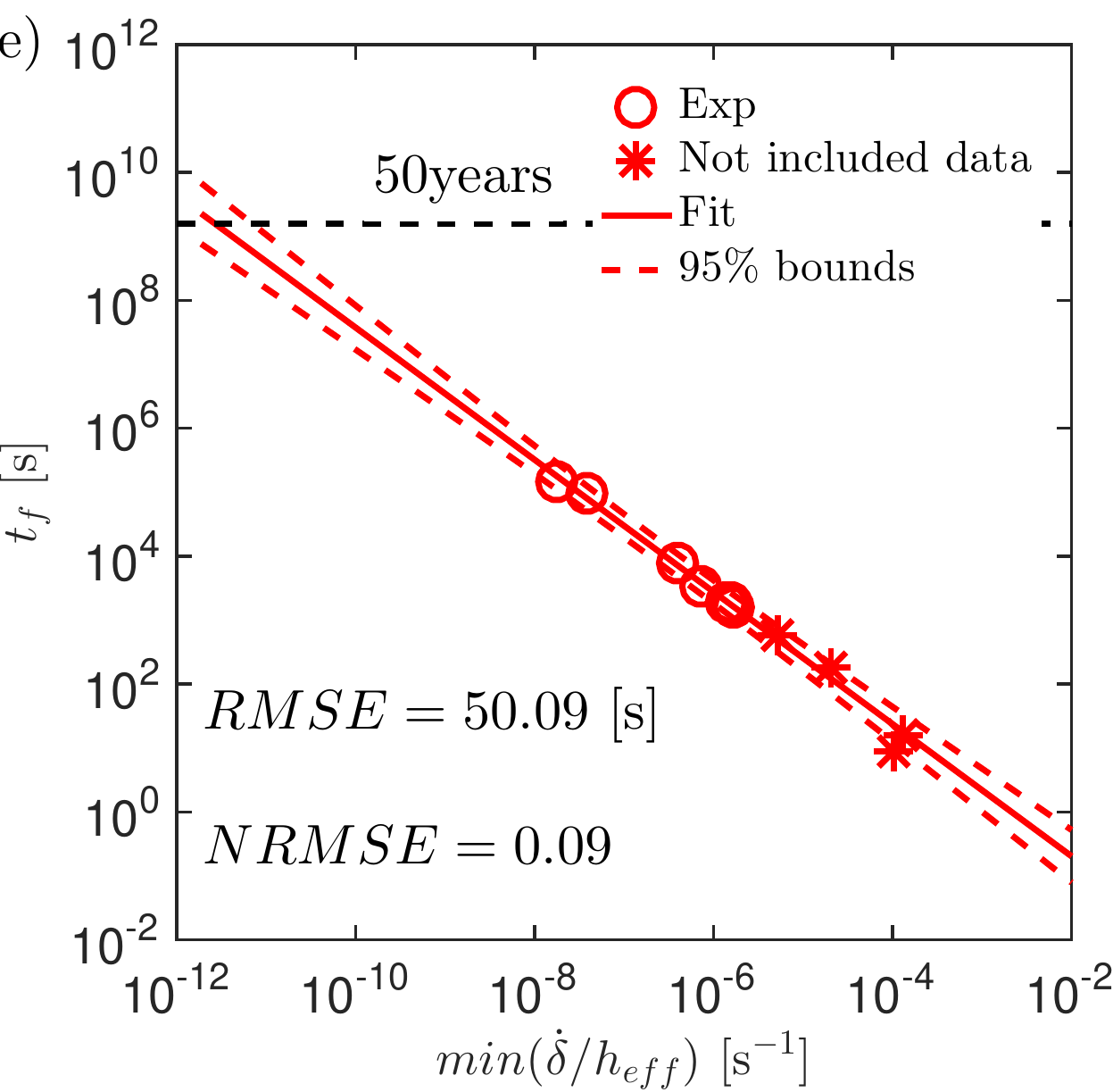}
  \end{subfigure}
  \begin{subfigure}[b]{0.3\textwidth}
%   VELowRates_190315
       \includegraphics[width=1\textwidth]{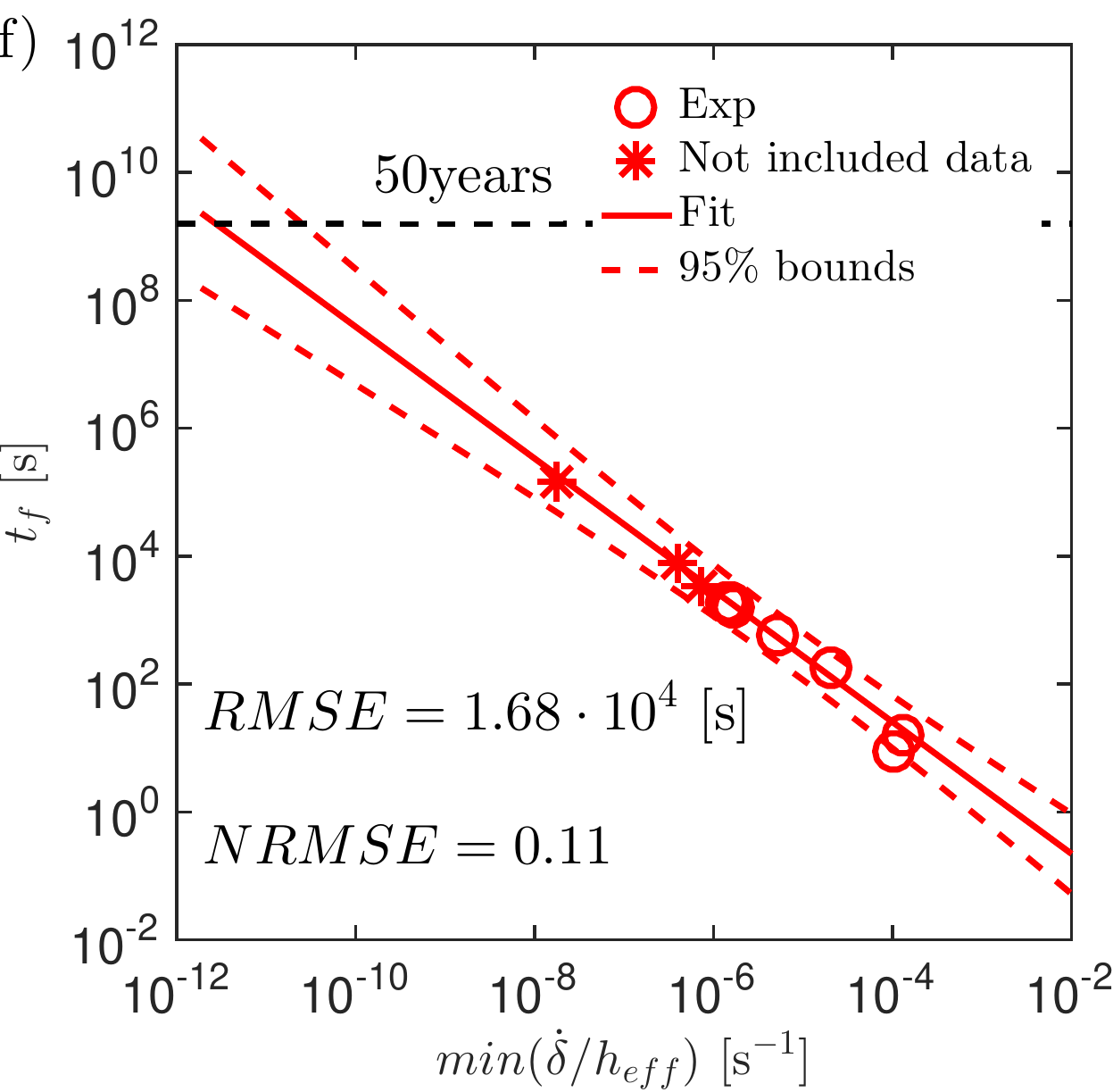}
  \end{subfigure}
\caption{Evaluation of data point sensitivity based on selected data following the three cases (I), (II), and (III): P1 (a-c), and P2 (d-f).} 
\label{fig:PrTest}
\end{figure}

The robustness of the Monkman-Grant relation is also studied in this manuscript in order to determine the method's predictive capability and establish the required range of experimental data for the fit. 
For that reason, the data is resampled to reflect three different cases, (i) creep rates between the minimum and maximum are neglected, (ii) the higher creep rates are neglected, and (iii) the lower creep rates are neglected. 
The first scenario is expected to provide the best fits, since it is a typical interpolation problem. On the other hand, cases (ii) and (iii) serve as a test of the accuracy of the extrapolation. 
In particular, case (iii) is the most interesting, since the failure times of specimens of lower load levels, and lower creep rates, need to be predicted. The results of this test are shown in Fig.~\ref{fig:PrTest}. 
It is obvious that the quality of predictions is high for all cases, although case (i) provides the best results for both products. 
The quality is estimated by two widely used statistical estimates, the root mean squared error (RMSE) \cite{HYNDMAN2006679} and the normalized squared error (NRMSE) which is defined as the fraction of the RMSE over the range of the observed failure times,  NRMSE=RMSE$/(t_{f_{max}}-t_{f_{min}})$. 
The closer to zero is the value of NRMSE, the better the prediction of the model.  
The evaluation of the fit on the data of P1 shows that case (i) gives the best results in terms of prediction uncertainty. 
However the NRSME found to be the larger among the 3 cases.
This can be explained by the fact that the range of the $(t_{f_{max}}-t_{f_{min}})$ is quite small, vertical distance in Fig.\ref{fig:PrTest} (a).
The best prediction based on the NRSME is obtained in case (ii), see Fig. \ref{fig:PrTest} (b), while also the prediction of case (iii) is quite good. 
However, the uncertainty bounds are much wider in case (iii), see Fig. \ref{fig:PrTest}(c). 
This fact reveals that this method requires also lower creep rates, or failure times at least until $10^5$~s,  in order to extrapolate more accurately the failure times. 
On the other hand, for the adhesive P2, the best prediction is obtained for the case (i), Fig. \ref{fig:PrTest}(d), while the fits of the cases (ii) and (iii), predict with high accuracy the excluded failure times, Fig. \ref{fig:PrTest}(e) (f). 
The fit without lower creep rates, for the data of P2, also exhibits an increased uncertainty in far extrapolations. 
Nevertheless, this is not as pronounced as in the case of P1. 
This can be explained by the fact that the data obtained from P2 are more equally distributed, while the data of P1 are more clustered. 
Thus, larger uncertainties have to be expected if for some regions of the curve data points are missing. 
In either case, the method performs quite well as a prediction method, i.e. the not considered data points lie on the mean value prediction (solid line). 
A recommendation would be that the fit of Eq.~\ref{eq:MGL} should be performed in a sufficient wide range of creep rates, depending on the inherent experimental scatter. 
The selection of this range should be based on predicted uncertainty bounds for a given extrapolation time. 
For example the extrapolation in Fig.~\ref{fig:PrTest}(c) results in confidence bounds on the failure time between $200$~days and $600,000$~years, for the same creep rate, making clear that the fit needs to take into account also lower creep rates.
In the latter case the uncertainty bounds shrink down to $4$ and $600$~years. % IB place here the right numbers please. 
%RWW Done
   
\begin{figure}[h!]
  \includegraphics[width=0.5\textwidth]{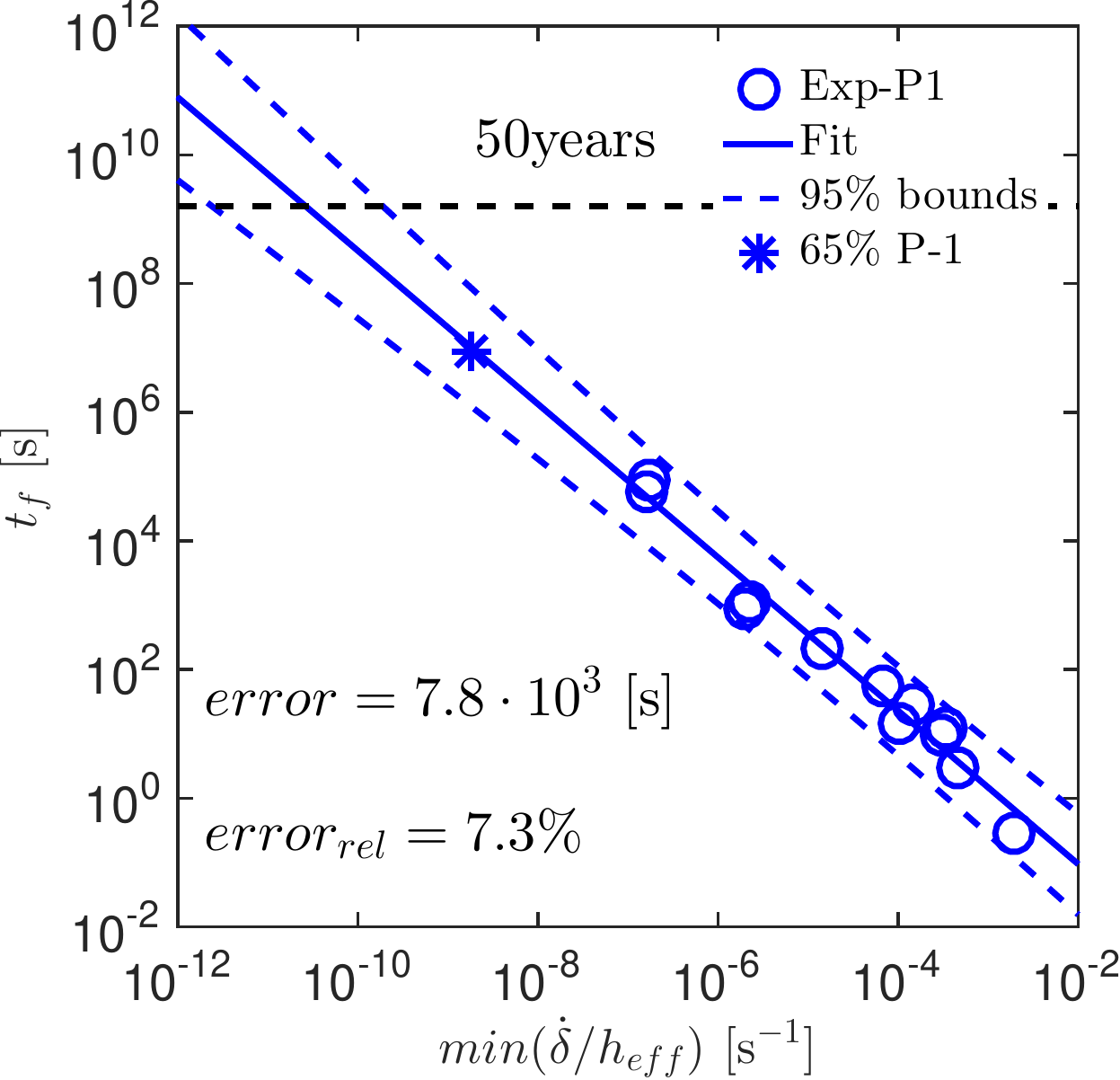}

 \caption{Validation of MG relation by the failure time of a specimen of P1, not taken into account in the analysis.}
 \label{fig:65PC}
 \end{figure}

The MG criterion is also validated on the failure of one specimen of P1, loaded at $65\%$ which failed after the main analysis was completed. 
In the beginning two of the specimens loaded at $65\%$ failed after $56233$~s and $85425$~s, respectively. 
% IB make sure that we have , for numbers larger than 1,000
% RWW Checked
The third specimen failed after a much longer time of $107$~days. 
This high scatter ($170\%$) complicates the estimation of a mean failure time from a stress versus time to failure curve. 
% IB we should plot the data points in the usual way of a TTF plot to illustrate the large scatter
% RWW done I placed the figure in experimental section
However, the MG criterion, being formulated as function of the stable creep rate is not affected and manages to predict the failure of the anchor with high accuracy, corresponding to an error of only $7.3\%$. 
The reason, both for the large scatter in failure time and the success of the MG criterion, stems from the large differences in observed creep for the same relative load level. 
The creep rate of the long lasting specimen ($1.862\cdot 10^{-9}$~1/s) was 2 orders of magnitude lower than that of the other two anchor specimens, ($1.608\cdot 10^{-7}$, and $1.669\cdot 10^{-7}$~1/s), a fact that was taken into account by the MG relation. 
The MG criterion predicts for the observed stable creep rate a failure time of $114.8$~days, while the actual failure time was $107$~days. 
The comparison of the MG prediction with the actual failure time of this specimen (asterix) is shown if Fig.~\ref{fig:65PC}. 

In all presented cases, for both investigated products, the MG relation manages to predict well data points that have been excluded from the analysis even though the uncertainty bounds open up considerably in some cases. 
This clearly proves the criterion's high predictive quality. 
It is also interesting to note that the MG criterion is relatively insensitive to the experimental scatter in failure times which represents a large obstacle for more traditional analyses of TTF data.

\subsection{Geometry dependence} 
 
\begin{figure}[h!]

\begin{subfigure}[b]{0.5\textwidth}
       \includegraphics[width=1\textwidth]{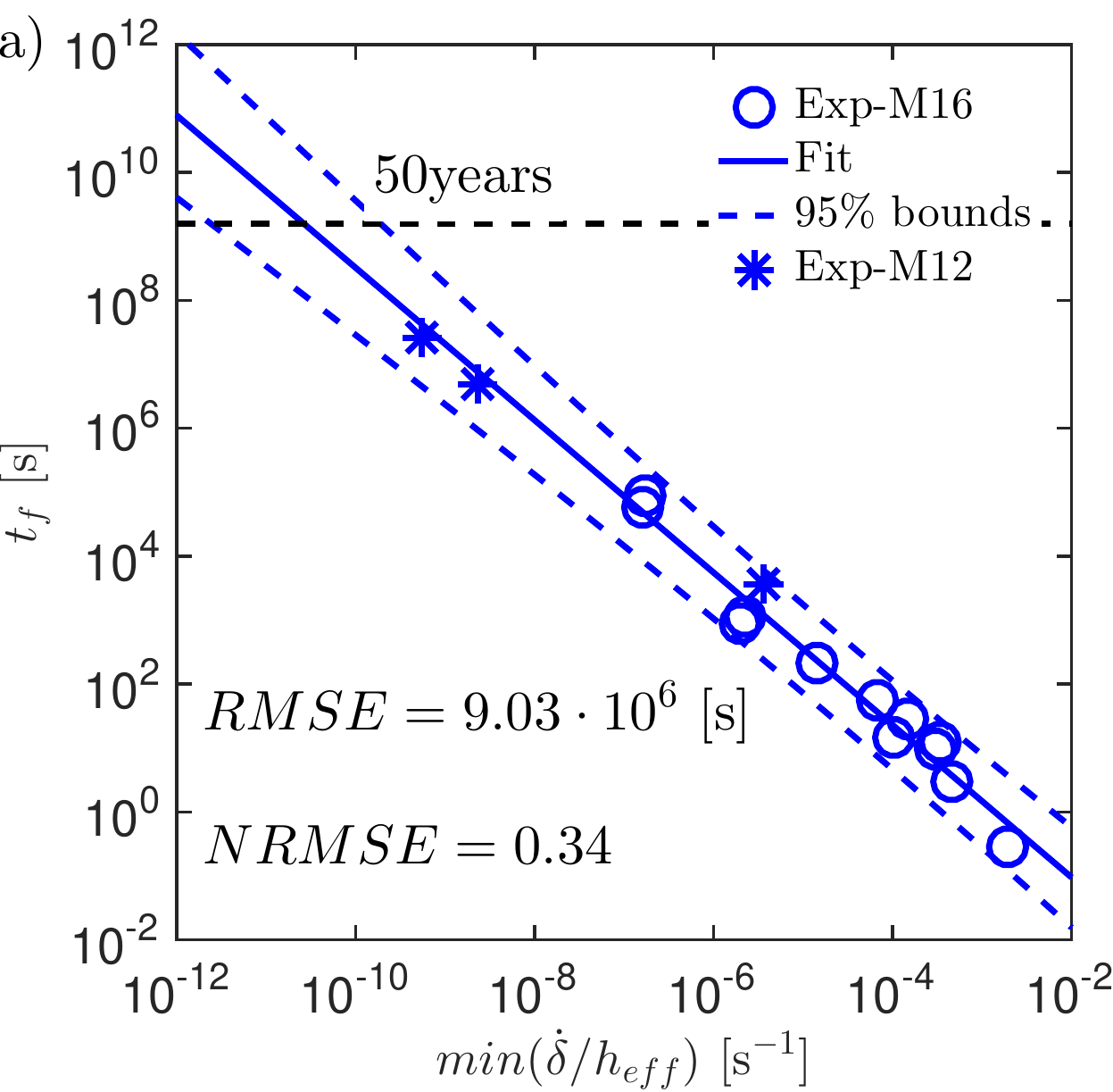}
  \end{subfigure}
   \begin{subfigure}[b]{0.5\textwidth}
      \includegraphics[width=1\textwidth]{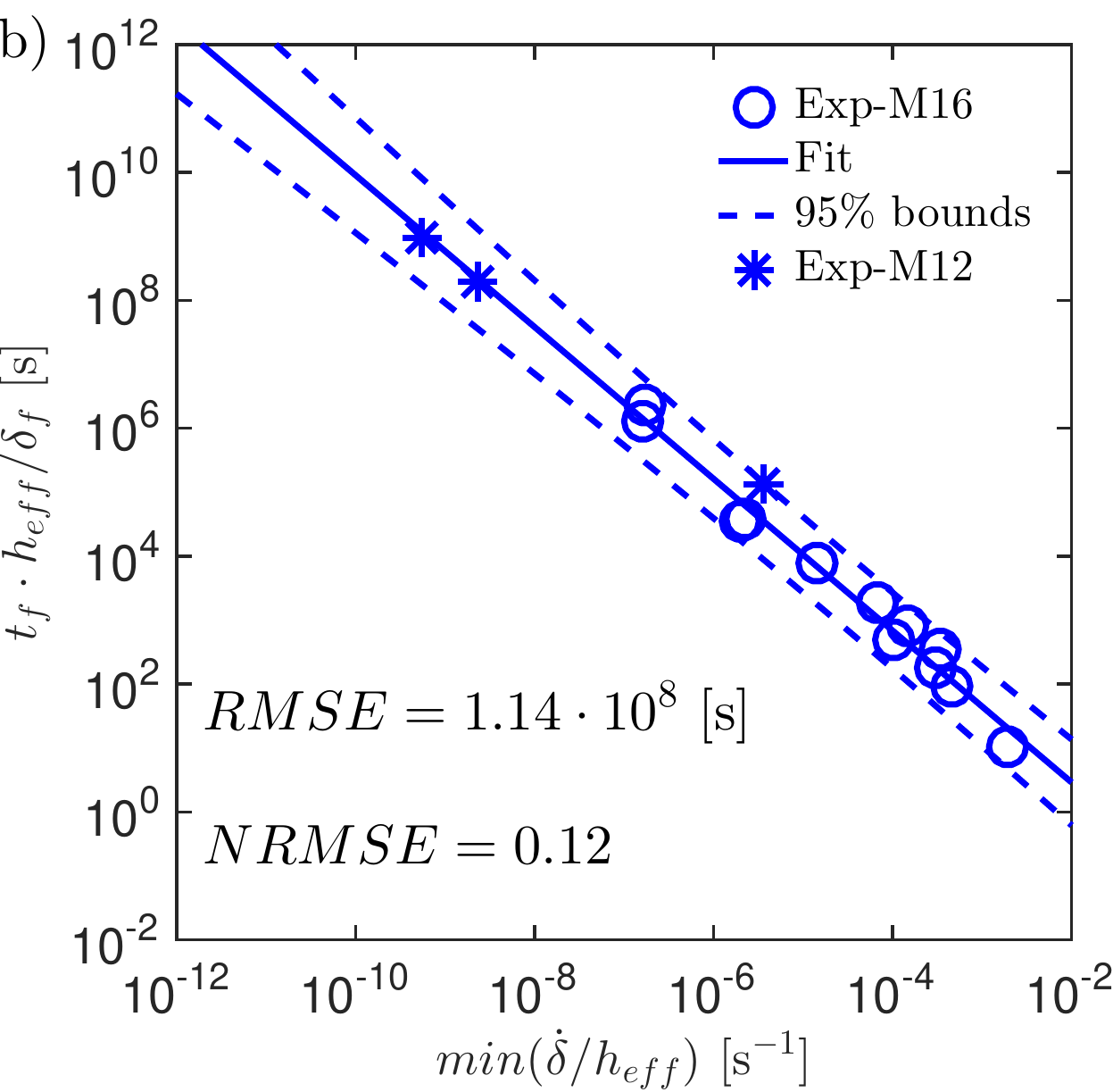}
  \end{subfigure}
 
\caption{Comparison of the fitted MG (a) and MMG (b) relation to minimum creep rates versus failure times for adhesive anchor systems of P1, tested on different geometry \cite{WendPoda, WendPodb}.}
\label{fig:HiltiCheck}
\end{figure}

After establishing the MG criterion's predictive quality for consistent data sets on a single concrete, anchor rod diameter, and embedment depth the investigation is extended to different anchor geometries. 
In particular, the time-displacement curves of $3$ bonded anchor systems, found in \cite{WendPoda, WendPodb} are used together with the previously introduced data. 
The anchors of these tests had an embedment depth of $72$~mm, a diameter of $12$~mm, and were bonded to concrete with adhesive mortar P1. 
Fig.~\ref{fig:HiltiCheck} shows the comparison of the actual data and the prediction (solid line) based on the Monkman-Grant equation. 
The error in the prediction (vertical difference between marker and line) is relatively small, considering that the fit can correctly predict the order of magnitude of the failure times, especially in case of the MMG predictions, see Fig.~\ref{fig:HiltiCheck}(b)

Based on the limited evidence presented it seems that the MG relation can correctly predict failure times associated with different anchor geometries and concretes, at least as long as the failure mechanism does not change. 
This indicates that a product specific Monkman-Grant constant could be potentially established for each adhesive anchor system.
% IB please make sure that we use everywhere adhesive anchor
The only parameters that could vary the constant are, of course, the adhesive itself -- product dependency -- and the annular gap between anchor rod and bore hole, i.e. the thickness of the adhesive mortar layer. 
In the studied case the mortar had the same thickness, of $1$~mm, allowing the direct comparison of the different cases. 

This argument becomes stronger when additional independent data points for the same adhesive P1 are considered. 
In a recent publication, O\v{z}bolt et al. \cite{Ozbolt} present the displacement time data for 6 bonded anchors (P1) loaded at $65\%$ and $85\%$, respectively, until failure. 
The tests were performed on adhesive anchors of $12~$mm diameter and $h_{ef}=72~$mm at a controlled temperature of $23^o$C, while the bond strength was reported as $30$~MPa. 
Fig.~\cite{Ozbolt} shows the comparison of two independently obtained MG fits on P1 of adhesive anchors M12 and M16. 
It is clear that both data sets provide virtually identical results, especially concerning the mean 50-year prediction.

\begin{figure}[h!]
 
       \includegraphics[width=0.5\textwidth]{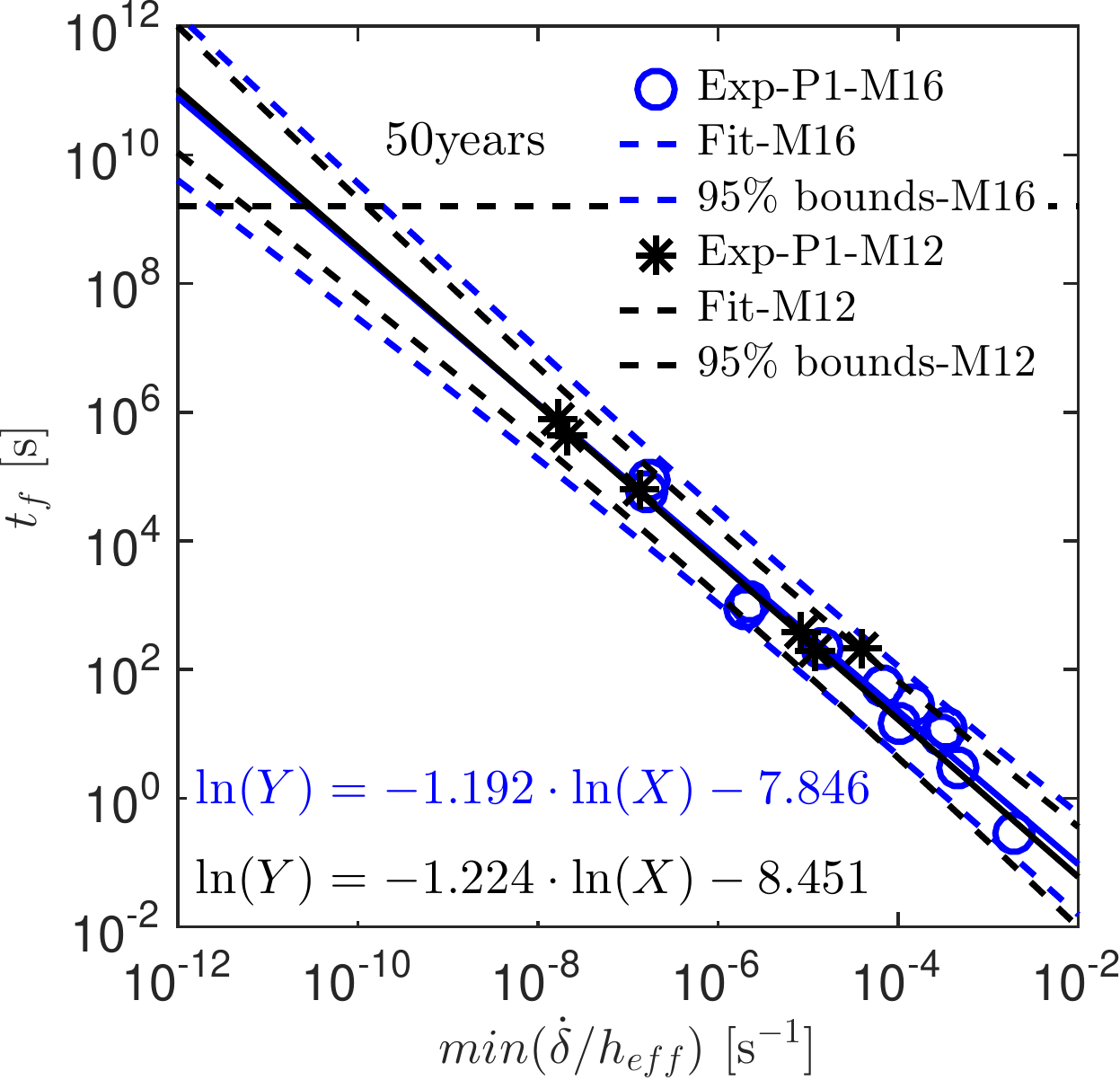}
        
\caption{Comparison of the independent MG fits on data of P1 of M16 with $h_{eff}=75$~mm, and data of P1 of M12 with $h_{eff}=72$~mm (data extracted from \cite{Ozbolt})  } 
\label{fig:Ozbolt}
\end{figure}

\section{Creep rate - stress relation } % IB please make sure we use everywhere in the paper the same word "relation"
\label{sec:State:MODELCAL}
After a master curve of the failure times as a function of minimum creep rate is established, a relationship between the creep rate and the applied shear stress $\tau$ is required in order to recover a functional relationship between allowable stress and desired life-time. 
For metals it is widely accepted that the relationship between minimum creep rate and applied stress follows a power-law \cite{Cao}. 
However, the exponent is found not to be constant but a function of applied stress. 
Additionally, for high load levels the power-law % IB please check if everywhere the hyphen is used
% RWW checked power law is used everywehere with hyphen
function fails to fit the experimental data, i.e. there is no exponent that could describe the experimental data. 

In this investigation two approaches are investigated -- a power-law fit and the functional form proposed previously for the rate-effect on strength of concrete.
For the purpose of this investigation, the applied shear stress is determined based on the uniform bond law according to Equ.~\ref{eq:BL}. 
All fits are obtained on the natural logarithms of the creep rate and the applied stress.

\subsection{Power-law relation}

\begin{figure}[h!]

\begin{subfigure}[b]{0.5\textwidth}
       \includegraphics[width=1\textwidth]{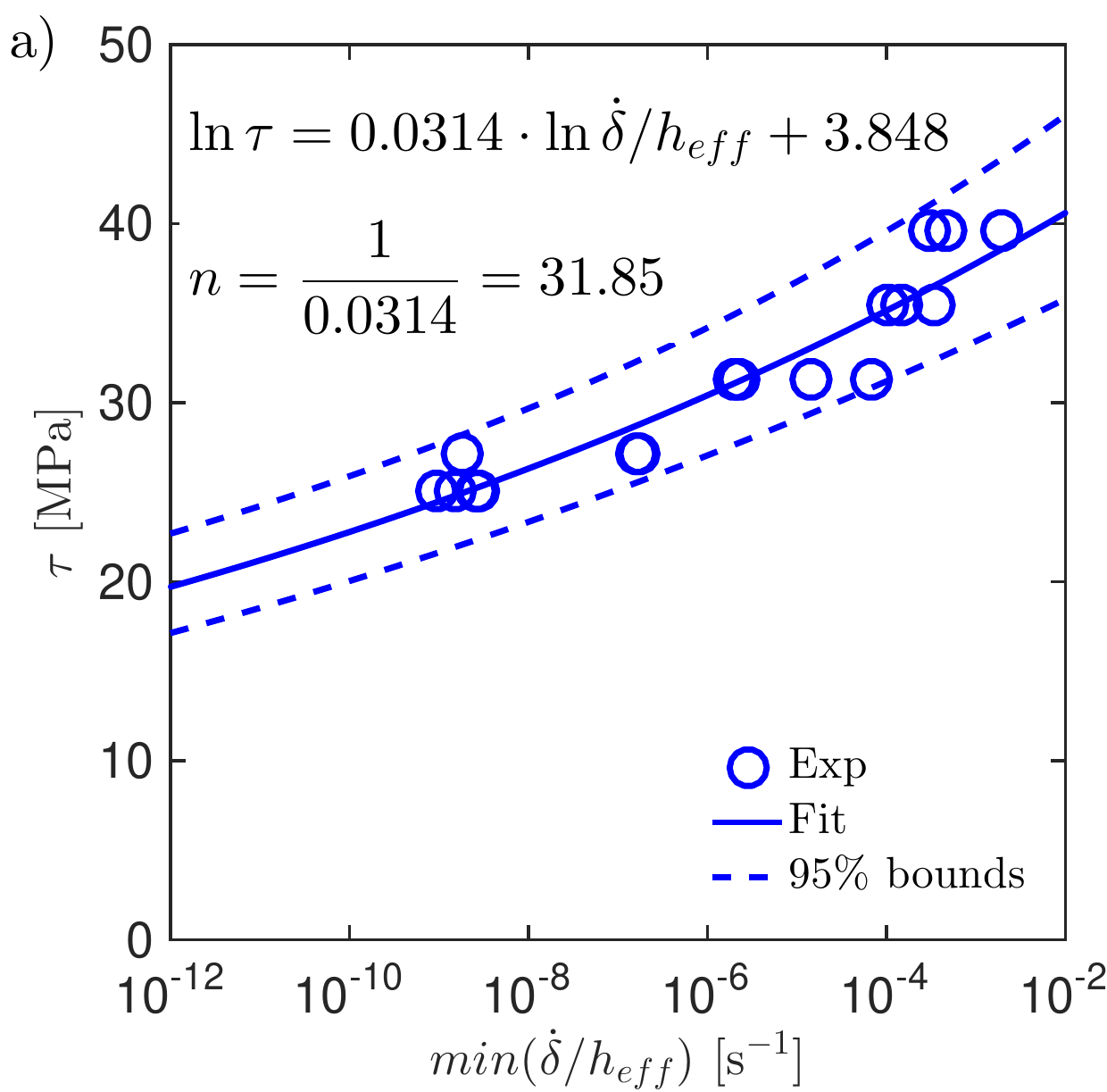}
  \end{subfigure}
   \begin{subfigure}[b]{0.5\textwidth}
      \includegraphics[width=1\textwidth]{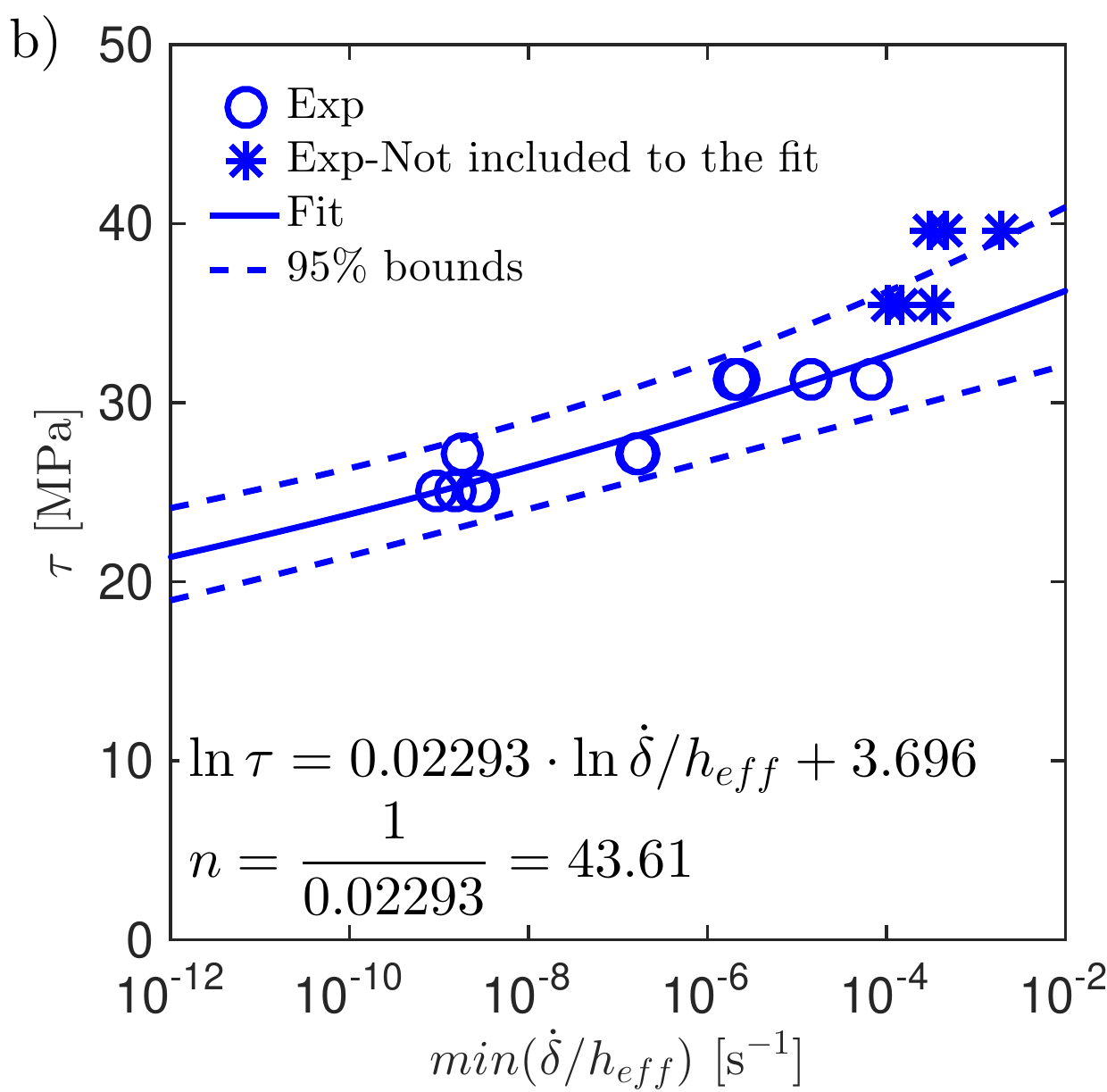}
  \end{subfigure}
 
\begin{subfigure}[b]{0.5\textwidth}
       \includegraphics[width=1\textwidth]{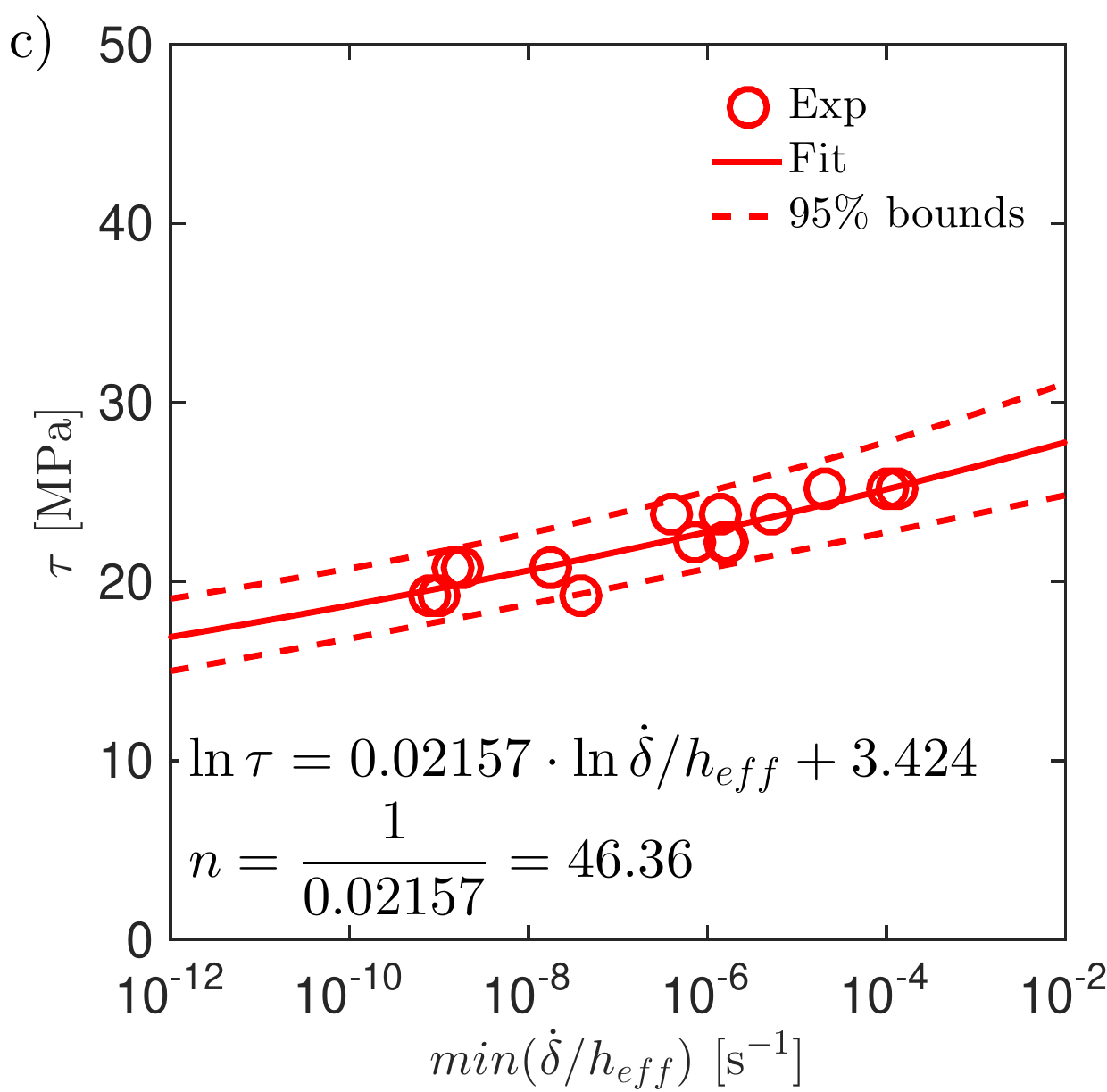}
  \end{subfigure}
   \begin{subfigure}[b]{0.5\textwidth}
      \includegraphics[width=1\textwidth]{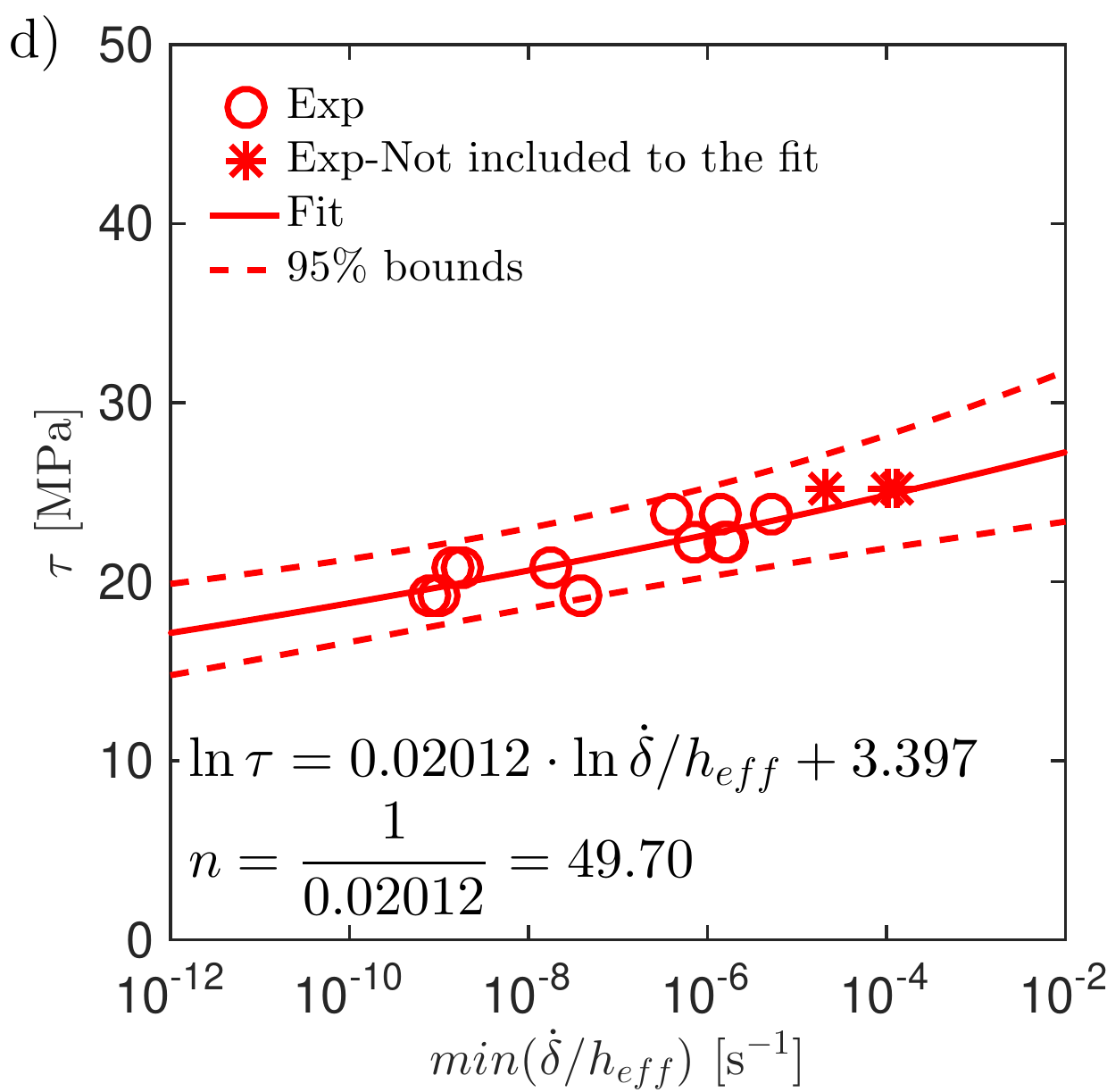}
  \end{subfigure}
\caption{Power-law relationship fitted on the entire data (a,c) and on the reduced data of load levels $\leq70\%$ (b,d): P1 system (a,b), and P2 system (c,d) respectively.}
\label{fig:StresvsRatePL}
\end{figure}

Initially, the suitability of a power-law is studied on data presented in Table \ref{tab:LLv}.
% IB which ones are these ? link to an earlier figure
% RWW I linked them to the table where the load levels are give.
First a fit of the entire data set was performed, see Fig.~\ref{fig:StresvsRatePL}(a,c).
The best fit in this range was found for an exponent of $31.85$ for P1, and $43.61$ for P2. % IB show the fitted exponents in the subplots
% RWW done
Additionally, a power-law was fitted to the data neglecting the higher load levels ($\geq 80\%$). 
These fits are presented in Fig.~\ref{fig:StresvsRatePL}(b,d). 
In the latter case the best fit is found with an exponent of $46.36$ for P1 and $49.70$ for P2.

\subsection{Rate-theory based relation}

Due to the severe stress dependence of the power-law an alternative formulation is compared -- the functional form derived by Ba\v{z}ant \cite{BazRat} for the rate-effect on concrete strength. 
This concept is based on the bond distribution of the microstructure of the material, which is assumed to follow a Maxwell-Boltzmann distribution, and their breakage and restoration rates. 
% IB add here a short statement what this means -- is there a strss level for which the restoration rate is equal to the rate of breakege - no damage. <> polymer chains reconnecting ?
More specifically, the equation derived by Ba\v{z}ant scales vertically the cohesive behavior of the material and has been successfully used by Di Luzio \cite{diluzio-nonloinearcreep} and Boumakis et al. \cite{BouTert18} for modeling failure under low strain rates as observed for tertiary creep, and by Smith and Cusatis \cite{SmithCus} for failure at the high strain rates characteristic of e.g. projectile impact in concrete. 

The corresponding minimum creep rate versus sustained stress relationship applied to the studied problem reads

\begin{equation}\label{eq:RateBaz}
\tau = \tau_{0}\left[1+c_{1}\cdot\sinh^{-1}\left(\dot{\epsilon}/c_{2}\right)\right]
\end{equation}

where $\tau_{0}$ is a strain-rate independent stress, $c_{1}$ and $c_{2}$ are material parameters that have to be fitted and describe the increase in stress with increasing strain rate.

\begin{figure}[h!]

\begin{subfigure}[b]{0.5\textwidth}
       \includegraphics[width=1\textwidth]{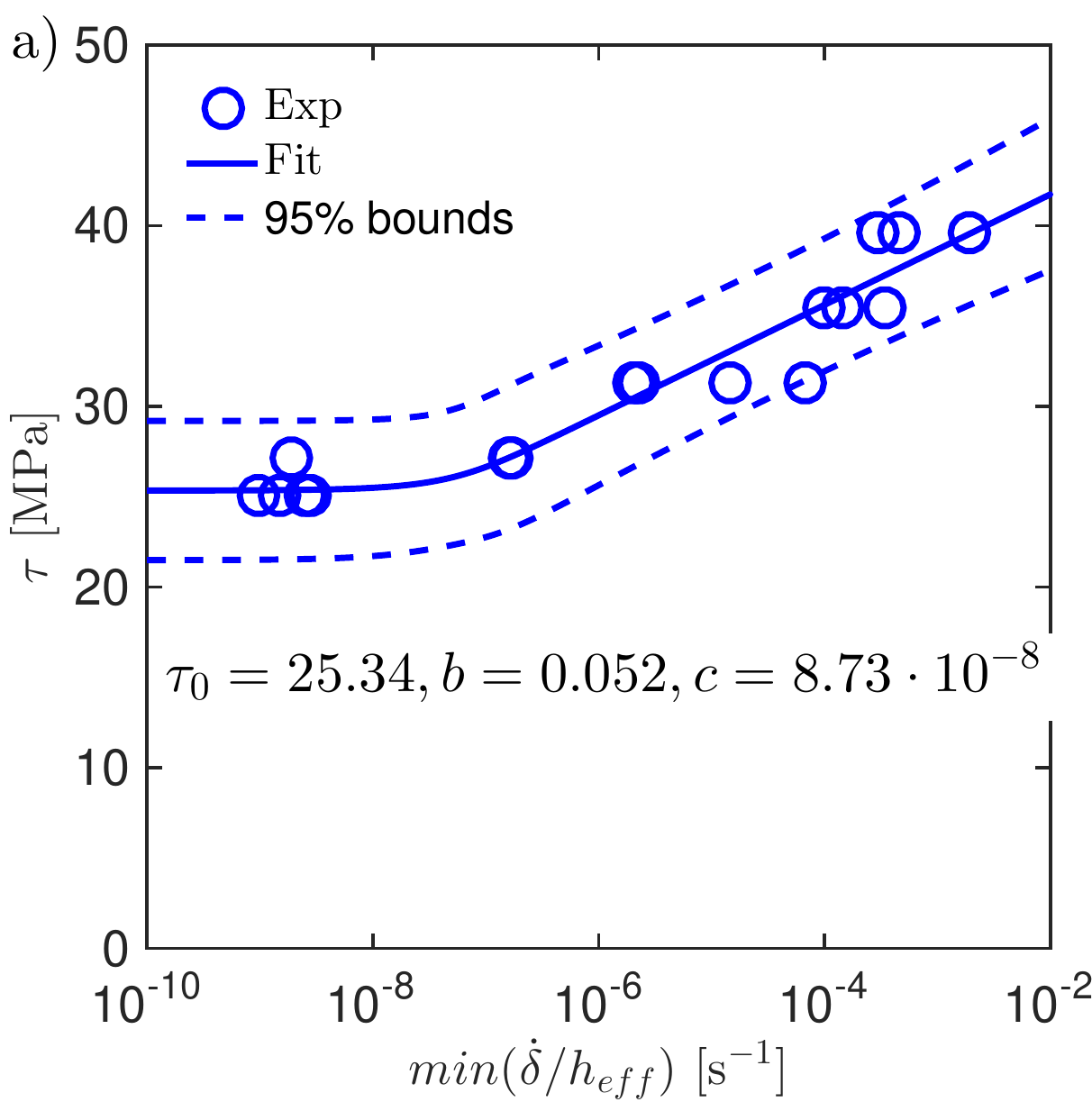}
  \end{subfigure}
   \begin{subfigure}[b]{0.5\textwidth}
      \includegraphics[width=1\textwidth]{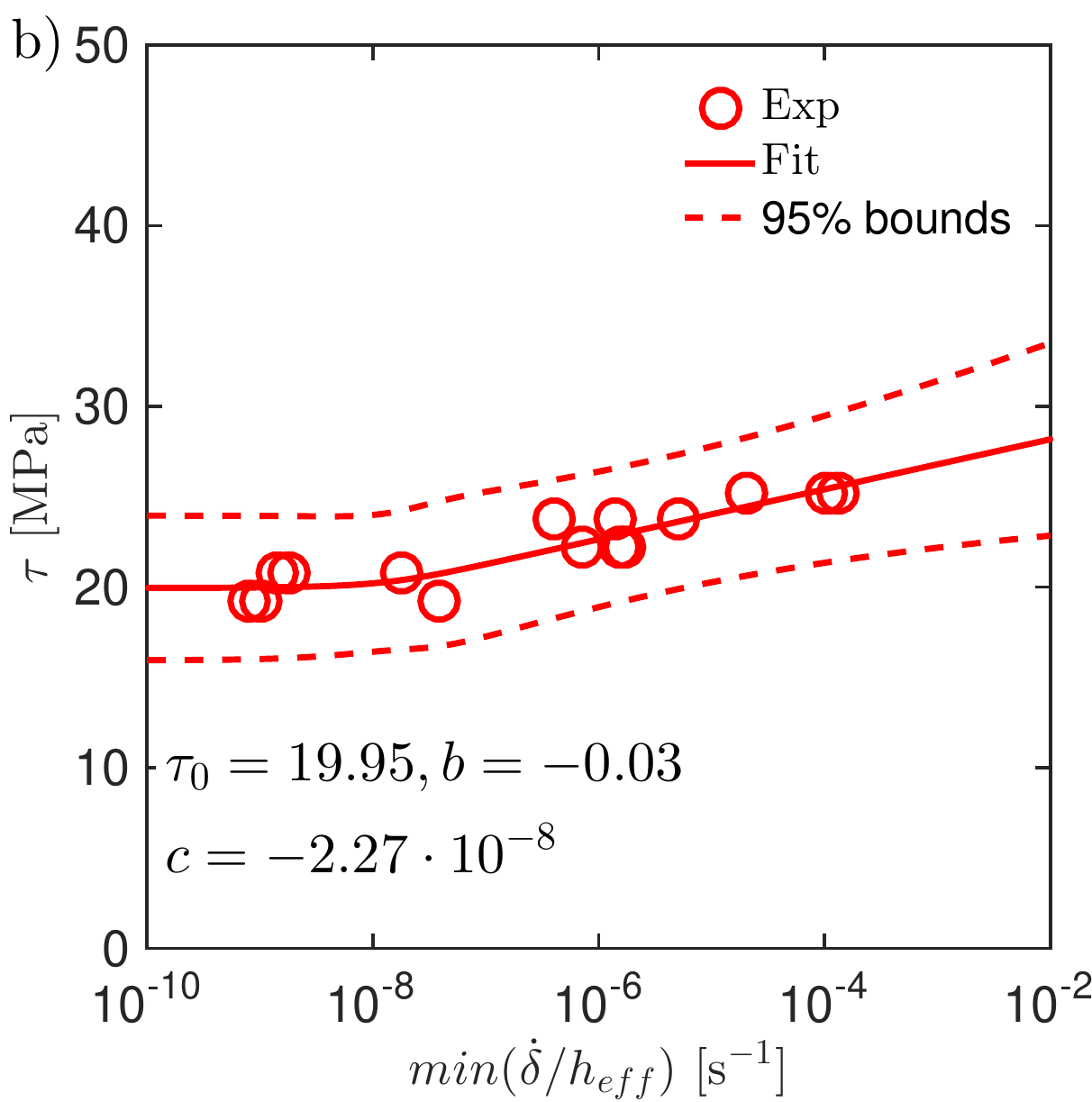}
  \end{subfigure}
 
\caption{Relationship of shear stress and minimum creep rate for the P1 (a) and the P2 (b) system.}
\label{fig:StresvsRate}
\end{figure}

It has to be noted that the above equation is of the same form as the Prandtl-Garofalo equation \cite{Garofalo}. 
The latter was introduced to overcome the problems of the power-law function. 
It finds a wide application to various materials \cite{Nabaro} and has been also justified based on the breakage and restoration of bonds \cite{RAGHAVAN19981673}. 
Fig.~\ref{fig:StresvsRate} shows the resulting fits for the previously investigated two products including the 95\% confidence bounds.
It can be seen that the data indeed seems to transition to a horizontal stress plateau for low strain rates, represented by $\tau_{0}$. 

However, this formulation does not allow the prediction of creep rates for stress values $\tau\leq\tau_{0}$ although the visco-elastic response of bonded anchors can be observed at quite low load levels, lower than the fitted values of Fig.~\ref{fig:StresvsRate}. 
This apparent inconsistency can only be reconciled if we assume that a certain stress level exists below which no damage is initiated and preexisting micro-damage is unable to propagate. 
Such a stress level can be explained by the rate-theory of bond breakage and restoration \cite{RAGHAVAN19981673}.
In such a case creep still exists but is characterized by a purely decaying creep rate that approaches asymptotically zero and lacks the otherwise progressively growing damage contribution. 
Consequently, the creep rate does not accelerate again, no finite minimum creep rate exits, and no creep failure.
%
%In that case, the progressively growing damage contribution to the overall creep rate does not exist.
%Consequently, the creep rate simply decays according to a power-law and approaches asymptotically zero without ever reversing direction and, thus, exhibiting a finite minimum creep rate.
%
In other words, Eq.~\ref{eq:RateBaz} directly gives access to a predictor of the sustained load strength $\tau_{0}$.
Lower sustained stress levels will not result in tertiary creep failure but merely cause growing creep deformations.  
For that reason, the Monkman-Grant relation is only valid and meaningful above this stress level.

% HONZA, IB: this section bugs me. we should try to find a better explanation here. stress and strength, ...
% do we see an actual min. creep rate for low stress or did we simply not reach it yet .... so which number do we use in the plots?
% we can try to fit two exp. functions (growing+ decaying) to the creep strain vs. time data > find minimum
% we could also define our own relation that combines a linear function up to tau_0 and then smoothly transitions to a power-law; 
% can we add linear creep and damage to non-linear creep > formualte the rates  
% superposition of two phenomena - decaying creep rate and growing damage rate ...? this would make sense in a plot of deformation rate versus time ...

% IB what are the creep rates for stresses << tau_0 in our other tests? 

\section{Time to failure functional form}
In the following section the two previously introduced approaches are combined and the stress versus time to failure curve is reconstructed. 
The failure times are estimated through the Monkman-Grant equation as introduced in section~\ref{sec:MG} for the minimum creep rates that are predicted by either the power-law or the rate-theory based model for a given applied stress (section~\ref{sec:State:MODELCAL}).

\subsection{TTF prediction based on power-law stress-rate relation}

\begin{figure}[h!]
%\vskip -1cm
\begin{subfigure}[b]{0.5\textwidth}
       \includegraphics[width=1\textwidth]{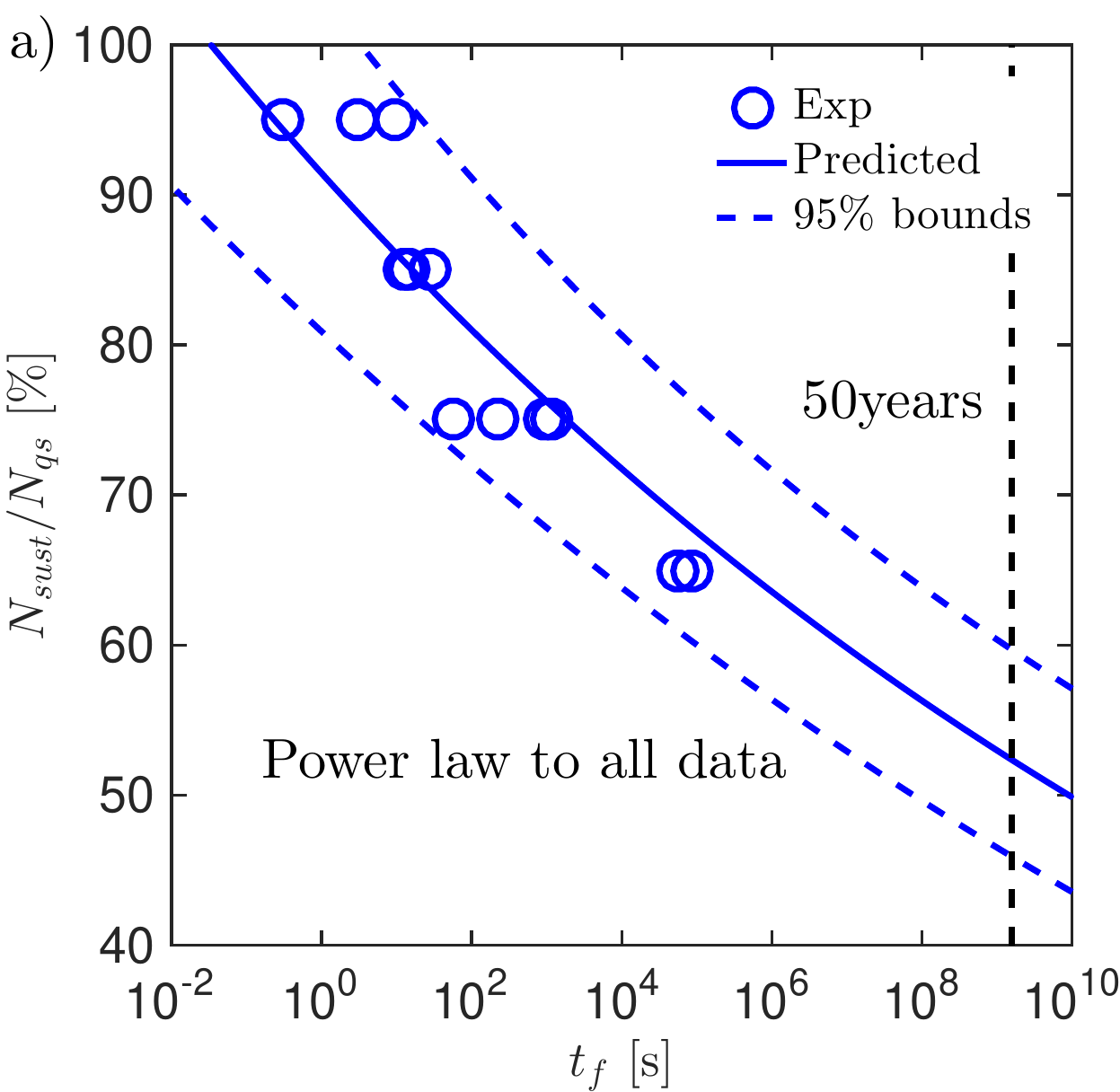}
  \end{subfigure}
%   \begin{subfigure}[b]{0.5\textwidth}
%%   
%      \includegraphics[width=1\textwidth]{./figures/Figxb_190301}
%%       
%  \end{subfigure}
  \begin{subfigure}[b]{0.5\textwidth}
       \includegraphics[width=1\textwidth]{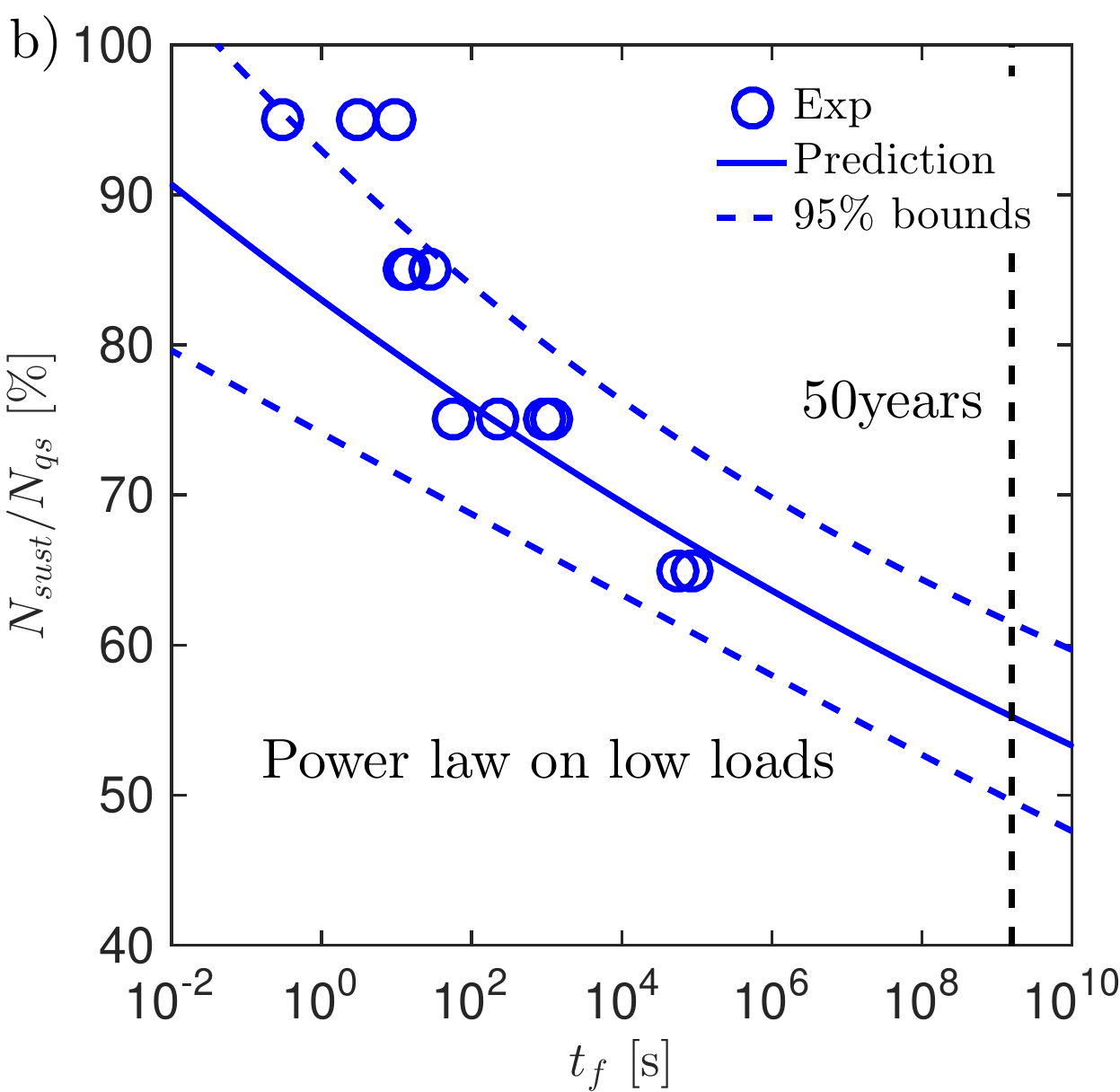}
  \end{subfigure}
  
  \begin{subfigure}[b]{0.5\textwidth}
       \includegraphics[width=1\textwidth]{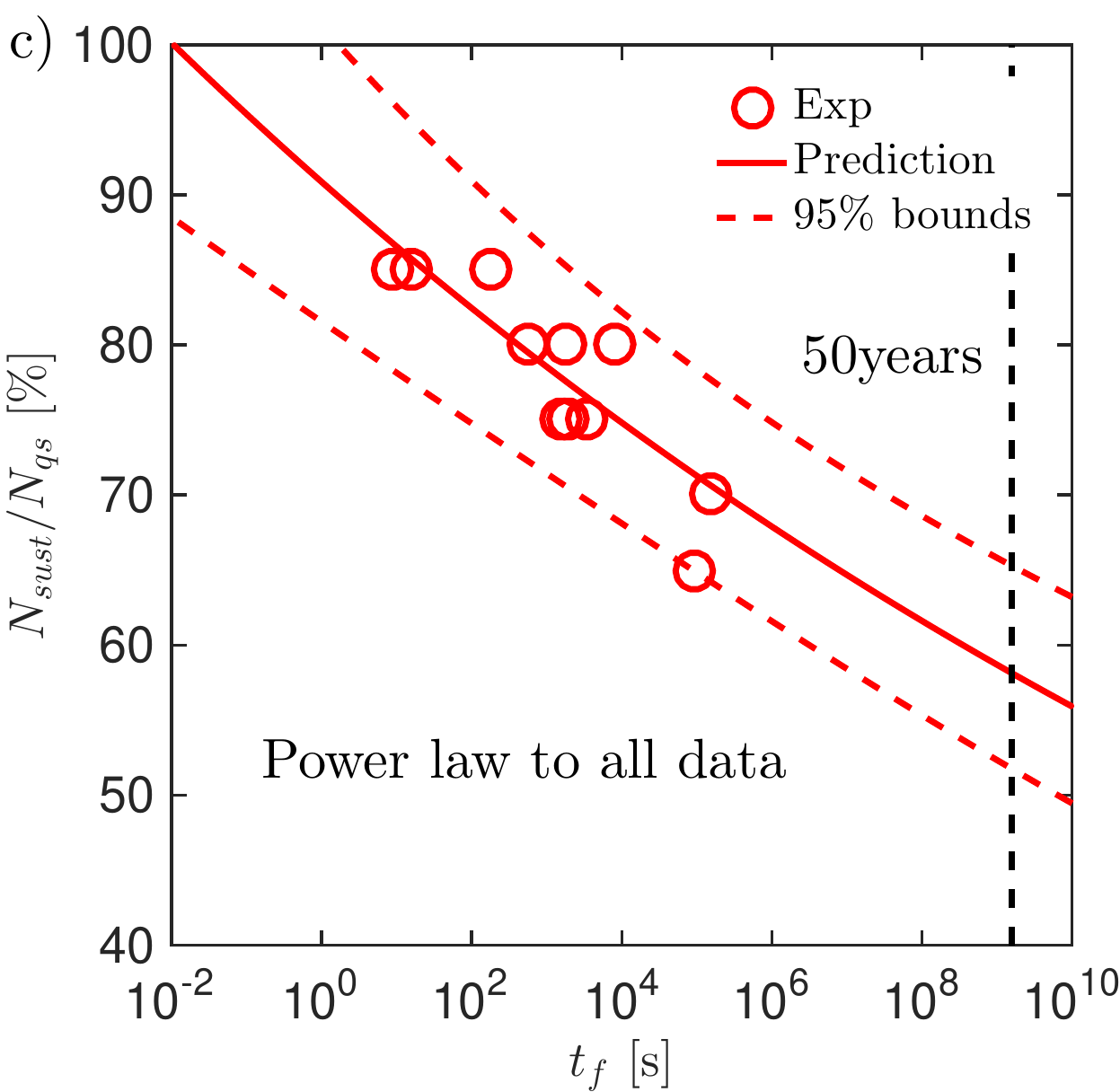}
  \end{subfigure}
%   \begin{subfigure}[b]{0.5\textwidth}
%%   
%      \includegraphics[width=1\textwidth]{./figures/Figxb_190301}
%%       
%  \end{subfigure}
  \begin{subfigure}[b]{0.5\textwidth}
       \includegraphics[width=1\textwidth]{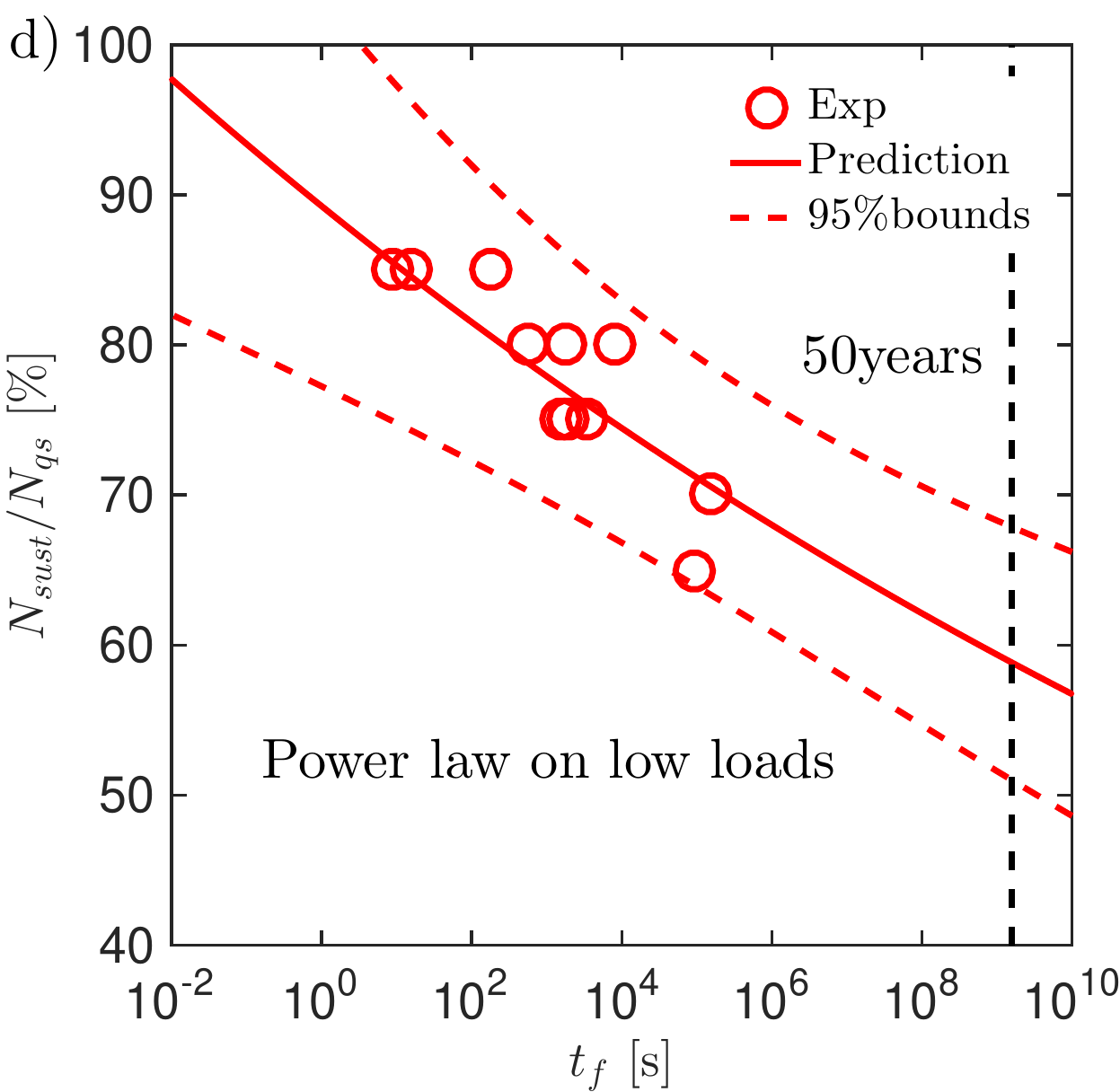}
  \end{subfigure}
\caption{Time to failure response derived by the failure criterion and a power-law stress-rate relationship, for the P1 (a-b) and for the P2 (c-d) adhesive anchor system: power-law fit to all loads (a,c), and low loads (b,d)} 
\label{fig:TTFPL}
\end{figure}

First the power-law function relating stress and creep rates is pursued.
Fig.~\ref{fig:TTFPL} shows the predicted time to failure curves for the $2$ different power-law fits that were studied in section~\ref{sec:State:MODELCAL} (the one for all data and the one ignoring the high stress level data) if combined with the MG relation.
All experimental data points lie within the 95\% confidence bounds.
Note the large experimental scatter of approximately two orders of magnitude between the minimum and maximum failure time at each load level.

If the power function is fitted to all available data it yields a mean allowable stress level of $52.4\% (\pm 12\%)$ for P1, and $58.2\% (\pm 12\%)$ for P2, given a target life-time without failure of $50$~years. 
On the other hand, when the power-law function is fitted only to the lower load levels the 50-year sustained load strength was found to be $55.2\% (\pm 11\%)$ for P1, and $58.2\% (\pm 14\%)$ for P2, respectively.
Although the prediction has a functional form that resembles the power-law proposed by Cook et al. \cite{cook_nchrp_2009} the proposed approach has two advantages: 
First, it removes the prohibitively large scatter in the TTF data from the fitting problem in time; and second, it allows the approximate use of lower load levels (that have not failed yet) to determine the relationship between stress and creep rate.
% IB: I don't understand the next sentence
%Thus the extrapolation is still conservative, but it can be used to predict a range of optimized load levels that will not lead to failure in the life-time of the system. .
 
\subsection{TTF prediction based on rate-theory stress-rate relation} 
 
\begin{figure}[h!]
%\vskip -1cm
\begin{subfigure}[b]{0.5\textwidth}
       \includegraphics[width=1\textwidth]{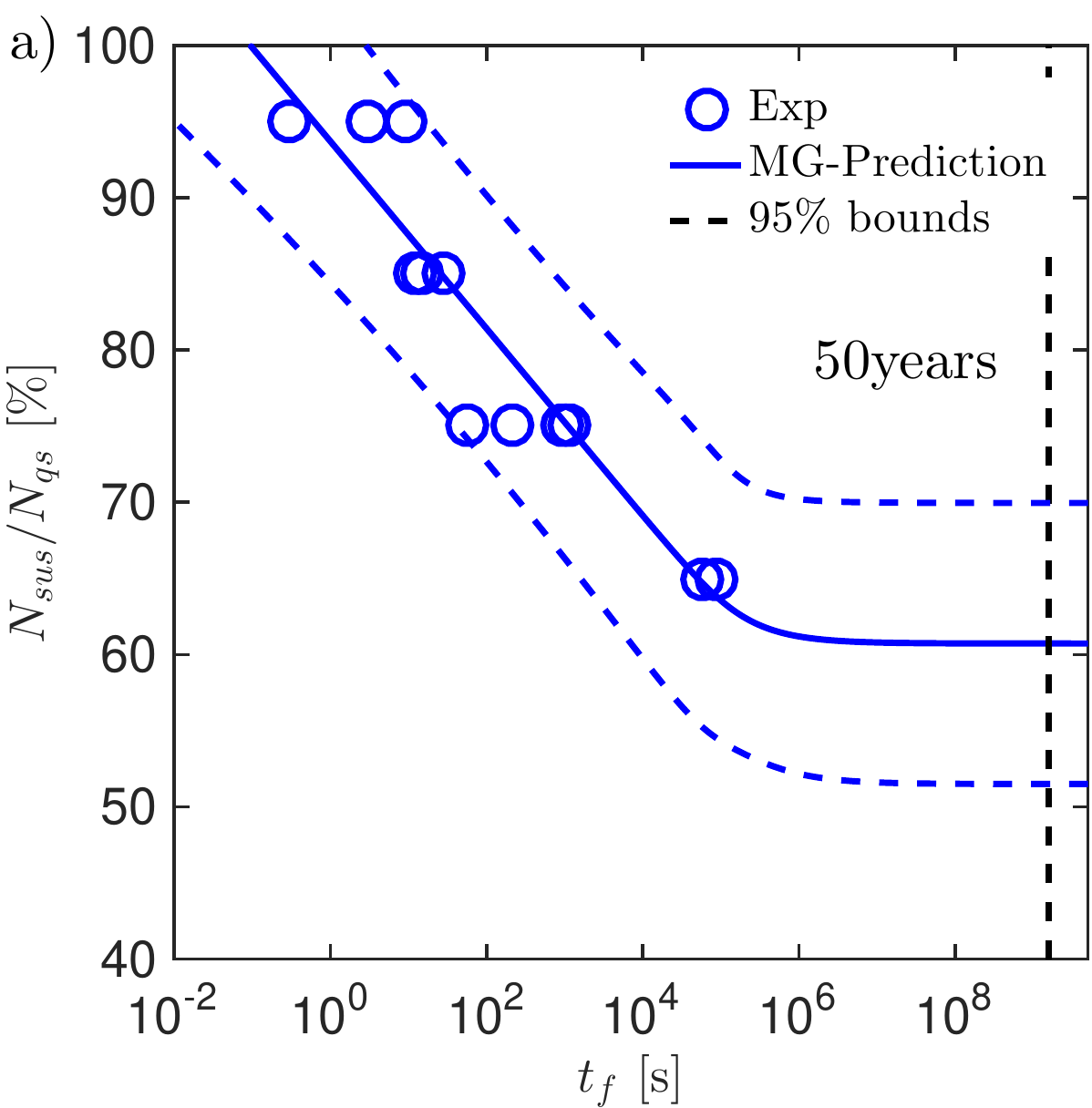}
  \end{subfigure}
%   \begin{subfigure}[b]{0.5\textwidth}
%%   
%      \includegraphics[width=1\textwidth]{./figures/Figxb_190301}
%%       
%  \end{subfigure}
  \begin{subfigure}[b]{0.5\textwidth}
       \includegraphics[width=1\textwidth]{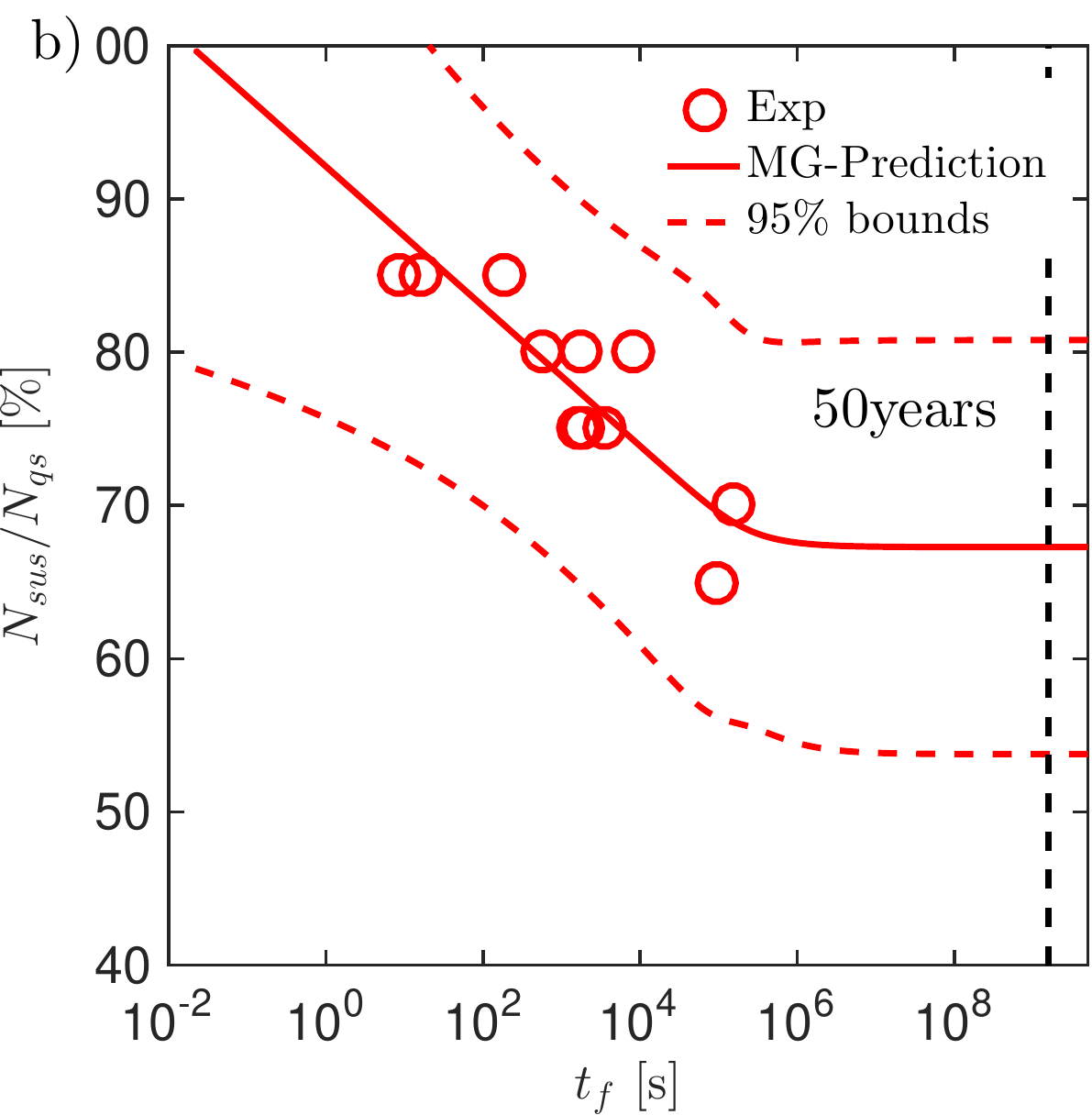}
  \end{subfigure}
\caption{Time to failure response derived by the failure criterion and the stress-rate relationship for the P1 (a) and the P2 (b) adhesive anchor system.} 
\label{fig:TTFRate}
\end{figure}

The same method is applied using the rate-theory based relationship between minimum creep rate and applied stress according to Eq.~\ref{eq:RateBaz}. 
 The results are presented in Fig.~\ref{fig:TTFRate}. 
 It is clear that the combination of this stress function with the MG relation yields excellent prediction of time to failure curves.
 The pure predictions recover the inherent features of the data that are an approximately linear response, in logarithmic times, for the higher load levels and relatively wide scatter bands. 
 The predicted transition point to a domain that is no longer controlled by propagating damage, i.e. a horizontal asymptote is consistent with the available data. 
 
% IB is that what you meant with transition point ?
% RWW I meant the following
% An additional advantage is that it is also capable of predicting the transition point, i.e. the load where the slope of the curve changes, to a domain that is not longer controlled by propagating damage. 
 For P1 a sustained load strength of $60.7\% (\pm 15\%)$ was found, while for P2 this limit is estimated as  $67.3\% (\pm 20\%)$. 
 The prediction of this equation can be validated also on the specimens that are still running at the time this paper is written ($60\%$). 
 Even if the specimens fail in the foreseeable future, their failure times will be inside the prediction bounds. 
 % iB please provide the time range where this is valid
 Nevertheless, until longer test series are available it remains questionable if the estimated sustained load strengths are reliable. 
 %A potential use of such a curve could be limited to the prediction of short failure times, and for the determination of the transition point. Then the asymptotic value could be better estimated by using the data of this section of the curve, if they are available. 
 
\subsection{Discussion} 
 
 This section illustrated the pure prediction of time to failure curves including the associated confidence bounds using the independently fitted Monkman-Grant criterion and two formulations linking the creep rates to sustained stresses -- a power-law and a rate-theory based hyperbolic sine function. 

 While the latter allows the direct determination of a sustained load strength at this point there is insufficient theoretical proof and no experimental confirmation.
 
 On the other hand, the power-law function can be fitted on creep rates at all load levels including those obtained by current approval tests (with and) without failure following the Findley approach \cite{findley_creep_1976, cook_nchrp_2013}. 
 Together with the Monkman-Grant relation it allows a probably conservative estimation of allowable sustained load levels for a given service life.
 
 In both cases tests without failure can be considered.
 To that end the minimum creep rate needs to be approximated e.g. based on the stabilized creep rate for a pre-determined strain.
 Such an approximation is probably conservative as the creep rates are always overestimated, ultimately leading to shorter life-time predictions.

 %IOANNIS here I need to rephrase
%    \begin{table}[h!]
%    \centering
%    \begin{tabular}{p{5.51 cm}cccc}
%    \hline\noalign{\smallskip}
%      $~$ & \textbf{P-1} & $~$ \\ 
%        \hline\noalign{\smallskip}
%         
%        \textbf{Function}  & \textbf{RMSE [s]} & \textbf{NRSME [-]} \\         
%        \noalign{\smallskip}\hline\noalign{\smallskip}
%         Power Function ($n=2.5$)      & $448.39$  & $0.0052$ \\
%         Power Function ($n=14.5$)    & $2.88\cdot10^{4}$  & $0.3374$\\
%         Rate  Function               & $1.138\cdot10^{4}$  & $0.1332$\\
%% BL2    & Confined     & $65$  & $74.95$ & $30.59$ \\
%% BL2    & Unconfined    & $65$  & $37.76$ & $-$ \\
%         \noalign{\smallskip}\hline
%         $~$ & \textbf{P-2} & $~$ \\ 
%        \hline\noalign{\smallskip}
%           \textbf{Function}  & \textbf{RMSE [s]} & \textbf{NRSME [-]} \\    
%         Power Function (All)       & $2.323\cdot10^{6}$  & $0.0197$ \\
%         Power Function (only lower loads)    & $3.106\cdot10^{6}$  & $15.4$\\
%        
%         Rate  Function               & $7.956\cdot10^{3}$  & $0.053$\\
%            \noalign{\smallskip}\hline
%    \end{tabular}
%    \caption{Root mean square error and normalized root mean square error for all cases.}
%    \label{tab:TestsSTA}
%\end{table}
       
%

\section{Application on literature time to failure data}
Finally, the presented approach is evaluated on six additional experimental time to failure data sets for different adhesives and anchor geometries found in the literature \cite{cook_nchrp_2013, cook_nchrp_2013A}. 
Additionally, the data sets contain tests performed at different temperatures. 
Hence, the hypothesis of the temperature independence of the Monkman-Grant relation, mentioned in \cite{Kim2002}, can also be tested.

The overview of the selected data is shown in Table~\ref{tab:TestsCook}. 
The time to failure tests were performed at load levels between $50\%$ and $80\%$, and the failure times were reported together with the failure displacements. 
The test results scattered significantly in terms of failure time. Additionally, some of the tests did not reach the design load level, and failed during the loading part. 
These inconsistencies could be for example explained by inconsistencies in the loading rates used during the determination of reference loads and in the actual creep test. 
A more extensive discussion on errors in time to failure tests is reported in Nin\v{c}evi\'{c} et al. \cite{Nincevic}, who show that the sustained load tests have to be performed using the same loading rate as in the short term tests. 
Fortunately, this high scatter is not an obstacle for the current analysis. 

% Need an explanation of the high scatter on B
\begin{table}[h!]
\centering
	\begin{tabular}{ccccc}
    %\begin{tabular}{p{2.5cm}p{1.6cm}p{2.1cm}}
    \textbf{Dataset} & \textbf{Adhesive} & \textbf{Diameter [mm]} & $\mathbf{h_{ef}}~$ 
   \textbf{[mm]} & \textbf{T [$^o$C]} \\
\hline
\centering
A1  & A & $16$ & $80$ & $43$\\
\centering
B1 &  B & $16$ & $80$ & $43$  \\
\centering
C1 &  C & $16$ & $80$ & $43$ \\
\centering
B2 &  B & $12$ & $80$ & $43$\\
\centering
B3  & B & $12$ & $80$ & $49$  \\
\centering
B4  & B & $12$ & $80$ & $21$ \\
\hline
\end{tabular}
\caption{Details of time to failure datasets obtained from \cite{cook_nchrp_2013,cook_nchrp_2013A}.}
\label{tab:TestsCook}
\end{table}

\subsection{Fit of MG and stress-rate relation}

\begin{figure}[h!]
%\vskip -1cm
\begin{subfigure}[b]{0.3\textwidth}
       \includegraphics[width=1\textwidth]{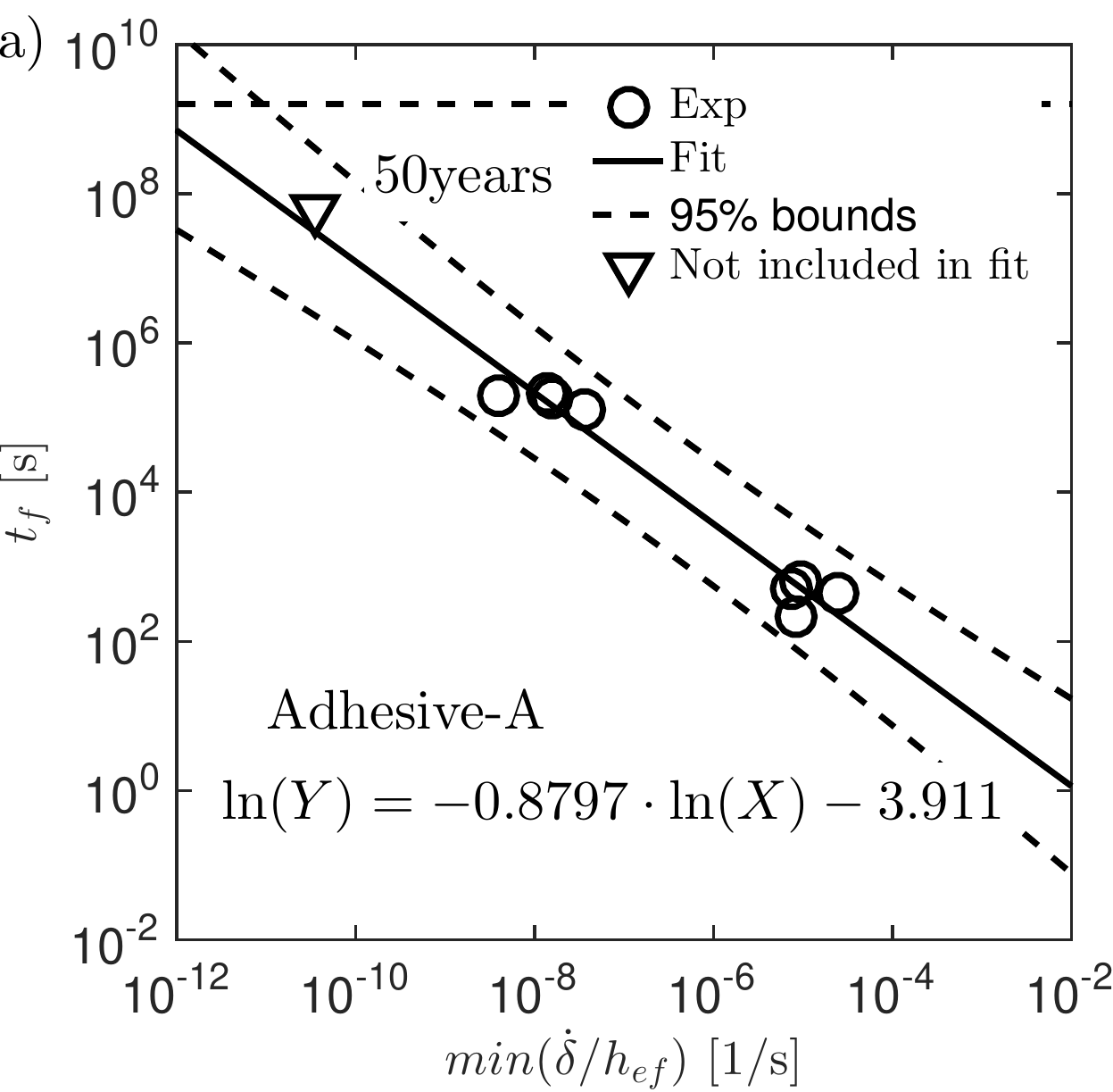}
  \end{subfigure}
%   \begin{subfigure}[b]{0.5\textwidth}
%%   
%      \includegraphics[width=1\textwidth]{./figures/Figxb_190301}
%%       
%  \end{subfigure}
  \begin{subfigure}[b]{0.3\textwidth}
       \includegraphics[width=1\textwidth]{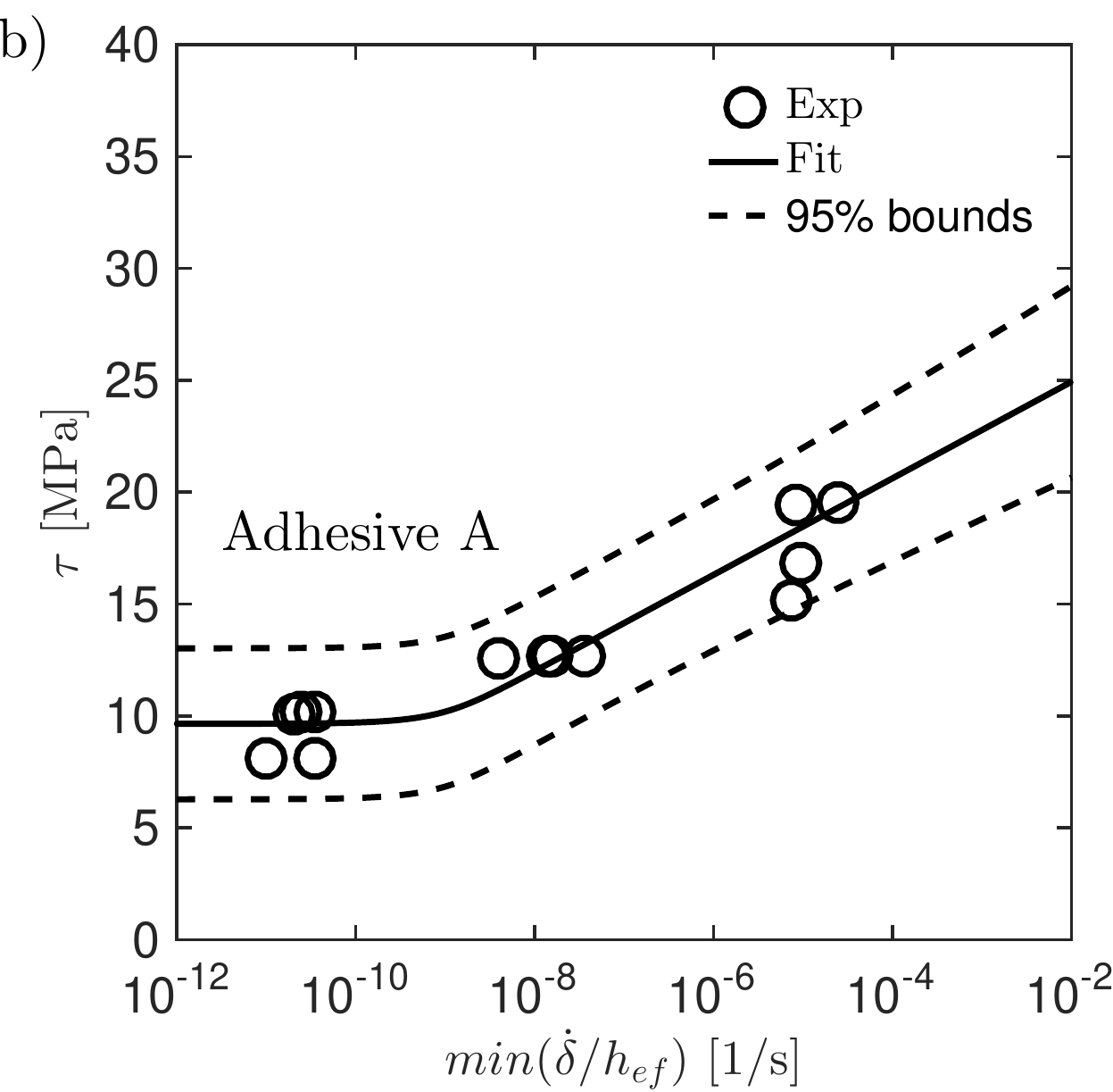}
  \end{subfigure}
    \begin{subfigure}[b]{0.3\textwidth}
       \includegraphics[width=1\textwidth]{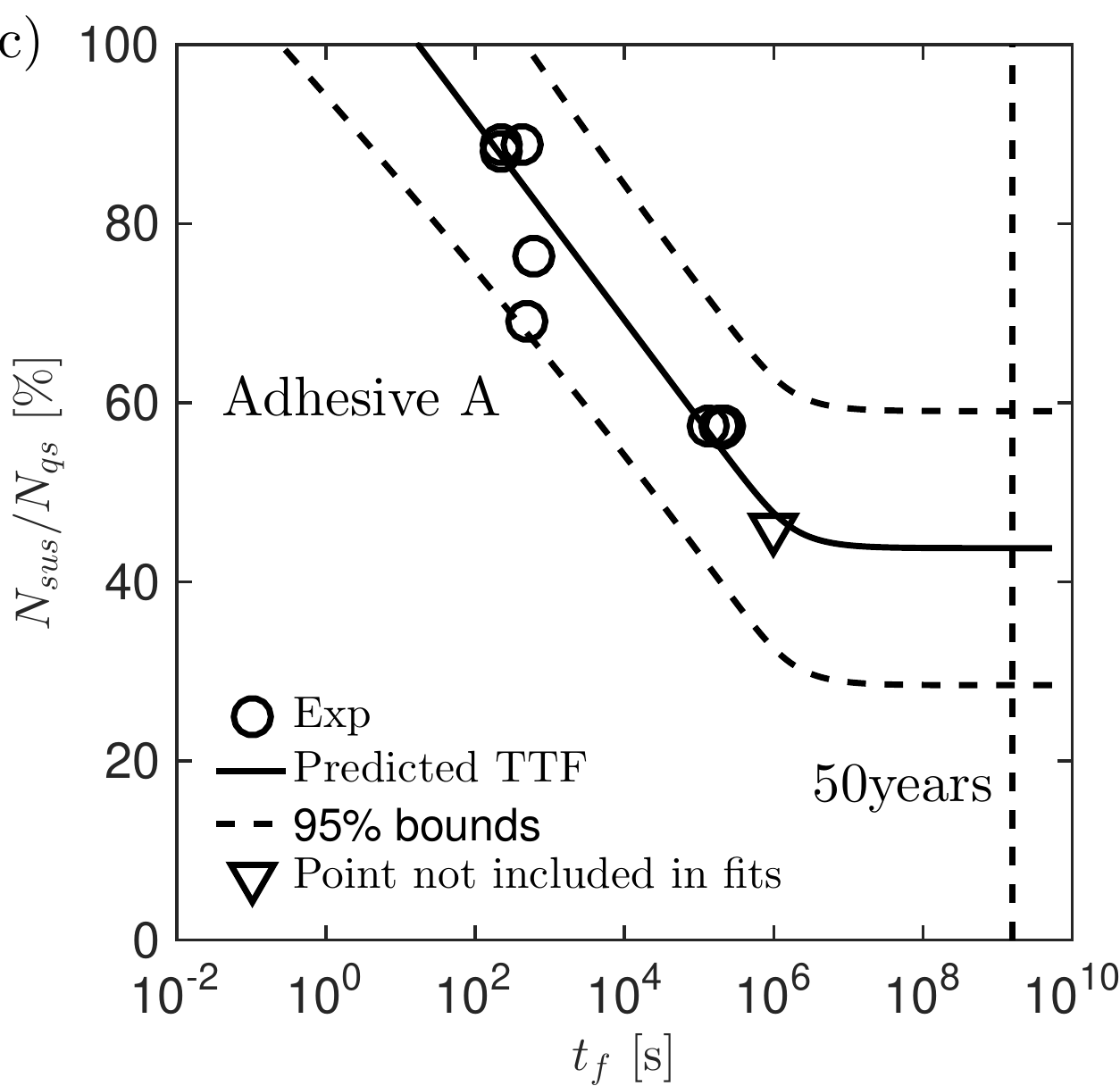}
  \end{subfigure}
%   \begin{subfigure}[b]{0.5\textwidth}
%%   
%      \includegraphics[width=1\textwidth]{./figures/Figxb_190301}
%%       
%  \end{subfigure}

  \begin{subfigure}[b]{0.3\textwidth}
       \includegraphics[width=1\textwidth]{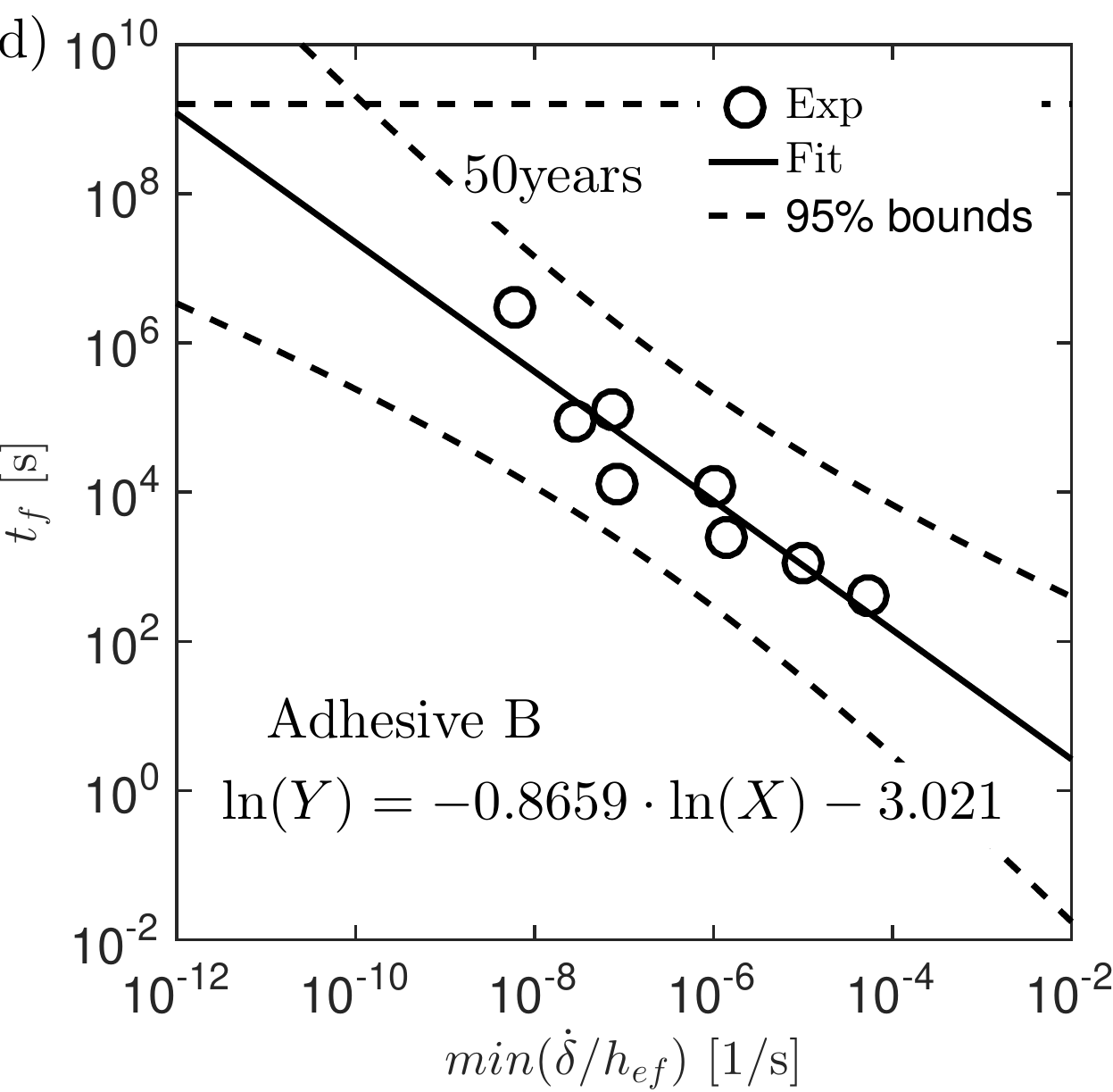}
  \end{subfigure}
    \begin{subfigure}[b]{0.3\textwidth}
%   VELowRates_190315
       \includegraphics[width=1\textwidth]{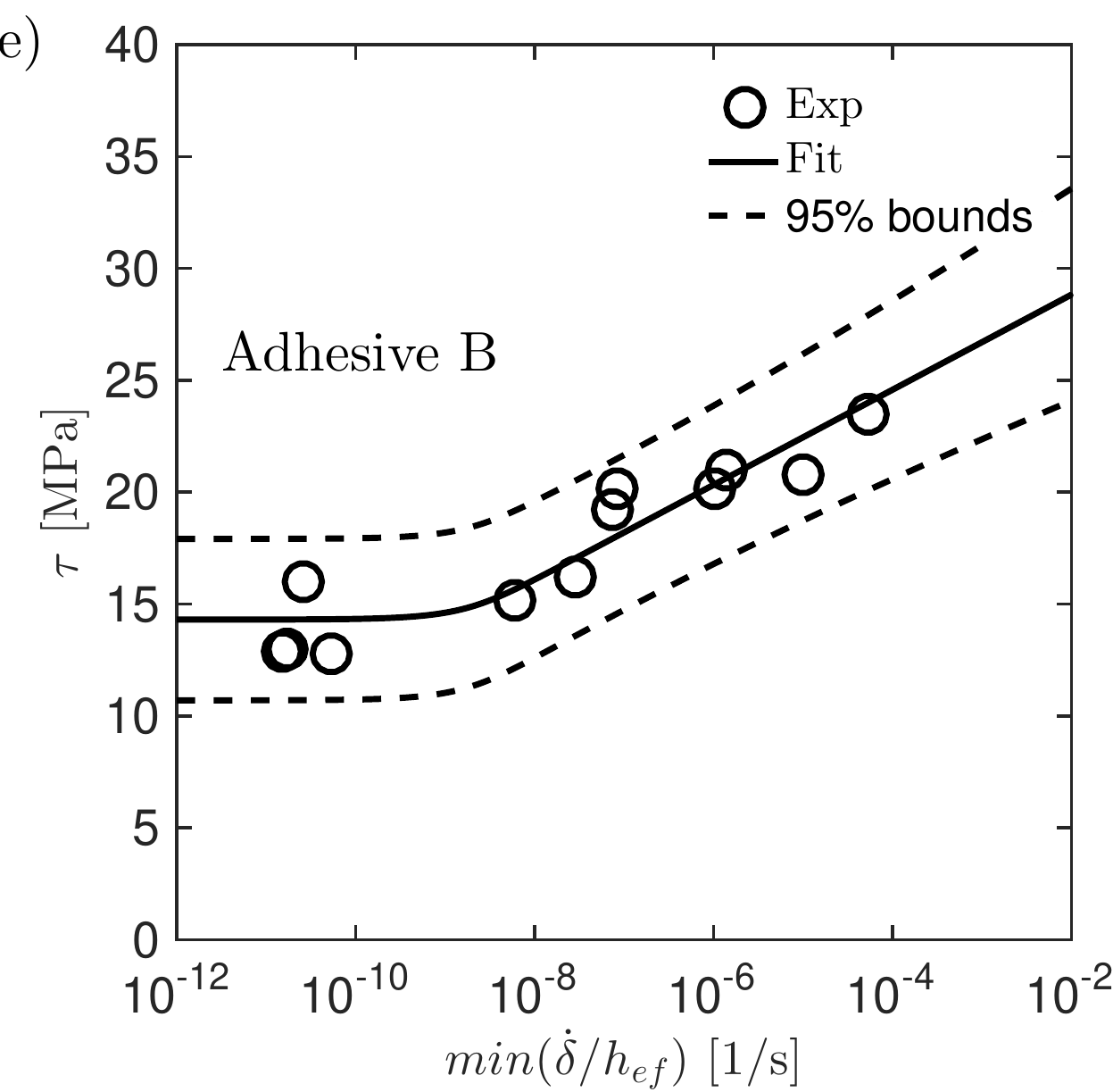}
  \end{subfigure}
  \begin{subfigure}[b]{0.3\textwidth}
%   VELowRates_190315
       \includegraphics[width=1\textwidth]{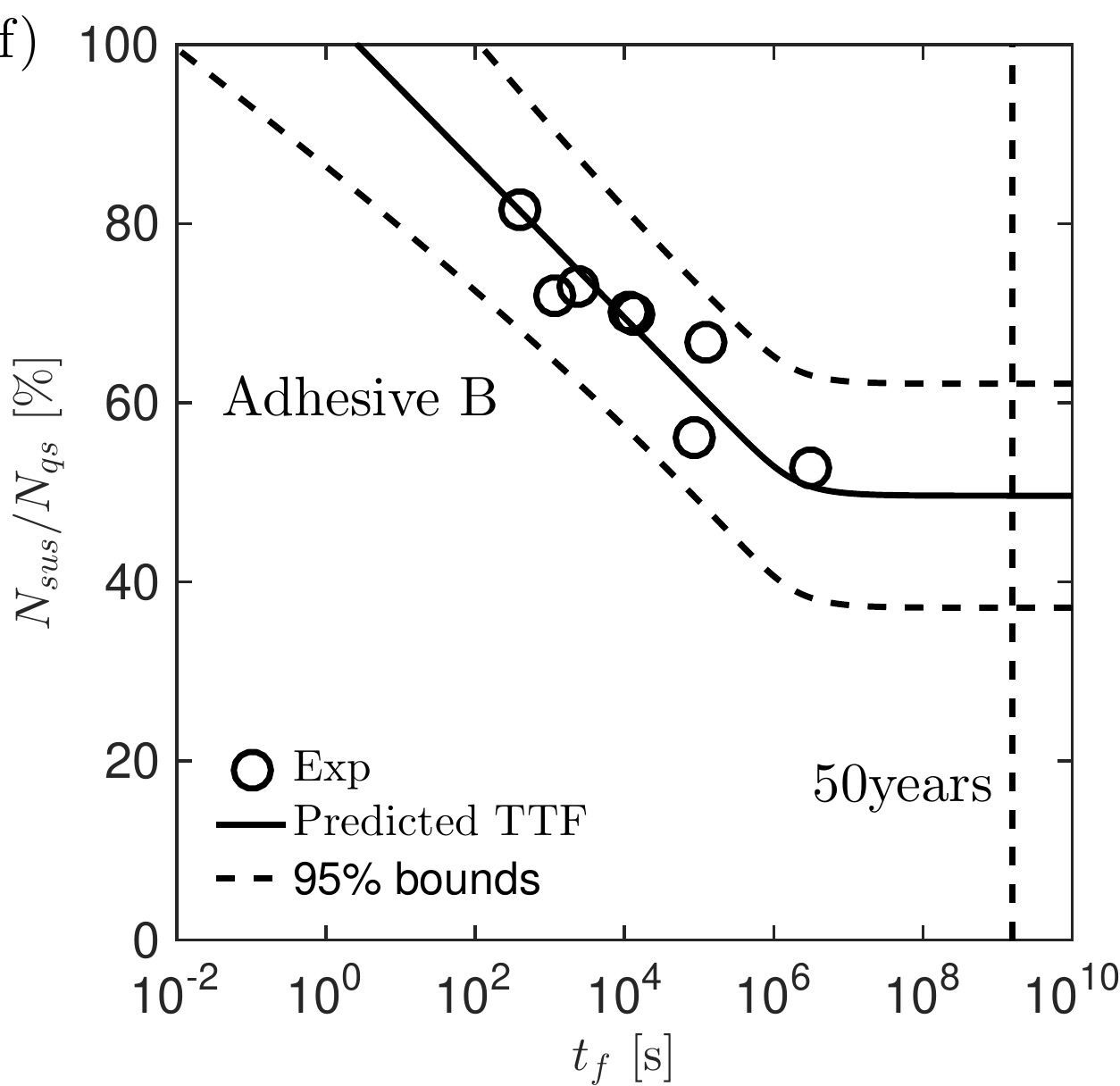}
  \end{subfigure}

  \begin{subfigure}[b]{0.3\textwidth}
       \includegraphics[width=1\textwidth]{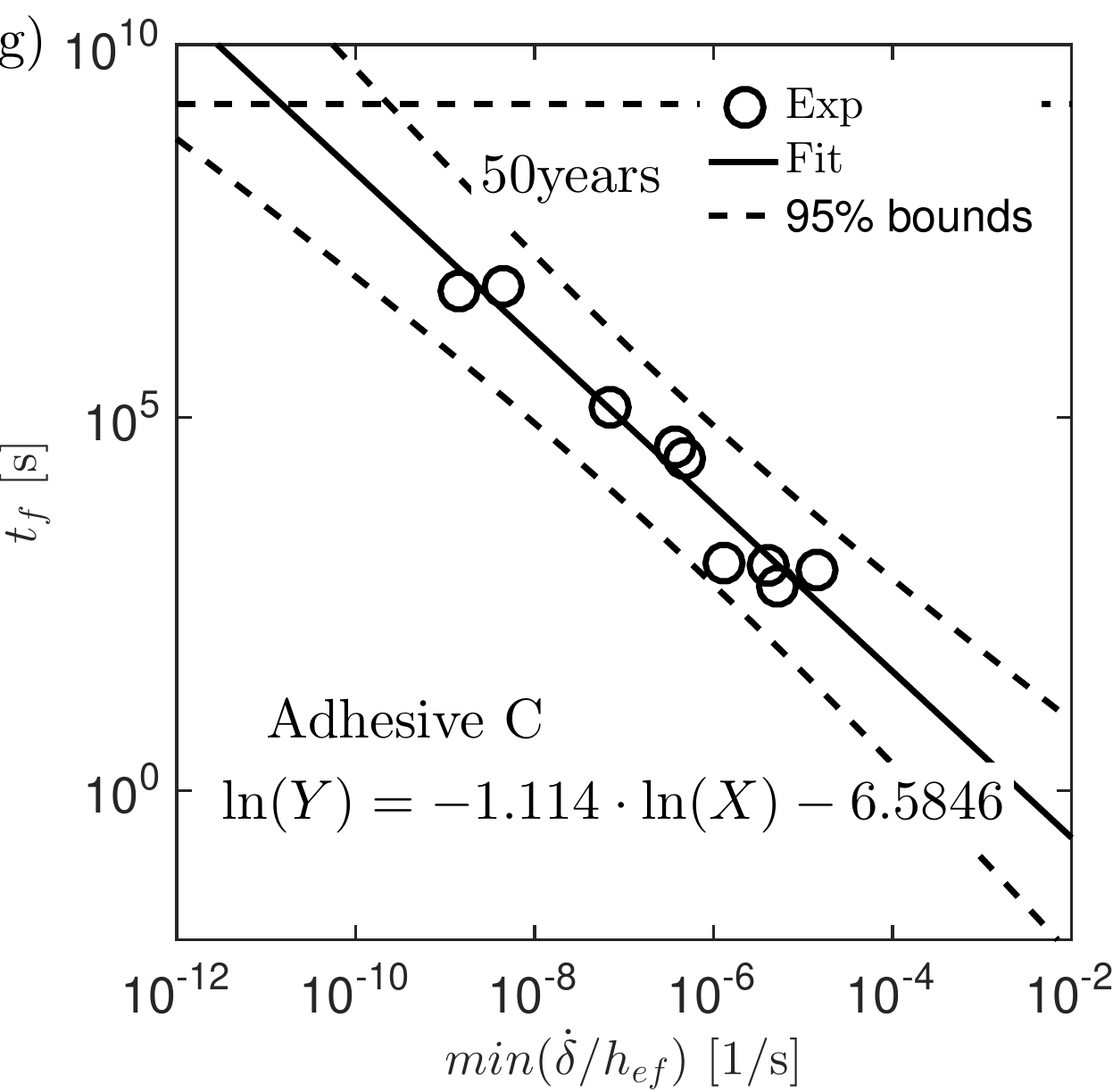}
  \end{subfigure}
    \begin{subfigure}[b]{0.3\textwidth}
%   VELowRates_190315
       \includegraphics[width=1\textwidth]{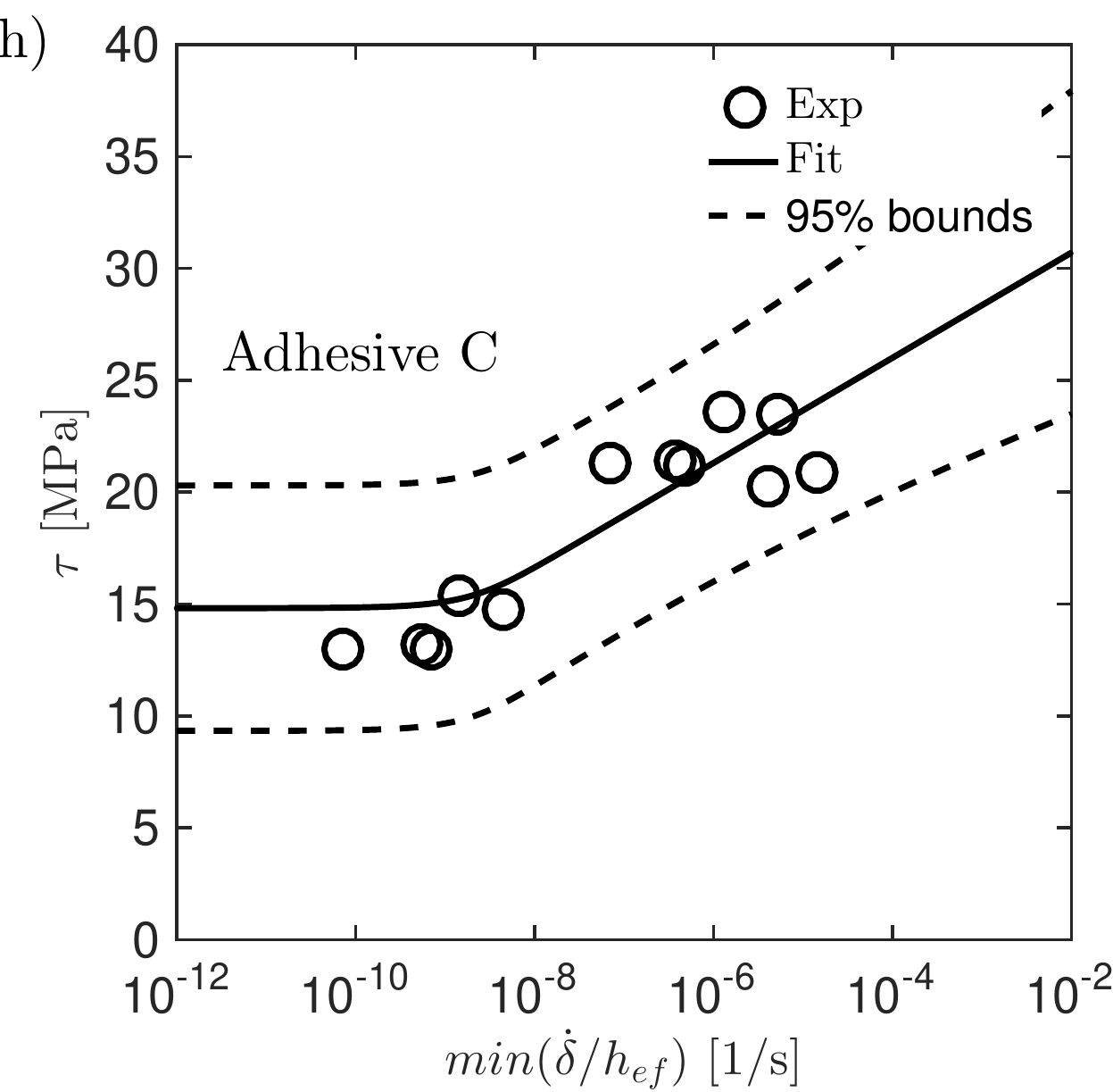}
  \end{subfigure}
  \begin{subfigure}[b]{0.3\textwidth}
%   VELowRates_190315
       \includegraphics[width=1\textwidth]{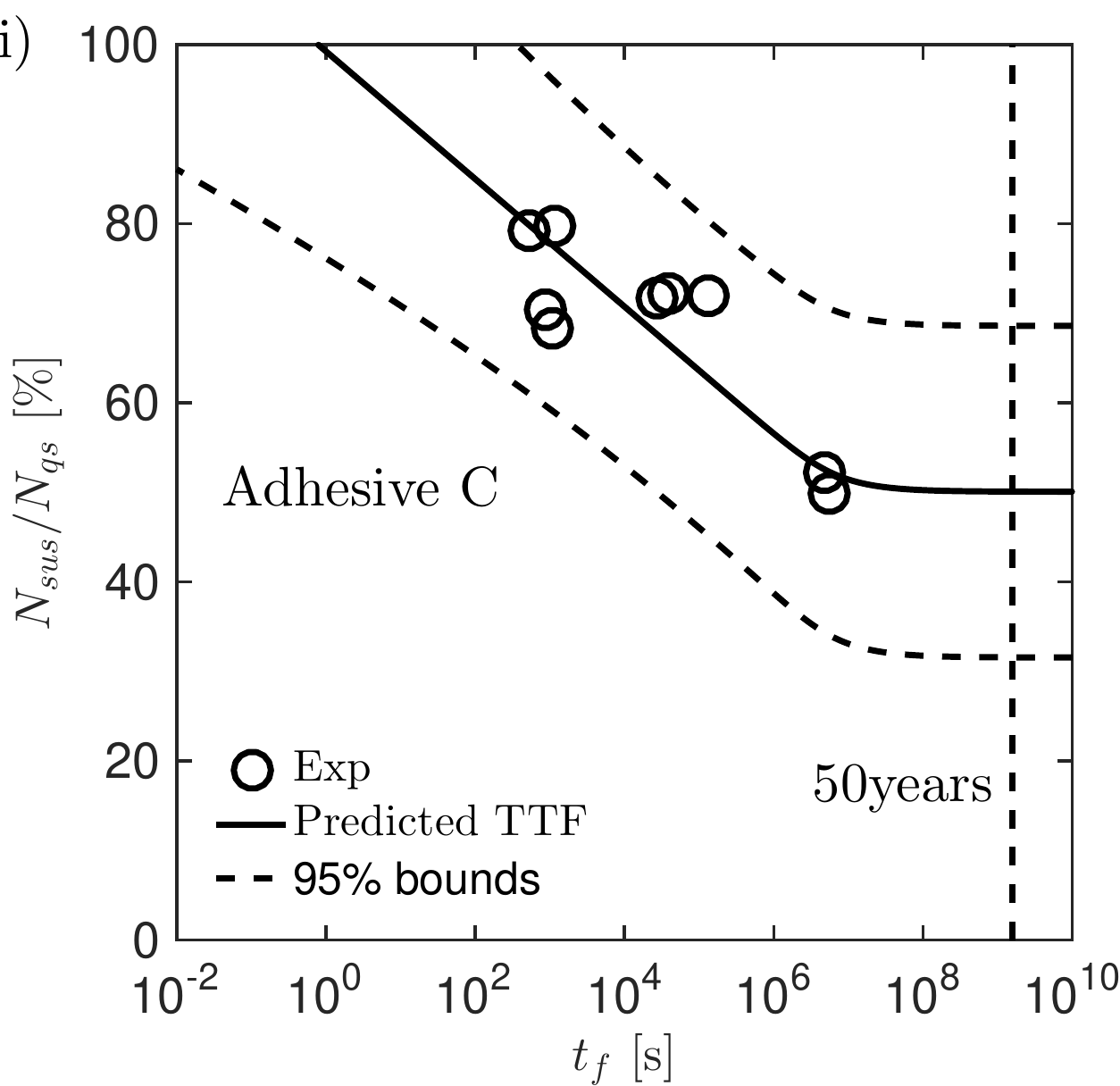}
  \end{subfigure}
\caption{MG fit, stress versus creep rate and reconstructed stress versus time to failure for time to failure data obtained from \cite{cook_nchrp_2013, cook_nchrp_2013A} for adhesive A (a)-(c),  adhesive B (d)-(e), and adhesive C, (g)-(i). } 
\label{fig:CookAB}
\end{figure}

First the same type of analysis, as performed for the P1 and the P2  adhesive anchor systems, is carried out. 
For both adhesive A, B, and C, for the adhesive anchors of $8$mm radius (M16), the Monkman-Grant equation is fitted as is shown in Fig.~ \ref{fig:CookAB}(a), (d) and (g). 
It has to be noted that for adhesive A, a test that failed at later time is not considered in the fit, triangle marker. 
The test is used as verification of the fit of the MG. 
The predicted failure time has a value of $357~$days, while the real failure time was $692$~days. 
Although the relative error is quite high $48\%$, the fit is capable to predict correctly the order of magnitude of the failure time. 

Next the relation between stress and creep rate is fitted, for all adhesives, as it is shown in Fig.~ \ref{fig:CookAB}(b), (e), and (h). 
For this fit also the lower load levels without failure can be used, allowing a better estimation of the stress plateau. 
Note, the stabilized creep rates serve here as a conservative approximation of the actual minimum creep rates that would be smaller.

Finally the stress versus time to failure curve is constructed based on the previous fits, see Fig.~ \ref{fig:CookAB}(c), (f) and (i).    
This approach resulted in sustained load strengths of $43.8\% \pm 35\%$ for adhesive A, $49.6\% \pm 25\%$ for adhesive B, and $50.1\% \pm 37\%$. Although the fit and the prediction uncertainty is high, certainly in the case of adhesive B, the MG criterion is valid also for this three additional adhesives.

\subsection{Temperature and diameter dependence}

At last the potential temperature independence of the MG relation as proposed by Kim et al. \cite{Kim2002} is evaluated.
Three data sets of adhesive anchor system~B \cite{cook_nchrp_2013, cook_nchrp_2013A} are used. 
The only difference of these tests compared to the previously discussed results was the radius of the anchor, $r=6$ mm (M12). 
The specimens of the three data sets were loaded under sustained load, at different load levels, $7$, $8$, and $8$ for adhesive anchor systems A, B and C respectively. 
% IB how many ? 3 ?
% RWW I added them
The temperature during the tests was kept constant at three different levels.  
First a baseline data set was created, with tests at $43^{o}$C. 
Then two additional data sets were tested at temperatures of $49^o$~C and $21^{o}$~C, respectively. 

As expected the lower temperature resulted in only two failed specimens during the test duration of $2118$~days. 
% IB please add the number
% RWW I added it
Of course, this is explained by the fact that temperature and stress have a similar effect on the creep rate, i.e. the higher the temperature or the stress, the higher the creep rates and, thus, the shorter the failure times. 
If the MG relation indeed were temperature independent then the same creep rates would always be associated with the same failure time, regardless of the temperature and applied stress combination causing the observed minimum creep rate.
 
Fig.~\ref{fig:CookB} shows the results of this study. 
The MG relation is fitted on the data set of the baseline temperature at $43^{o}$C. 
Then the error of the prediction for other temperatures is determined. 
For the case of the higher temperature $T=49^{o}$~C the RMSE~$=4.544\cdot 10^5$~s, and the NRMSE~$=0.5563$. 
This can be also seen in Fig.~\ref{fig:CookB}(a), where the points represented by the crosses are in very good agreement with the prediction. 
However, the data from the tests at lower temperature, $T=21.1^{o}$~C exhibit a high prediction error, with RMSE~$=1.365\cdot 10^4$~s, and NRMSE~$=9.4802$, and lie outside of the prediction bounds. 

Then the MG relation is fitted on the higher temperature (T$=49^o$C), see Fig.~\ref{fig:CookB}(b). 
% IB using a different color for the fit is nice but I would not change the markers
% RWW I updated the figures keeping same the markers
The fitted MG relation for this case differs significantly from that of the previous case, see Fig.~\ref{fig:CookB}(a). 
This could be explained by the clustering of the fitted points or result from other unknown experimental differences. 
However, it could be also an indication that the MG relation is not temperature independent for this adhesive anchor system.   
In this case the prediction of the baseline tests at T$=43^o$~C have a RMSE $=1.3028 \cdot 10 ^6$~s and a NRMSE=$0.17 27$, while the data of T=$21^o$~C exhibit again a high error with RMSE $=8421.5$~s and NRMSE=$5.8483$.  

If the data of M12 are compared to the fit of MG on the adhesive B of M16, as it is shown in Fig~\ref{fig:CookB}(c), then all the points are lying inside the $95\%$ bounds of the prediction. 
Of course this is due to the high uncertainty of the MG fit on the M16.
Nevertheless the fitted line is found to have a slope similar to that from the fit on M12 of the same temperature. 
Moreover, the constant term, expressing the MG constant, changes for the different geometries, with the variation equal to $\pm11.6\%$ of the constant obtained by the fit of M16. 
Fig.~ \ref{fig:CookB}(d) shows the three different fits. A first conclusion could be that the different geometries, as in the case of P1, see section \ref{sec:State:Robust}, provide similar fits. 
Furthermore, the temperature independence of the MG seems not to be validated for this case. 
However, for more solid conclusions further tests at more temperature and stress levels are necessary in which all previously encountered influence factors such as loading rate \cite{Nincevic} are properly controlled.

\subsection{Normalization by modified Monkman-Grant relation}

\begin{figure}[h!]
%\vskip -1cm
\begin{subfigure}[b]{0.5\textwidth}
       \includegraphics[width=1\textwidth]{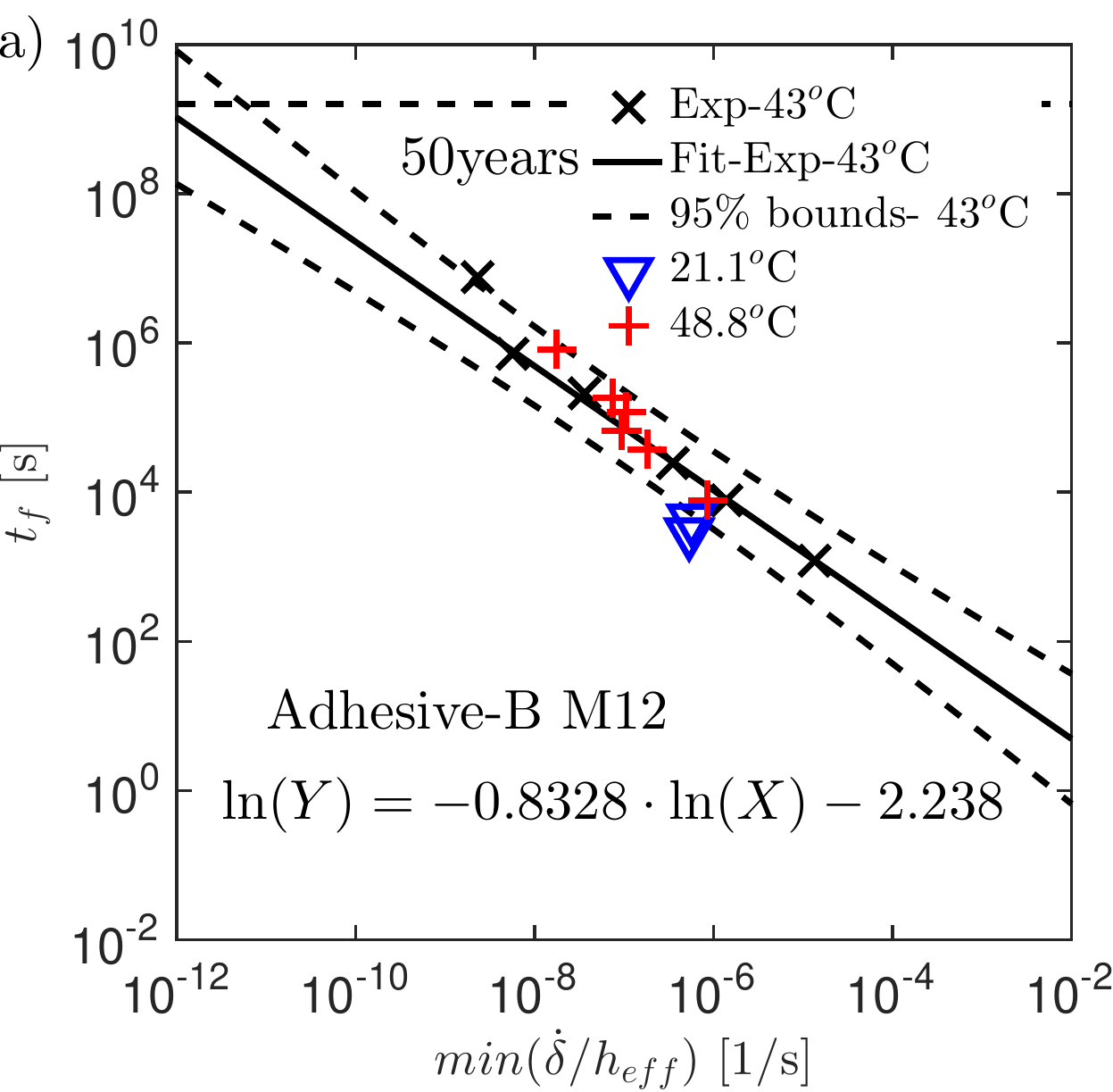}
  \end{subfigure}
%   \begin{subfigure}[b]{0.5\textwidth}
%%   
%      \includegraphics[width=1\textwidth]{./figures/Figxb_190301}
%%       
%  \end{subfigure}
  \begin{subfigure}[b]{0.5\textwidth}
       \includegraphics[width=1\textwidth]{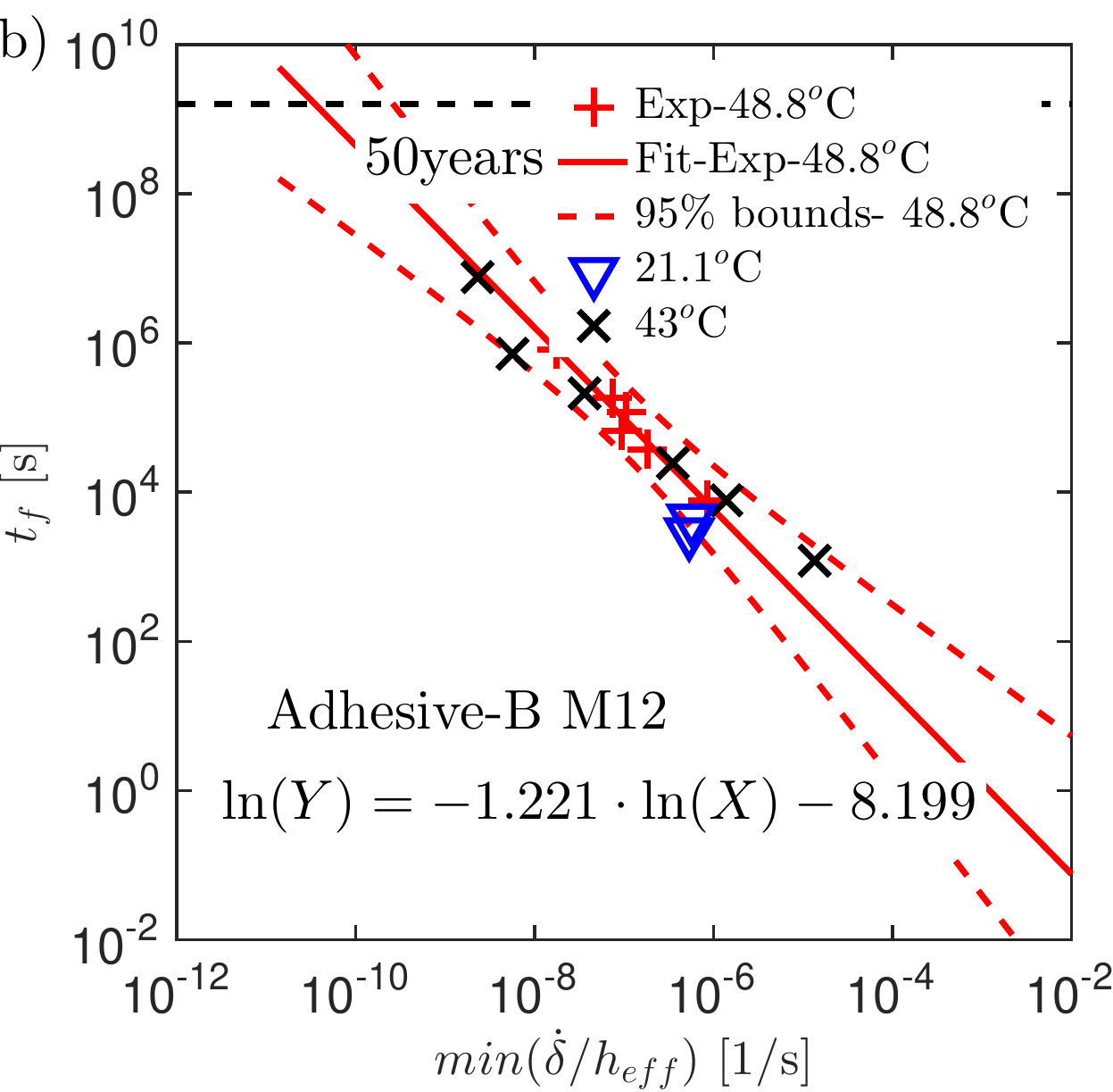}
  \end{subfigure}
  
  \begin{subfigure}[b]{0.5\textwidth}
       \includegraphics[width=1\textwidth]{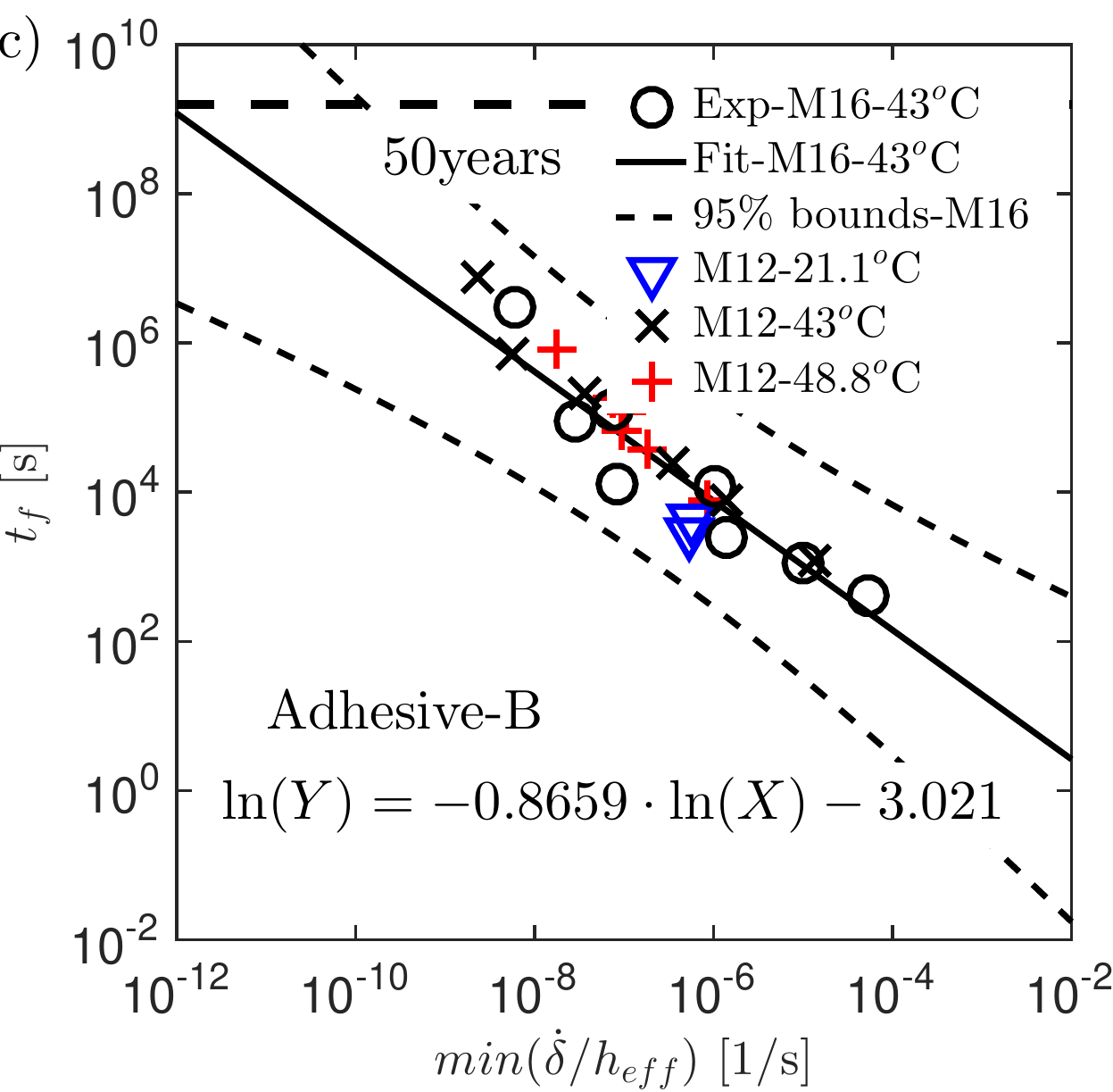}
  \end{subfigure}
%   \begin{subfigure}[b]{0.5\textwidth}
%%   
%      \includegraphics[width=1\textwidth]{./figures/Figxb_190301}
%%       
%  \end{subfigure}
  \begin{subfigure}[b]{0.5\textwidth}
       \includegraphics[width=1\textwidth]{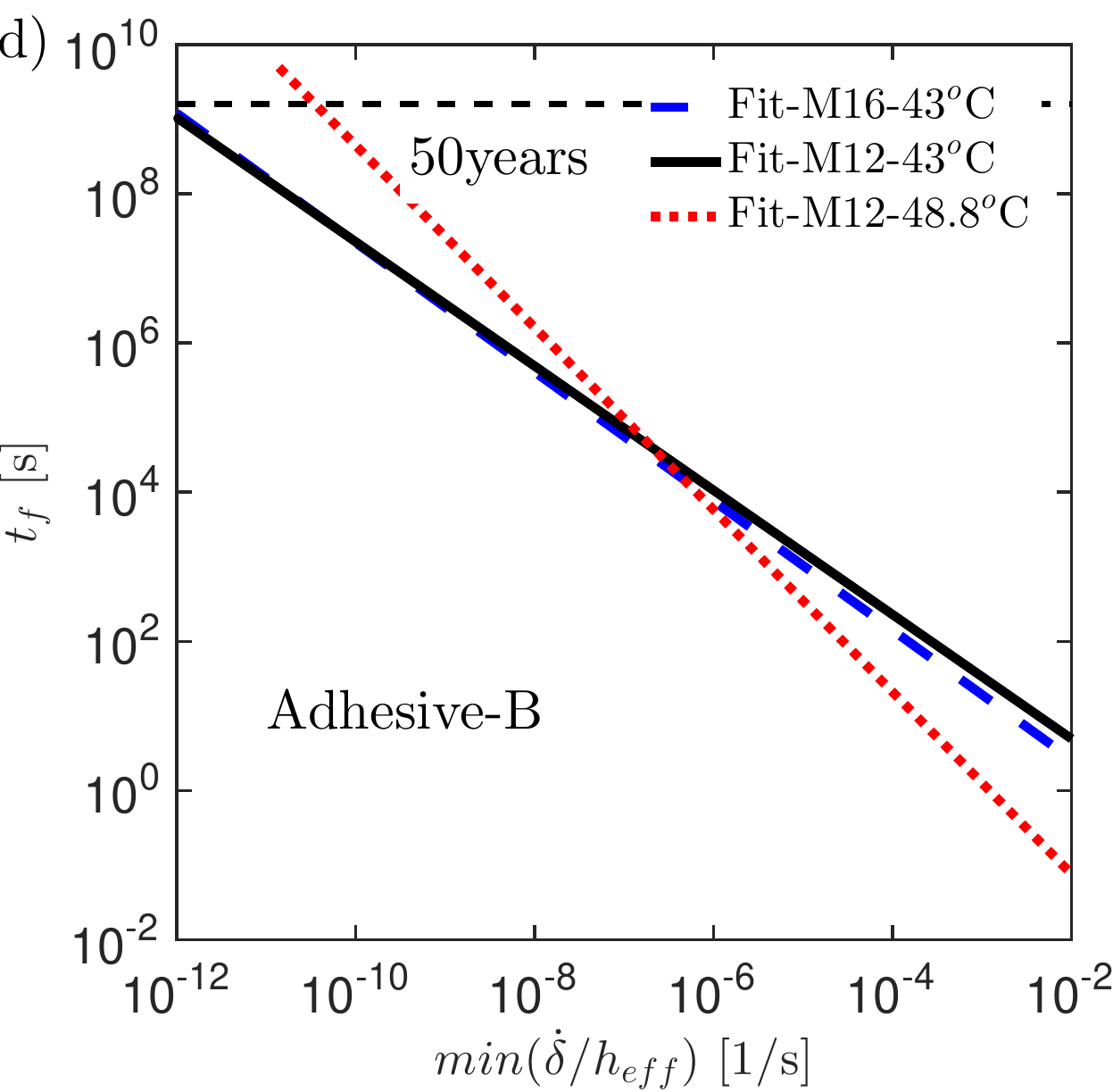}
  \end{subfigure}
\caption{MG fit to different data sets of adhesive~B \cite{cook_nchrp_2013, cook_nchrp_2013A} (a) on M12  tested at $43^o$C, (b) on M12 tested at $48.8^o$C, (c) on M16 tested at $43^o$C, and (d) the comparison of the three fits.}  
\label{fig:CookB}
\end{figure}

\begin{figure}[h!]
%\vskip -1cm
\begin{subfigure}[b]{0.5\textwidth}
       \includegraphics[width=1\textwidth]{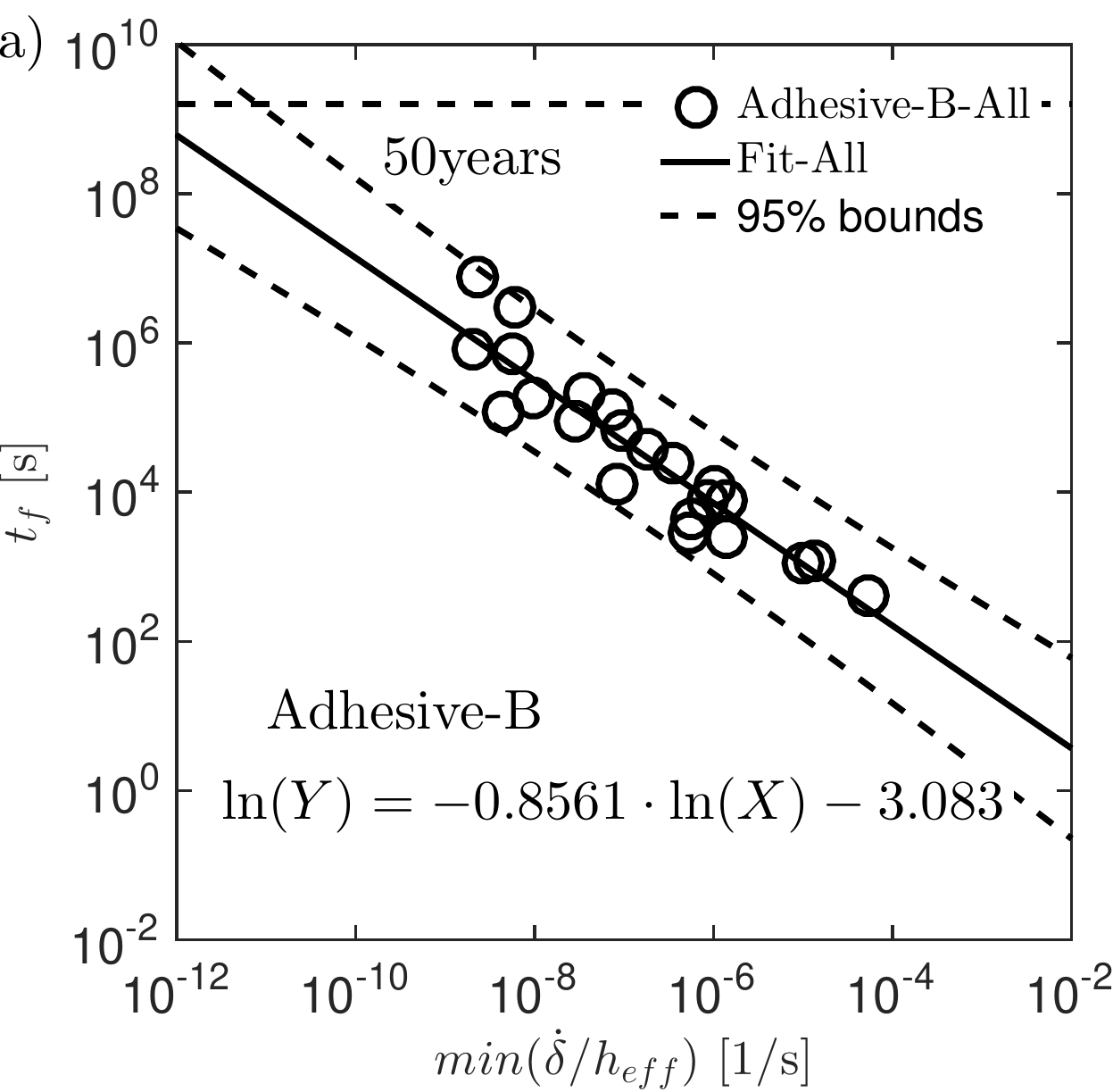}
  \end{subfigure}
%   \begin{subfigure}[b]{0.5\textwidth}
%%   
%      \includegraphics[width=1\textwidth]{./figures/Figxb_190301}
%%       
%  \end{subfigure}
  \begin{subfigure}[b]{0.5\textwidth}
       \includegraphics[width=1\textwidth]{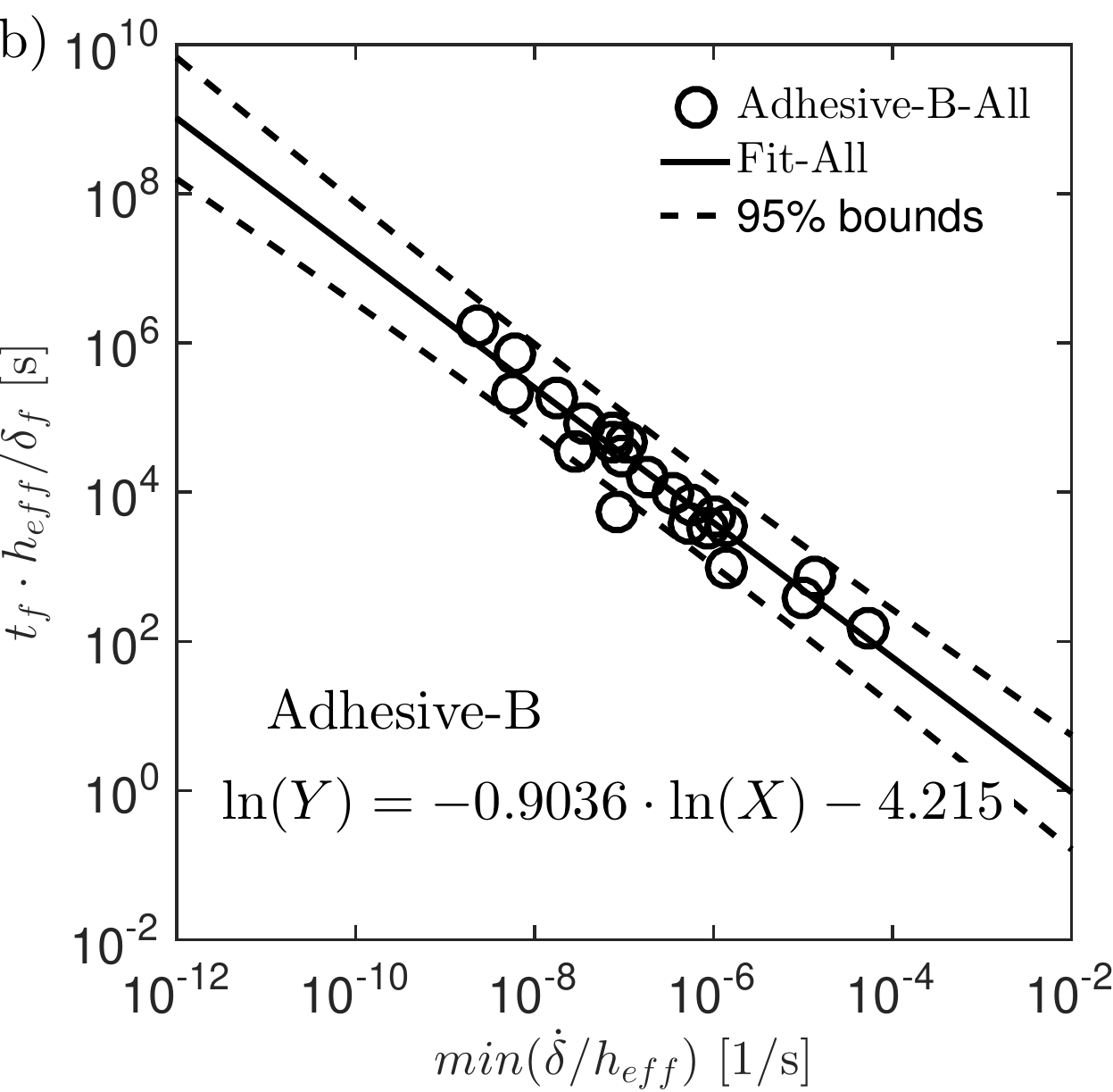}
  \end{subfigure}
\caption{Fit to all data of adhesive B \cite{cook_nchrp_2013, cook_nchrp_2013A} the MG (a) and MMG (b).}  
\label{fig:CookBMG}
\end{figure}

Finally, the data for adhesive B, for all specimens regardless of temperature and geometry are fitted together, as it is shown in Fig~\ref{fig:CookBMG}(a). 
The width of confidence bounds in this case is reduced in comparison to that in Fig~\ref{fig:CookB} (b),(c) but is increased significantly in comparison with the scatter of the fit in Fig~\ref{fig:CookB}(a). 
However the slope and the constant term of the fit do not change significantly, if it is compared to the fit of M16. 
%That could be an indication that the parameters of the MG for the bonded anchor systems are, for a range of geometries, only product dependent.%
Finally, a fit of the MMG relation is performed to all data if adhesive~B, see Fig~\ref{fig:CookBMG} (b). 
It is clear that this normalization is indeed able to remove the differences between the data sets. 
Considering the small remaining scatter, the constant of the modified Monkman-Grant relation could be considered an only product-dependent parameter.
Unfortunately, the MMG relation requires also an estimation of the total rupture displacement which makes it's application more cumbersome as discussed in section~\ref{sec:MG}.
                           

%\begin{figure}[h!]
%%\vskip -1cm
%%\begin{subfigure}[b]{0.3\textwidth}
%%  
%       \includegraphics[width=1\textwidth]{./figures/AdhesiveBTemperature_190322-eps-converted-to.pdf}
%%        
%\caption{Monkmann-Grant criterion fitted on data of baseline temperature and compared to data from two different temperatures, for adhesive B. } 
%\label{fig:CookB}
%\end{figure} 
\section{Conclusions}
This work investigates the applicability of a failure criterion that is frequently used for metals and polymers -- the Monkman-Grant (MG) relation -- to adhesive anchor systems. 
The criterion implies a linear relationship between the logarithm of the minimum creep rate and the logarithm of failure time.
Additionally, two model formulations linking the creep rate to the applied stress are investigated -- a power-law model and a formulation based on the rate-theory of strength.
Both allow the prediction of time to failure curves if combined with the MG relation.

The investigated approaches were successfully applied to time to failure tests on five different adhesives, including tests of two different diameters and three different temperature levels. 
Two of the test series were performed by the authors, while the other three were retrieved from the literature \cite{cook_nchrp_2013A}. 
In all cases the Monkman-Grant relation is found to perform very well.

The main findings can be summarized as follows:
\begin{itemize}
	\item A linear relationship between logarithm of the minimum creep rate and the logarithm of failure time exists. 
	The minimum creep rate can be observed only shortly before failure but can be predicted based on stabilized creep rate measurements.
	
	\item MG fits predict with high accuracy the failure times of adhesive anchors not taken into account during the determination of the MG relation constants. 
	\item A stable determination of MG parameters requires data points spanning at least 4 orders of magnitude in creep rate.
	\item In the investigated cases the modified Monkman-Grant relation was able to predict the failure times of other embedment depths and anchor diameters, tested on the same product.
	\item A power-law can be used to relate applied stress to creep rate reasonably well. 
	The exponent is stress-level dependent, though. 
	Nevertheless, reasonable and probably conservative predictions of time to failure curves can be obtained.
	\item The rate-theory based model linking applied stress to creep rate matches well the experimental data and yields directly a sustained stress level below which no tertiary creep failure is possible. 
	However, experimental validation is difficult to impossible.
	\item Preliminary results indicate a potential temperature independence of the MG relation.
	Conclusive data and further analysis are missing

\end{itemize}

Ultimately the findings of this paper could be used to formulate  alternative design and approval test methods for adhesive anchor systems as function of the required service life time. 
As a matter of fact the proposed method in ways brings together the two existing approaches --  the Findley and the stress versus time to failure methods. 

% IB I don't know if we want to put this there, yet.
% RWW fine for me I didn't remove it yet.
%The bonded anchor systems should be tested under sustained loads selected in such a way that both the linear viscoleastic parameters, and the failure times are measured. Then a lifetime criterion can be used, i.e. the minimum creep rate after $2000$ hours is less than a critical rate that leads to failure. Additionally by combining the MG criterion and a functional form for the stress-creep rate relation a stress versus time to failure curve can be created for the anchor system. Thus the selection of the optimal load level that will ensure a service life of no failure, can be achieved.     

%\end{enumerate}
%\section{Bibliography styles}
%
%There are various bibliography styles available. You can select the style of your choice in the preamble of this document. These styles are Elsevier styles based on standard styles like Harvard and Vancouver. Please use Bib\TeX\ to generate your bibliography and include DOIs whenever available.
%
%Here are two sample references: \cite{Feynman1963118,Dirac1953888}.
\section*{Acknowledgements}

The financial support by the Austrian Federal Ministry of Economy, Family and Youth and the National Foundation for Research, Technology and Development is gratefully acknowledged, as is the additional support by our industrial partners.

\section*{References}

\bibliography{mybibfile}

\end{document}